\pdfoutput=1

\documentclass[11pt]{article}
\usepackage{jheppub}

\usepackage{amsmath}
\usepackage{amssymb}
\usepackage{graphicx,xcolor,slashed}
\usepackage{enumerate}
\usepackage{ifpdf}
\usepackage[english]{babel}
\usepackage{url}
\usepackage{booktabs}

\notoc

\definecolor{darkgreen}{rgb}{0.0, 0.26, 0.15}
\definecolor{darkred}{rgb}{0.65,0.15,0}

\usepackage{float}

\hypersetup{bookmarksnumbered,bookmarksdepth=3}

\allowdisplaybreaks[1]

\DeclareFontFamily{U}{mathx}{\hyphenchar\font45}
\DeclareFontShape{U}{mathx}{m}{n}{
      <5> <6> <7> <8> <9> <10>
      <10.95> <12> <14.4> <17.28> <20.74> <24.88>
      mathx10
      }{}
\DeclareSymbolFont{mathx}{U}{mathx}{m}{n}
\DeclareFontSubstitution{U}{mathx}{m}{n}
\DeclareMathAccent{\widecheck}{0}{mathx}{"71}

\font\tenshuffle=shuffle10 \font\sevenshuffle=shuffle7 \font\fiveshuffle=shuffle7 at 5pt
\def\shuffle{{%
\def\Dshuffle{\mathbin{\hbox{\tenshuffle\char'001}}}%
\def\Sshuffle{\mathbin{\hbox{\sevenshuffle\char'001}}}%
\def\SSshuffle{\mathbin{\hbox{\fiveshuffle\char'001}}}%
\mathchoice{\Dshuffle}{\Dshuffle}{\Sshuffle}{\SSshuffle}}}

\restylefloat{figure}
\definecolor{dgreen}{rgb}{0,0.70,0.30}
\definecolor{gold}{rgb}{0.85,.66,0}
\definecolor{purple}{rgb}{1.0,0.3,0.6}

\usepackage{tikz}
\usetikzlibrary{calc} \usetikzlibrary{patterns} \usetikzlibrary{decorations.pathreplacing} \usetikzlibrary{decorations.markings} \usetikzlibrary{decorations.pathmorphing} \usetikzlibrary{positioning} \usetikzlibrary{bending}

\newcommand{\bea}{\begin{eqnarray}}
\newcommand{\eea}{\end{eqnarray}}

\newcommand{\WAS}{Y}

\let\Re\relax
\let\Im\relax
\DeclareMathOperator{\Re}{Re}
\DeclareMathOperator{\Im}{Im}

\newcommand{\cx}{\mathbb{C}}
\newcommand{\ints}{\mathbb{Z}}
\newcommand{\nn}{\nonumber}
\newcommand{\sv}{\mathrm{sv}}

\newcommand{\dd}{\mathrm{d}}
\newcommand{\te}{\textrm}
\newcommand{\ap}{{\alpha'}}

\newcommand{\mC}{{\cal C}}

\newcommand{\dgcr}{\nabla_{\! \scriptscriptstyle \rm DG}}
\newcommand{\dgcrbar}{\overline{\nabla}_{\! \scriptscriptstyle \rm DG}}

\DeclareMathOperator{\KN}{KN}

\newcommand{\cformS}[1]{\,{\cal C}[\protect\begin{smallmatrix}#1\protect\end{smallmatrix}]}

\newcommand{\cform}[1]{\,{\cal C}\!\left[\protect\begin{smallmatrix}#1\protect\end{smallmatrix}\right]}
\newcommand{\cformtri}[3]{\,{\cal C}\!\left[\protect\begin{smallmatrix}#1\protect\end{smallmatrix}\middle|\protect\begin{smallmatrix}#2\protect\end{smallmatrix}\middle|\protect\begin{smallmatrix}#3\protect\end{smallmatrix}\right]}

\newcommand{\rspoonarrowin}{\raisebox{0.1\height}{\scalebox{.8}{\begin{tikzpicture}[>=latex]
\draw [fill=black] (0.4,0) circle (1.5pt);
\draw[thick,arrows=->] (0,0)--(0.4,0);
\end{tikzpicture}\,}}}
\newcommand{\rspoonarrowout}{\raisebox{0.1\height}{\scalebox{.8}{\begin{tikzpicture}[>=latex]
\draw [fill=black] (0.4,0) circle (1.5pt);
\draw[thick,arrows=->] (0.4,0)--(0,0);
\end{tikzpicture}\,}}}

\newcommand{\RR}{\mathbb R}

\newcommand{\NN}{\mathbb N}
\newcommand{\ZZ}{\mathbb Z}

\newcommand{\QQ}{\mathbb Q}

\usepackage[format=plain,
            labelfont=bf,
            textfont=it]{caption}

\title{All-order differential equations for one-loop closed-string integrals
and modular graph forms}

\author[a]{Jan E.\ Gerken,}
\author[a,b]{Axel Kleinschmidt,}
\author[c]{Oliver Schlotterer}

\affiliation[a]{Max--Planck--Institut f\"ur Gravitationsphysik,
Albert--Einstein--Institut,
DE-14476 Potsdam, Germany}
\affiliation[b]{International Solvay Institutes ULB-Campus Plaine CP231, BE-1050 Brussels, Belgium}
\affiliation[c]{Department of Physics and Astronomy, Uppsala University, SE-75108 Uppsala, Sweden}

\emailAdd{jan.gerken@aei.mpg.de}
\emailAdd{axel.kleinschmidt@aei.mpg.de}
\emailAdd{oliver.schlotterer@physics.uu.se}

\date{\today}

\abstract{We investigate generating functions for the integrals over world-sheet tori appearing in closed-string one-loop
amplitudes of bosonic, heterotic and type-II theories. These closed-string integrals are shown to 
obey homogeneous and linear differential equations in the modular parameter of the torus. 
We spell out the first-order Cauchy--Riemann and second-order Laplace equations for the 
generating functions for any number of external states. The low-energy expansion of such torus integrals 
introduces infinite families of non-holomorphic 
modular forms known as modular graph forms. Our results generate homogeneous first- and second-order 
differential equations for arbitrary such modular graph forms and can be viewed as a step towards all-order 
low-energy expansions of closed-string integrals.}

\preprint{UUITP--45/19}

\begin{document}

\maketitle{}

\newpage

\setcounter{page}{1}
\pagenumbering{roman}

\setcounter{tocdepth}{3}

\tableofcontents

\numberwithin{equation}{section}

 \newpage


\section{Introduction}
\label{sec:intro}

\setcounter{page}{1}
\pagenumbering{arabic}

The low-energy expansion of string amplitudes has become a rewarding subject that carries
valuable input on string dualities and that has opened up fruitful connections with number theory and
particle phenomenology. The key challenge of low-energy expansions resides in the integrals over
punctured world-sheets that are characteristic for string amplitudes and typically performed order-by-order in the inverse string tension $\ap$. The $\ap$-expansion of such string integrals then generates large classes of special 
numbers and functions~--~the periods of the moduli space ${\cal M}_{g,n}$
of $n$-punctured genus-$g$ surfaces.

In particular, one-loop closed-string amplitudes are governed by world-sheets with the topology of
a torus and the associated modular group ${\rm SL}_2(\ZZ)$. The $\ap$-expansion of such torus
integrals introduces a fascinating wealth of non-holomorphic modular forms known as (genus-one) ``modular
graph forms'' which have been studied from a variety of perspectives \cite{Green:1999pv, Green:2008uj, Green:2013bza, DHoker:2015gmr, DHoker:2015sve, Basu:2015ayg, Zerbini:2015rss, DHoker:2015wxz, DHoker:2016mwo, Basu:2016xrt, Basu:2016kli, Basu:2016mmk, DHoker:2016quv, Kleinschmidt:2017ege, DHoker:2017zhq, Basu:2017nhs, Broedel:2018izr, Ahlen:2018wng,Zerbini:2018sox,Gerken:2018zcy, Gerken:2018jrq, DHoker:2019txf, Dorigoni:2019yoq, DHoker:2019xef, DHoker:2019mib, DHoker:2019blr, Basu:2019idd, Zagier:2019eus, Berg:2019jhh}.\footnote{Higher-loop generalizations of modular graph forms have been studied in~\cite{DHoker:2013fcx, DHoker:2014oxd, DHoker:2017pvk,DHoker:2018mys,Basu:2018bde}.} These modular graph forms satisfy an intricate web of differential equations with respect to the modular parameter 
$\tau$ of the torus, namely: first-order Cauchy--Riemann equations relating modular graph forms to
holomorphic Eisenstein series \cite{DHoker:2016mwo, DHoker:2016quv, Broedel:2018izr, Gerken:2018jrq} and (in-)homogeneous Laplace eigenvalue equations \cite{DHoker:2015gmr, Basu:2016xrt, Kleinschmidt:2017ege, Basu:2019idd}. 

As the main result of this work, we derive homogeneous Cauchy--Riemann and Laplace equations for generating
series of $n$-point one-loop closed-string integrals and the associated modular graph forms. In contrast to earlier
approaches in the literature, our results are valid to all orders in $\ap$ and do not pose any restrictions
on the topology of the defining graphs. As part of our construction, we also propose a basis of closed-string 
integrals in one-loop amplitudes of the bosonic, heterotic and type--II theories.

More specifically, we study torus integrals over doubly-periodic Kronecker--Eisenstein series and Koba--Nielsen factors which 
are shown to close under the action of the Maa\ss\ and Laplace operator in $\tau$. These
Kronecker--Eisenstein-type integrals are shown to generate modular graph forms at each order
of their $\ap$-expansion, and they additionally depend on formal variables $\eta_2,\eta_3,\ldots,\eta_n$. 
The expansion of the generating series in the $\eta_k$-variables retrieves the specific torus integrals that enter the massless 
$n$-point one-loop amplitudes of bosonic, heterotic and type--II strings. At each order in $\ap$, the 
differential equations of individual modular graph forms can be extracted from elementary 
operations -- matrix multiplication, differentiation in $\eta_j$ and extracting the coefficients 
of suitable powers in 
$\ap$ (or dimensionless Mandelstam invariants) and $\eta_k$.

We stress that the terminology ``closed-string integrals'' or ``open-string integrals'' in this work 
only refers to the integration over the world-sheet punctures. The resulting modular graph forms 
in the closed-string case are still functions of the modular parameter $\tau$ of the torus world-sheet
and need to be integrated over $\tau$ in the final expressions for one-loop amplitudes. A variety of modular
graph forms have been integrated using the techniques of \cite{Lerche:1987qk, Green:2008uj, DHoker:2015gmr, Basu:2017nhs,DHoker:2019mib, DHoker:2019blr}, and the differential equations in this work are hoped to be instrumental for
integrating arbitrary modular graph forms over $\tau$.

The motivation of this work is two-fold and connects with the analogous expansions 
of one-loop open-string integrals:
\begin{itemize}
\item The $\ap$-expansion of one-loop open-string integrals can be expressed via functions that depend on the modular parameter $\tau$ of the cylinder or M\"obius-strip world-sheet. These functions need to be integrated over $\tau$ to obtain the full one-loop string amplitude and were identified~\cite{Broedel:2014vla, Broedel:2017jdo} as Enriquez' elliptic multiple zeta values (eMZVs) \cite{Enriquez:Emzv}. 
A systematic all-order method to generate the eMZVs in open-string $\alpha'$-expansions 
\cite{Mafra:2019ddf, Mafra:2019xms} is based on generating functions of Kronecker--Eisenstein type, similar to the ones we shall introduce in a closed-string setting.

More specifically, the $\ap$-expansions in \cite{Mafra:2019ddf, Mafra:2019xms} are driven by the open-string 
integrals' homogeneous linear differential equations in $\tau$ and their solutions in terms of iterated integrals over holomorphic Eisenstein series~\cite{Enriquez:Emzv,Broedel:2015hia,Broedel:2019vjc}. Similarly, the new differential equations obtained 
in the present work 
are a first step\footnote{
Closed-string integrals pose additional challenges beyond the open string
in solving their differential equations order-by-order in $\ap$. These challenges stem from the expansion of
modular graph forms around the cusp and the interplay of holomorphic and anti-holomorphic eMZVs, which will be addressed in follow-up work. See Section \ref{outlook} for an initial discussion of this point.}
towards generating the analogous closed-string 
$\ap$-expansions to all orders.
The resulting expressions for modular forms will be built from  
iterated Eisenstein integrals and their complex conjugates.
\item Closed-string tree amplitudes were recently identified as single-valued open-strings trees \cite{Schlotterer:2012ny, Stieberger:2013wea, Stieberger:2014hba, Schlotterer:2018abc, Brown:2018omk,Vanhove:2018elu, Brown:2019wna},
with the single-valued map of \cite{Schnetz:2013hqa, Brown:2013gia} acting at the level of the (motivic) multiple zeta values (MZVs) in the respective $\ap$-expansions. Accordingly, similar relations are expected between one-loop amplitudes involving open and closed strings, and a growing body of evidence and examples has been assembled 
from a variety of perspectives \cite{Zerbini:2015rss, DHoker:2015wxz, Broedel:2018izr, Gerken:2018jrq, DHoker:2019xef, Zagier:2019eus}. 

The first-order differential equations for closed-string integrals in this work turn out to closely resemble their
open-string counterparts \cite{Mafra:2019ddf, Mafra:2019xms}. This adds a crucial facet to the tentative 
relation between closed strings and single-valued open strings at genus one. The resulting connections 
between modular graph forms and eMZVs and a link with the non-holomorphic modular forms of 
Brown \cite{Brown:2017qwo, Brown:2017qwo2} will be discussed in the future~\cite{toappsoon}.
\end{itemize}

In summary, the long-term goal is to obtain a handle on the connection between modular graph forms and iterated integrals in the context of the $\alpha'$-expansion of closed-string one-loop amplitudes. The new results reported in this paper arise from the strategy to study the modular differential equations satisfied by suitable generating functions of modular graph forms. Together with appropriate boundary conditions they will imply representations of amplitudes in terms of iterated integrals.


\subsection{Summary of main results}
\label{sec:int.1}

This section aims to give a more detailed preview of the main results and key equations in
this work. The driving force in our study of new differential equations for modular graph forms
is the matrix of generating series
\begin{align}
  W^\tau_{\vec{\eta}}(\sigma|\rho) = \int \Big( \prod_{j=2}^n \frac{ \dd^2 z_j}{\Im \tau} \Big)  \sigma \big[ \overline{ \varphi^\tau_{\vec{\eta}}(\vec{z}) } \big] \rho\big[ \varphi^\tau_{\vec{\eta}}(\vec{z}) \big]  \exp \Big( \sum_{1\leq i<j}^n s_{ij} G(z_{i}{-}z_{j},\tau) \Big)
  \label{intr1}
\end{align}
of one-loop closed-string integrals. The integration domain for the
punctures $z_2,z_3,\ldots,z_n$ is a torus with modular parameter $\tau$, and translation invariance
has been used to fix $z_1=0$. 
The integrand involves doubly-periodic functions $\varphi^\tau_{\vec{\eta}}(\vec{z})$ of the punctures that 
are built from Kronecker--Eisenstein series and depend meromorphically on $n-1$ bookkeeping variables 
$\eta_2,\eta_3,\ldots,\eta_n$. The rows and columns of $ W^\tau_{\vec{\eta}}(\sigma|\rho)$ 
are indexed by permutations $\rho,\sigma \in {\cal S}_{n-1}$ that
act on the labels $2,3,\ldots,n$ of both the $z_j$ and $\eta_j$. Note in particular that the permutations $\rho$ and
$\sigma$ acting on $\varphi^\tau_{\vec{\eta}}(\vec{z})$ and the complex conjugate 
$\overline{ \varphi^\tau_{\vec{\eta}}(\vec{z})}$ may be chosen independently, so
(\ref{intr1}) defines an $(n{-}1)!\times (n{-}1)!$ matrix of generating integrals. Finally, $G(z,\tau) $ denotes
the standard closed-string Green function on the torus to be reviewed below, and the 
Mandelstam invariants are taken to be dimensionless throughout this work:
\begin{align}
  s_{ij} = -\frac{\ap}{2} k_i \cdot k_j \, , \ \ \ \ \ \
  1\leq i<j\leq n
  \label{intr3}
\end{align}
As will be detailed in Section~\ref{sec:sub23}, the Kronecker--Eisenstein-type integrands
$\varphi^\tau_{\vec{\eta}}(\vec{z})$ will be viewed as Laurent series in the $\eta_j$ variables. The
accompanying coefficient functions from the $\eta_j$-expansion of the Kronecker--Eisenstein series 
are building blocks for correlation functions of massless vertex operators on a torus 
\cite{Dolan:2007eh, Broedel:2014vla, Gerken:2018jrq}. By independently 
expanding (\ref{intr1}) in the $\eta_j$ and $\bar \eta_j$ 
variables, one can flexibly extract the torus integrals in one-loop closed-string amplitudes 
with different contributions from the left- and right movers -- including those 
of the heterotic string, see Appendix~\ref{app:string} for more details.

The component integrals at specific $(\eta_j,\bar \eta_j)$-orders of (\ref{intr1}) in turn generate modular graph
forms upon expansion in $\ap$, i.e.\ in the dimensionless Mandelstam invariants (\ref{intr3}).
The modular graph forms in such $\ap$-expansions have been actively studied in recent years, and their differential
equations in $\tau$ were found to play a crucial role to understand the systematics of their relations \cite{DHoker:2016mwo, DHoker:2016quv}. 
The generating functions (\ref{intr1}) will be used to streamline the differential equations for infinite families of
arbitrary modular graph forms, without any limitations on the graph topology.
As will be demonstrated in Sections~\ref{sec:CR} to~\ref{sec:Lap}, the 
$W_{\vec{\eta}}^\tau$-integrals close under the action of first-order Maa\ss\ operators and the Laplacian.
Like this, one can extract differential equations among the $n$-point component integrals and therefore for
modular graph forms at all orders in $\ap$.


\subsubsection{Open-string integrals and differential equations}
\label{sec:intr1.1.1}

The definition and Cauchy--Riemann equations of the closed-string integrals $W^\tau_{\vec{\eta}}$
are strongly reminiscent of recent open-string analogues \cite{Mafra:2019ddf, Mafra:2019xms}:
The generating functions 
\begin{align}
  Z^\tau_{\vec{\eta}}(\sigma |\rho) = \int_{{\cal C}(\sigma)}\dd z_2\, \dd z_3 \ldots \dd z_n \, \rho\big[ \varphi^\tau_{\vec{\eta}}(\vec{z})  \big]
  \exp \Big(  \sum_{1\leq i<j}^n s_{ij} G_A(z_{i}{-}z_{j},\tau) \Big)
  \label{intr2}
\end{align}
of cylinder integrals in one-loop open-string amplitudes involve the same doubly-periodic 
$\varphi^\tau_{\vec{\eta}}(\vec{z})$ as seen in their closed-string counterpart (\ref{intr1}).
However, instead of the complex conjugate $\overline{\varphi^\tau_{\vec{\eta}}(\vec{z})}$
in the integrand of $W^\tau_{\vec{\eta}}$, the open-string integrals (\ref{intr2}) are characterized by an
integration cycle ${\cal C}(\sigma),\, \sigma \in {\cal S}_{n-1}$ which imposes a cyclic ordering of the punctures
on the world-sheet boundaries. In case of a planar cylinder amplitude, the punctures can be taken to
be on the $A$-cycle of an auxiliary torus\footnote{The $A$-cycle integrals associated with non-planar cylinder diagrams satisfy the same differential
equations (\ref{intr4}) as in the planar case \cite{Mafra:2019ddf, Mafra:2019xms}. The 
details on the non-planar integration cycles and the associated Green function can be found
in the reference, following the standard techniques in \cite{Polchinski:1998rq}.}, that is why we will refer to the open-string 
quantities (\ref{intr2}) as 
``A-cycle integrals''. Accordingly, the open-string Green function $G_A(z_{i}{-}z_{j},\tau) $ is essentially
obtained from the restriction of the closed-string Green function $G(z,\tau)$ in (\ref{intr1}) to 
the $A$-cycle $z \in (0,1)$.

The collection of open-string integrands $\rho [\varphi^\tau_{\vec{\eta}}(\vec{z})]$ in (\ref{intr2})
with permutations $\rho \in {\cal S}_{n-1}$ ensures that the $Z^\tau_{\vec{\eta}}$ close under $\tau$-derivatives
\cite{Mafra:2019ddf, Mafra:2019xms}
\begin{align}
  2\pi i \partial_\tau Z^\tau_{\vec{\eta}}(\sigma|\rho) =  \sum_{\alpha \in {\cal S}_{n-1}} D^\tau_{\vec\eta}(\rho| \alpha)Z_{\vec{\eta}}^\tau(\sigma|\alpha)\, .
  \label{intr4}
\end{align}
The $(n{-}1)!\times (n{-}1)!$ matrix-valued differential operators $D^\tau_{\vec\eta}$ relates different permutations
of the integrands. Its entries are linear in Mandelstam invariants (\ref{intr3}) and comprise
derivatives w.r.t.\ the auxiliary variables $\eta_j$ as well as Weierstra\ss\ functions of the latter.
In fact, the entire $\tau$-dependence in the $\eta_j$-expansion of $ D^\tau_{\vec\eta}$ is carried
by holomorphic Eisenstein series, see Section~\ref{outlook} for further details. That is why (\ref{intr4}) 
manifests the appearance of iterated Eisenstein integrals in the $\ap$-expansion of open-string integrals, 
a canonical representation of eMZVs exposing all their relations over $\QQ$, MZVs and $(2\pi i)^{-1}$ 
\cite{Enriquez:Emzv, Broedel:2015hia}.


\subsubsection{Cauchy--Riemann equations}
\label{sec:intr1.1.2}

The first main result in this work is the closed-string counterpart of the first-order equation (\ref{intr4})
for open-string integrals. The closed-string component integrals in $W^\tau_{\vec{\eta}}$ are non-holomorphic modular forms
whose holomorphic and anti-holomorphic weights depend on the orders in the $(\eta_j,\bar \eta_j)$-expansion. We therefore extend the $\tau$-derivative in (\ref{intr4}) to the Maa\ss{} operator $\nabla_{\vec\eta}^{(k)}=(\tau{-}\bar \tau) \partial_\tau+\dots$, where the connection terms in the ellipsis can be found in~\eqref{eq:MRLn}. Similar to (\ref{intr4}), the 
Maa\ss\ operators close on the $(n{-}1)!$ permutations $\rho [ \varphi^\tau_{\vec{\eta}}(\vec{z}) ]$ in (\ref{intr1})
\begin{align}
  2\pi i \nabla_{\vec\eta}^{(n-1)} W^\tau_{\vec{\eta}}(\sigma|\rho) =   (\tau-\bar\tau) \sum_{\alpha \in {\cal S}_{n-1}} \sv D^\tau_{\vec\eta}(\rho| \alpha)W_{\vec{\eta}}^\tau(\sigma|\alpha) + 2\pi i \sum_{j=2}^n \bar{\eta}_j\partial_{\eta_j} W_{\vec\eta}^\tau(\sigma|\rho) \, .
  \label{intr5}
\end{align}
The closed-string differential operator $\sv D^\tau_{\vec\eta}(\rho| \alpha)$ is obtained from the
$(n{-}1)!\times (n{-}1)!$-matrix $D^\tau_{\vec\eta}(\rho| \alpha)$ in the open-string analogue (\ref{intr4})
by removing the term $\sim\! 2 \zeta_2 \delta_{\rho, \alpha}$ on its diagonal. Hence, the ``sv''-notation
instructs to formally discard the contribution $\sim\! 2\zeta_2 \delta_{\rho, \alpha}$ from
$D^\tau_{\vec\eta}(\rho| \alpha)$ and alludes to the fact that the single-valued map of MZVs 
\cite{Schnetz:2013hqa, Brown:2013gia} annihilates $\zeta_2$.
The explicit expression for $\sv D^\tau_{\vec\eta}(\rho|\alpha)$ is given in~\eqref{eq:svCR} and our analysis proves this form for any number $n$ of points, confirming the conjecture of~\cite{Mafra:2019ddf, Mafra:2019xms}.


\subsubsection{Laplace equations}
\label{sec:intr1.1.3}

As a second main result of this work, we evaluate the Laplacian action on the generating series (\ref{intr1}) of
closed-string integrals. Our representation of the Laplacian $\Delta_{\vec\eta} =  \overline{\nabla}_{\vec\eta}^{(n-2)} \nabla_{\vec\eta}^{(n-1)} +\dots$, with ellipsis specified in \eqref{eq:lapW}, reduces to $- (\tau-\bar \tau)^2 \partial_\tau \bar \partial_\tau$ when acting on modular invariants. The $W_{\vec\eta}^\tau$-integrals
will be shown to close under the Laplacian
\begin{align}
  (2\pi i)^2 \Delta_{\vec\eta} W_{\vec\eta}^\tau (\sigma|\rho)  &= \! \! \sum_{\alpha,\beta\in {\cal S}_{n-1}}  \! \bigg\{
                                                                  2\pi i (\tau-\bar\tau) \Big[ \delta_{\beta,\sigma} \sum_{i=2}^n \eta_i \partial_{\bar\eta_i} \sv D_{\vec\eta}^\tau(\rho| \alpha) + \delta_{\alpha,\rho} \sum_{i=2}^n \bar\eta_i \partial_{\eta_i} \overline{\sv D_{\vec\eta}^\tau}(\sigma|\beta)\Big]\notag \\
                                                                &+ (\tau-\bar\tau)^2 \sv D_{\vec\eta}^\tau(\rho| \alpha) \, \overline{\sv D_{\vec\eta}^\tau}(\sigma|\beta)
                                                                  +\delta_{\alpha,\rho}\delta_{\beta,\sigma} {\cal O}(\eta,\bar \eta,\partial_\eta,  \partial_{\bar\eta})\bigg\} W_{\vec\eta}^\tau (\beta|\alpha)\, ,
                                                                  \label{intr6}
\end{align}
where the $(n{-}1)!\times (n{-}1)!$-matrix $ \sv D_{\vec\eta}^\tau$ is given by (\ref{intr4}) and (\ref{intr5}),
and ${\cal O}(\eta,\bar \eta,\partial_\eta,  \partial_{\bar\eta})$ denotes the following $\tau$-independent
combination of $\eta_j,\partial_{\eta_j}$ and their complex conjugates
\begin{align}
{\cal O}(\eta,\bar \eta,\partial_\eta,  \partial_{\bar\eta}) &=
(2\pi i)^2  (2-n) \Big(n-1+ \sum_{i=2}^n (\eta_i \partial_{\eta_i}+ \bar\eta_i \partial_{\bar\eta_i} ) \Big)\notag\\
& \ \ \ \ + (2\pi i)^2 \sum_{2\leq i<j}^n (\eta_i\bar\eta_j - \eta_j\bar\eta_i)(  \partial_{\eta_j}\partial_{\bar\eta_i} - \partial_{\eta_i}\partial_{\bar\eta_j})   \label{intr94} \\
& \ \ \ \ + 2\pi i (\tau-\bar\tau) \sum_{1\leq i<j}^{n} s_{ij} (\partial_{\eta_j}-\partial_{\eta_i}) (\partial_{\bar\eta_j}-\partial_{\bar\eta_i}) \, .
\notag
\end{align}
We emphasize that the differential equations (\ref{intr5}) and (\ref{intr6}) of $W_{\vec\eta}^\tau$-integrals
uniformly address all orders in $\ap$. The differential equations for arbitrary $n$-point modular graph forms 
in the $\ap$-expansion of $W_{\vec\eta}^\tau$ follow from elementary operations:
\begin{itemize}
\item matrix multiplication and differentiation w.r.t.\ auxiliary parameters
$\eta_j$ in (\ref{intr5}) and (\ref{intr6}) 
\item extracting the $(\eta_j,\bar \eta_j)$-order for the desired component integral from the 
right-hand side of (\ref{intr5}) and (\ref{intr6})
\end{itemize}
%


\subsection{Outline}
\label{sec:int.2}

This work is organized as follows: Section \ref{sec:rev} combines a review of 
background material with the introduction of the generating functions $W^\tau_{\vec{\eta}}$
of closed-string integrals. In Section~\ref{sec:CR}, we set the stage for the differential equations of 
$W^\tau_{\vec{\eta}}$-integrals by introducing the relevant differential operators and illustrating the
strategy by means of two-point examples. Then, Sections~\ref{sec:bigCRn} and~\ref{sec:Lap} are dedicated
to the $n$-point versions of the Cauchy--Riemann and Laplace equations, respectively.
In Section~\ref{outlook}, we comment on the problems and perspectives in uplifting
the differential equations of this work into all-order $\ap$-expansions of the $W^\tau_{\vec{\eta}}$-integrals. 
The concluding Section~\ref{summary} contains a short summary and outlook.
Several appendices provide additional background material, examples or technical aspects
of some of the derivations.

\section{Basics of generating functions for one-loop string integrals}
\label{sec:rev}

In this section, we review the basic properties of one-loop string integrals. The starting point is provided by the doubly-periodic version of a well-known Kronecker--Eisenstein series whose salient features we exhibit. Based on this we then define the generating integrals whose differential equations will be at the heart of our subsequent analysis.

\subsection{Kronecker--Eisenstein series}
\label{sec:KE}

The standard Kronecker--Eisenstein series is defined in terms of the (odd) Jacobi theta function as~\cite{Kronecker,BrownLev}
\begin{align}
  \label{eq:KEF}
  F(z,\eta,\tau) := \frac{\theta'(0,\tau) \theta(z+\eta,\tau)}{\theta(z,\tau)\theta(\eta,\tau)}\,,
\end{align}
where $\tau \in \cx$ lives on the upper half-plane ($\Im\tau>0$) and labels the world-sheet torus $\Sigma_\tau=\cx/(\ints[\tau]+\ints)$ and $z=u\tau +v$ with $u,v\in [0,1)$ is a point on that torus. The parameter $\eta$ can be used for a formal Laurent-series expansions
\begin{align}
  \label{eq:Fgexp}
  F(z,\eta,\tau) = \sum_{a\geq 0} \eta^{a-1} g^{(a)}(z,\tau)
\end{align}
starting with $g^{(0)}(z,\tau)=1$ and $g^{(1)}(z,\tau)=\partial_z \log \theta(z,\tau)$.
The Kronecker--Eisenstein series~\eqref{eq:KEF} is meromorphic in $z$ and $\eta$ due to the holomorphy of the Jacobi theta function.

The function~\eqref{eq:KEF} is not doubly-periodic on the torus $\Sigma_\tau$ but can be completed to a doubly-periodic function
\begin{align}
  \label{eq:Odef}
  \Omega(z,\eta,\tau) := \exp\left( 2\pi i \eta \frac{\Im z}{\Im \tau}\right) F(z,\eta,\tau)
\end{align}
with $\Omega(z+m\tau+n,\eta,\tau) = \Omega(z,\eta,\tau)$ for all $m,n\in\ints$. Expanding this function in $\eta$ defines doubly-periodic but non-holomorphic functions $f^{(a)}$ via 
\begin{align}
  \label{eq:Ofexp}
  \Omega(z,\eta,\tau) = \sum_{a\geq 0} \eta^{a-1} f^{(a)}(z,\tau)\,.
\end{align}
The first two instances are given explicitly by
\begin{align}
  f^{(0)}(z,\tau) &=1 \,,\nn\\
  f^{(1)}(z,\tau) &=\partial_z \log \theta(z,\tau) + 2\pi i \frac{\Im z}{\Im \tau}  = g^{(1)}(z,\tau) + 2\pi i \frac{\Im z}{\Im\tau}\,,
\end{align}
and $f^{(1)}$ is the only function among the $f^{(a)}$ with a pole, $f^{(1)}(z,\tau) = \frac1{z} + \mathcal{O}(z,\bar{z})$, and $f^{(2)}(z,\tau)$ is ill-defined at the origin $z=0$ as its expansion contains a term $\frac{\bar{z}}{z}$. 

For several insertion points $z_i$ we introduce $z_{ij} = z_i-z_j$ and use as well the short-hand $f^{(a)}_{ij} := f^{(a)}(z_{ij},\tau)$. The function $f^{(a)}(z,\tau)$ is even/odd in $z$ for even/odd $a$, such that $f^{(a)}_{ij} = (-1)^a f^{(a)}_{ji}$.

\subsubsection{Derivatives of Kronecker--Eisenstein series}

Since $F(z,\eta,\tau)$ is meromorphic in $z$, the derivative $\partial_{\bar{z}}$ of 
$\Omega$ is easy to evaluate and given by
\begin{align}
  \label{eq:dzbO}
  \partial_{\bar{z}} \Omega(z,\eta,\tau) = -\frac{2\pi i \eta}{\tau-\bar\tau} \Omega(z,\eta,\tau)+\pi\delta^{(2)}(z,\bar{z})\ .
\end{align}
The first contribution stems from the additional phase in~\eqref{eq:Odef} and 
 the $\delta^{(2)}$ contribution is due to the simple pole of $F(z,\eta,\tau)$ at $z=0$.\footnote{Our convention for the delta function on the torus $\Sigma_\tau$ is $\int\!\frac{\dd^2z}{\Im \tau}\, \delta^{(2)}(z,\bar{z}) = \frac{1}{\Im\tau}$ and we also have that $\delta(u)\delta(v)=\Im\tau\,\delta^{(2)}(z,\bar{z})$ so that $\int\!\dd u\,\dd v\, \delta(u)\delta(v)=1$.} Expanding~\eqref{eq:dzbO} in $\eta$ leads to
\begin{align}
\partial_{\bar{z}} f^{(a)}(z,\tau) = -\frac{2\pi i}{\tau-\bar{\tau}} f^{(a-1)}(z,\tau)+\pi\delta_{a,1} \delta^{(2)}(z,\bar z)\, ,\hspace{15mm} a\geq 1 \, .
\label{antipar}
\end{align}

When taking a derivative with respect to $\tau$, the meromorphic Kronecker--Eisenstein series satisfies 
the mixed heat equation~\cite{BrownLev}
\begin{align}
  2\pi i \partial_\tau F(z,\eta,\tau) = \partial_z\partial_\eta F(z,\eta,\tau) 
  \hspace{10mm}\text{($\partial_\tau$ at fixed $z$)\, .}
\end{align}
There are two different forms of the  corresponding equation for the doubly-periodic $\Omega$, keeping either $z$ or $u$ and $v$ fixed,
\begin{subequations}
  \begin{align}
    2\pi i \partial_\tau \Omega(z,\eta,\tau)  &=  \left(\partial_z{+}\partial_{\bar{z}}\right)\partial_\eta\Omega(z,\eta,\tau) - 2\pi i  \frac{\Im z}{\Im \tau}  \partial_z \Omega(z,\eta,\tau) &&\text{($\partial_\tau$ at fixed $z$)}\\
    \label{eq:Omixh}
\!\!    2\pi i \partial_\tau \Omega(u\tau{+}v,\eta,\tau)
&=  \partial_v\partial_\eta \Omega(u\tau{+}v,\eta,\tau) &&\text{($\partial_\tau$ at fixed $u,v$)}\,,  \end{align}
\end{subequations}
where $\partial_v = \partial_z + \partial_{\bar{z}}$. Noting the corollary
\begin{align}
  (\tau-\bar\tau) \partial_{\bar{z}} \partial_\eta \Omega (z,\eta,\tau)= -2\pi i (1+\eta \partial_\eta) \Omega(z,\eta,\tau)
\end{align}
of \eqref{eq:dzbO}, we can derive a third variant of the mixed heat equation from \eqref{eq:Omixh}:
\begin{align}
  \label{eq:Omixh2}
  2\pi i \big((\tau-\bar{\tau})\partial_\tau +1+\eta \partial_\eta\big) \Omega(u\tau+v,\eta,\tau) &= (\tau-\bar{\tau})  \partial_z\partial_\eta \Omega(u\tau+v,\eta,\tau) &&\text{($\partial_\tau$ at fixed $u,v$)}\, .
\end{align}
This form will be used to derive Cauchy--Riemann equations of Koba--Nielsen integrals in Section~\ref{sec:CR}.

\subsubsection{Fay identities}\label{sec:fay-ids}
The Kronecker--Eisenstein series obeys the Fay identity~\cite{Fay,BrownLev}
\begin{align}
  F(z_1,\eta_1,\tau)F(z_2,\eta_2,\tau) = F(z_{1}{-}z_{2},\eta_1,\tau)F(z_{2},\eta_1{+}\eta_2,\tau)+ F(z_2{-}z_{1},\eta_2,\tau)F(z_{1},\eta_1{+}\eta_2,\tau)\, ,\label{eq:24}
\end{align}
and the same identity also holds for the doubly-periodic version:
\begin{align}
  \label{eq:FayO}
  \Omega(z_1, \eta_1, \tau) \Omega(z_2,\eta_2,\tau) &= \Omega(z_{1}{-}z_{2},\eta_1,\tau)\Omega(z_2,\eta_1{+}\eta_2,\tau)  + \Omega(z_{2}{-}z_{1}, \eta_2,\tau)\Omega(z_1,\eta_1{+}\eta_2,\tau)\, .
\end{align}
Expanding \eqref{eq:FayO} in $\eta_{1}$ and $\eta_{2}$ generates relations between products of $f^{(a)}$ functions. For explicit expressions, see Appendix~\ref{app:Om}. Since the Koba--Nielsen integrals we are considering contain products of $f^{(a)}$ functions as integrands, Fay identities will be crucial in the derivation of Cauchy--Riemann equations.

By taking the limit $z_{1}\rightarrow z_{2}$ in the meromorphic~\eqref{eq:24} and then passing to the doubly-periodic version, one obtains
\begin{align}
  \Omega(z,\eta_{1},\tau)\Omega(z,\eta_{2},\tau)&=\Omega(z,\eta_{1}+\eta_{2},\tau)\left(g^{(1)}(\eta_{1},\tau)+g^{(1)}(\eta_{2},\tau)+\frac{\pi}{\Im \tau}(\eta_{1}+\eta_{2})\right)\notag\\
  &\quad-\partial_{z}\Omega(z,\eta_{1}+\eta_{2},\tau)\ .\label{eq:26}
\end{align}
As detailed in Appendix~\ref{app:Om}, this can be used to derive the following identity \cite{Mafra:2019xms}:
\begin{align}
\left( f^{(1)}(z,\tau) \partial_\eta - f^{(2)}(z,\tau)  \right) \Omega(z,\eta,\tau) =\left( \frac12 \partial_\eta^2 - \wp(\eta,\tau)\right) \Omega(z,\eta,\tau)\ ,\label{eq:25}
\end{align}
which is central in the derivation of Cauchy--Riemann equations. Here, $\wp(\eta,\tau)$ is the Weierstra\ss{} function, given by the following lattice sum over $\ZZ^2\setminus\{(0,0)\}$
\begin{align}
  \wp(\eta,\tau) = \frac{1}{\eta^{2}}+\hspace{-1ex}\sum_{(m,n)\neq (0,0)}\left[\frac{1}{(\eta+m\tau+n)^{2}}-\frac{1}{(m\tau+n)^{2}}\right]= \frac{1}{\eta^2}  + \sum_{k=4}^{\infty} (k{-}1) \eta^{k-2} {\rm G}_k(\tau)\, ,\label{eq:28}
\end{align}
and the functions ${\rm G}_k$ for $k\geq 4$ are the usual holomorphic Eisenstein series on the upper half-plane
\begin{align}
  \label{eq:Gk}
  {\rm G}_k (\tau)= \sum_{(m,n)\neq (0,0)} \frac{1}{(m\tau+n)^k}
\end{align}
that are only non-vanishing for even $k$. For $k=2$, \eqref{eq:Gk} is only conditionally convergent and we define  $\mathrm{G}_{2}$ to be the expression obtained from using the Eisenstein summation convention:
\begin{align}
  {\rm G}_2(\tau) = \sum_{n\neq 0} \frac{1}{n^2} + \sum_{m\neq 0} \sum_{n\in\ints} \frac{1}{(m\tau+n)^2}\,.
\end{align}
Since this function is not modular, we define for later reference also $\widehat{\rm G}_2$ as the non-holomorphic but modular function
\begin{align}
  \label{eq:G2hat}
  \widehat{\rm G}_2 (\tau) = \lim_{s\to0} \sum_{(m,n)\neq (0,0)} \frac{1}{(m\tau+n)^2 |m\tau+n|^{s}} = {\rm G}_2(\tau) - \frac{\pi}{\Im\tau}\ .
\end{align}

\subsubsection{Lattice sums and modular transformations}

Many of the properties of the Kronecker--Eisenstein series mentioned in the last sections can be checked conveniently using the formal lattice-sum representation
\begin{align}
  \label{eq:Olatt}
  \Omega(z,\eta,\tau) = \sum_{m,n\in\ints} \frac{e^{2\pi i (mv-nu)}}{m\tau+n+\eta}\,.
\end{align}
Note that the sum over the lattice is unconstrained. The term with $(m,n)=(0,0)$ corresponds to the singular term $\eta^{-1} f^{(0)}= \eta^{-1}$ in~\eqref{eq:Ofexp}. From the lattice form~\eqref{eq:Olatt} and the expansion~\eqref{eq:Ofexp} we see also that formally
\begin{align}
  \label{eq:fa}
  f^{(a)}(z,\tau) = (-1)^{a-1} \sum_{p\neq 0}\, \frac{e^{2\pi i \langle p, z\rangle}}{p^{a}}\,,\hspace{2em}a>0\, ,
\end{align}
where we have introduced the short-hand notation for the torus momentum $p$ and the real-valued pairing between the torus insertion $z$  and momentum $p$
\begin{align}
  \label{eq:not1}
  z=u\tau+v\,,\quad\quad p = m\tau+n\quad\quad\Longrightarrow\quad\quad \langle p,z\rangle = mv-nu=\frac{1}{\tau-\bar{\tau}}\left(p\bar{z}-\bar{p}z\right)\,.
\end{align}
For $a>2$, the sum~\eqref{eq:fa} converges absolutely at the origin $z=0$ and is given by
\begin{align}
  f^{(a)}(0,\tau)=-\mathrm{G}_{a}(\tau)\ .
\end{align}

We shall also encounter the complex conjugate functions
\begin{align}
  \overline{f^{(a)}(z,\tau)} = (-1)^{a-1} \sum_{p\neq 0} \frac{e^{-2\pi i \langle p, z\rangle}}{\bar{p}^{a}} = - \sum_{p \neq 0} \frac{e^{2\pi i \langle p, z\rangle}}{\bar{p}^a}
  \label{eq:facc}
\end{align}
and the following combined functions
\begin{align}
  \label{eq:Cabfunc}
  C^{(a,b)}(z,\tau) := \sum_{p\neq 0}  \frac{e^{2\pi i \langle p,z \rangle}}{p^a \bar{p}^b} \hspace{3em}a,b\geq0,\ a+b>0
\end{align}
which are in fact single-valued elliptic polylogarithms \cite{Ramakrish, DHoker:2015wxz, Broedel:2019tlz}
if both of $a,b$ are nonzero and include the special cases
\begin{align}
  C^{(a,0)}(z,\tau) = (-1)^{a-1} f^{(a)}(z,\tau)\,,\hspace{10mm}
  C^{(0,b)}(z,\tau) = -  \overline{ f^{(b)}(z,\tau) }\,,\hspace{10mm}a,b>0\, .\label{eq:21}
\end{align}
We also note that, using an auxiliary intermediate point $z_0$, we can write 
\begin{align}
  C^{(a,b)}_{ij} = (-1)^{a} \int \frac{\dd^2z_0}{\Im \tau} f^{(a)}_{i0} \overline{ f^{(b)}_{0j} } \,,\label{eq:1}
\end{align}
where we have used the same short-hand for $C^{(a,b)}_{ij} := C^{(a,b)}(z_i-z_j,\tau)$ and $\overline{f^{(a)}_{ij}} := \overline{f^{(a)}(z_{ij},\tau)}$ as for $f^{(a)}$. The integral is over the torus with a measure normalized to yield
$\int \frac{\dd^2z}{\Im \tau} = 1$. Note that the function $C^{(a,b)}(z,\tau)$ is even/odd in $z$ for even/odd $a{+}b$, i.e.\ $C^{(a,b)}_{ij} = (-1)^{a+b} C^{(a,b)}_{ji}$.

For later reference, we also define
\begin{align}
  C^{(0,0)}(z,\tau):=\Im\tau\,\delta^{(2)}(z,\bar{z})-1\ ,\label{eq:30}
\end{align}
which formally lines up with \eqref{eq:Cabfunc} but is incompatible with \eqref{eq:21} and \eqref{eq:1}.

The lattice-sum representation \eqref{eq:Olatt} of the Kronecker--Eisenstein series also manifests its properties
\begin{align}
  \Omega\left( \frac{z}{\gamma\tau+\delta} , \frac{\eta}{\gamma\tau+\delta} , \frac{\alpha\tau+\beta}{\gamma\tau+\delta} \right) = (\gamma\tau+\delta) \Omega(z,\eta,\tau)\, ,
  \hspace{5mm} \begin{pmatrix}\alpha&\beta\\\gamma&\delta\end{pmatrix}\in {\rm SL}_{2}(\ZZ)
  \label{eq:38}
\end{align}
under the modular group ${\rm SL}_{2}(\ZZ)$. Similarly, \eqref{eq:fa} and~\eqref{eq:Cabfunc} manifest the modular properties of the non-holomorphic functions $f^{(a)}$ and $C^{(a,b)}$
\begin{subequations}
\begin{align}
  f^{(a)} \left( \frac{z}{\gamma\tau+\delta} , \frac{\alpha\tau+\beta}{\gamma\tau+\delta}\right) &= (\gamma\tau+\delta)^{a} f^{(a)}(z,\tau)\ ,\\
  C^{(a,b)} \left( \frac{z}{\gamma\tau+\delta} , \frac{\alpha\tau+\beta}{\gamma\tau+\delta}\right) &= (\gamma\tau+\delta)^{a}(\gamma\bar\tau+\delta)^{b} C^{(a,b)}(z,\tau)\ .
\end{align}
\end{subequations}
Hence, $f^{(a)}$ transforms like a Jacobi form of vanishing index and weight $a$.

\subsection{Koba--Nielsen factor}
\label{sec:KNfactor}

A central ingredient in the study of closed-string amplitudes is the $n$-point Koba--Nielsen factor~\cite{Green:1987mn}
\begin{align}
  \label{eq:KNn}
  \KN^{\tau}_{n} := \prod_{1\leq i<j}^n \exp \left( s_{ij} G(z_{ij}, \tau) \right)\,,
\end{align}
where
\begin{align}
  G(z,\tau) = -\log \left| \frac{\theta(z,\tau)}{\eta(\tau)}\right|^2  + \frac{2\pi(\Im z)^2}{\Im \tau}
  \label{defcGF}
\end{align}
is the real scalar Green function on the world-sheet torus and  the dimensionless Mandelstam variables $s_{ij}$ for massless particles were defined in (\ref{intr3}).\footnote{We note that we are \textit{not} using momentum conservation at this stage and the variables $s_{ij}$ are symmetric in $i$ and $j$ but otherwise unconstrained.} The Green function can also be written as the (conditionally convergent) lattice sum~\cite{Green:1999pv}
\begin{align}
  \label{eq:Glatt}
  G(z,\tau) =  \frac{\Im\tau}{\pi} \sum_{p\neq 0} \frac{e^{2\pi i \langle p,z \rangle}}{|p|^2}\,,
\end{align}
using the notation~\eqref{eq:not1}. Hence, the Green function is a special case
\begin{align}
  G(z,\tau)= \frac{\Im\tau}{\pi}  C^{(1,1)}(z,\tau)\label{eq:22}
\end{align}
of the more general lattice sums 
in (\ref{eq:Cabfunc}) and related to the functions $f^{(1)}$ and $\overline{f^{(1)} }$ by 
\begin{align}
  \label{eq:dzG}
  f^{(1)}(z,\tau) =-\partial_z G(z,\tau)\,,\quad\quad \overline{f^{(1)}(z,\tau)} =-\partial_{\bar{z}} G(z,\tau)\, .
\end{align}
Moreover, its $\tau$-derivative satisfies
\begin{align}
  \label{eq:dtauG}
  2\pi i \partial_\tau G(u\tau+v,\tau)
  =   \sum_{p\neq 0} \frac{e^{2\pi i \langle p, z\rangle}}{p^2} = -  f^{(2)}(z,\tau)
  \ \ \ \ \ \
    \text{($\partial_\tau$ at fixed $u,v$)} \,.
\end{align}
The derivatives of the Koba--Nielsen factor following from~\eqref{eq:dzG} and~\eqref{eq:dtauG} are
\begin{subequations}
  \begin{align}
    \label{eq:dzKN}
    \partial_{z_j} \KN^{\tau}_{n} &= \sum_{i\neq j}  s_{ij} f^{(1)}(z_{ij},\tau)  \KN^{\tau}_{n}\,,\\
    \label{eq:dtauKN}
    2\pi i\partial_\tau  \KN^{\tau}_{n} &= -  \sum_{1\leq i<j}^n s_{ij} f^{(2)}(z_{ij},\tau) \KN^{\tau}_{n} \ \ \ \ \ \
    \text{($\partial_\tau$ at fixed $u_j,v_j$)}\,.
  \end{align}
\end{subequations}
For the $z$-derivative we have used the anti-symmetry of $f^{(1)}(z,\tau)$ in its argument $z$. We note that~\eqref{eq:dzKN} implies for any $1\leq k\leq n$ that
\begin{align}
  \label{eq:dzKN2}
  \left(\partial_{z_k} +\partial_{z_{k+1}} + \ldots + \partial_{z_n}\right) \KN^{\tau}_{n} = \sum_{j=k}^n \partial_{z_j} \KN^{\tau}_{n} = \sum_{i=1}^{k-1}\sum_{j=k}^n  s_{ij} f_{ij}^{(1)}\, \KN^{\tau}_{n}\,.
\end{align}


\subsection{Introducing generating functions for world-sheet integrals}
\label{sec:sub23}

The main results of this work concern the following $(n-1)!\times (n-1)!$ matrix of integrals over the punctures
\begin{align}
  \label{eq:Wdef}
  W_{\vec{\eta}}^\tau(\sigma|\rho) &:=
                                     W_{\vec{\eta}}^\tau (1, \sigma(2,\ldots, n) | 1,\rho(2,\ldots,n)) := \int  \Big( \prod_{j=2}^n\frac{\dd^2z_j}{\Im \tau}\Big) \KN^{\tau}_{n}\\
                                   &\hspace{4mm}\times 
                                     \rho\Big[ \Omega(z_{12} , \eta_{23\ldots n}, \tau) \,\Omega(z_{23},\eta_{34\ldots n}, \tau)\cdots \Omega(z_{n-2,n-1}, \eta_{n-1,n},\tau)\, \Omega(z_{n-1,n} ,\eta_n,\tau )\Big]\nn\\
                                   &\hspace{4mm}\times \sigma\Big[ \overline{\Omega(z_{12} , \eta_{23\ldots n}, \tau)} \,\overline{\Omega(z_{23},\eta_{34\ldots n}, \tau)} \cdots \overline{\Omega(z_{n-2,n-1}, \eta_{n-1,n},\tau) }\,\overline{\Omega(z_{n-1,n} ,\eta_n,\tau)} \Big]\,,\nn
\end{align}
which is defined by the Koba--Nielsen factor~\eqref{eq:KNn} and the doubly-periodic Kronecker--Eisenstein 
series (\ref{eq:Odef}). The matrix elements of $W_{\vec{\eta}}^\tau(\sigma|\rho)$ are parametrized by 
permutations $\rho,\sigma\in \mathcal{S}_{n-1}$ that act separately on the $\Omega(\ldots)$ 
and $\overline{\Omega(\ldots)}$. The parameters of the Kronecker--Eisenstein series are
\begin{align}
  \label{eq:etasum}
  \eta_{i,i+1\ldots n} = \eta_i + \eta_{i+1} +\ldots +\eta_n\,.
\end{align}
The permutations $\rho$ and $\sigma$ act on the points $z_i$ and parameters $\eta_i$ by 
permutation of the indices. The integrals~\eqref{eq:Wdef} are only over $n-1$ points 
$z_j = u_j \tau + v_j$ with $\int \frac{\dd^2z_j}{\Im \tau}=\int^1_0 \dd u_j \int^1_0 \dd v_j$
since translation invariance on the torus has been fixed to put $z_1=0$. 

Due to the phase $\exp(2\pi i \eta \frac{ \Im z}{\Im \tau} )$ in its definition (\ref{eq:Odef}), the doubly-periodic
Kronecker--Eisenstein series is not meromorphic in $z$ or $\tau$. Accordingly we will refer to
$\Omega(z,\eta,\tau)$ and $\overline{\Omega(z,\eta,\tau)}$ as {\it chiral} and {\it anti-chiral}, respectively.
Still, $\Omega(z,\eta,\tau)$ is a meromorphic function of its second argument $\eta$.

The Kronecker--Eisenstein series in the definition (\ref{eq:Wdef}) of $W$-integrals
specify the placeholders $ \varphi^\tau_{\vec{\eta}}(\vec{z})$ and $\overline{ \varphi^\tau_{\vec{\eta}}(\vec{z}) }$
in the schematic formula (\ref{sec:int.1}) in the introduction. Accordingly, the open-string analogues
(\ref{intr2}) of the $W$-integrals are given by \cite{Mafra:2019ddf, Mafra:2019xms}
\begin{align}
  Z^\tau_{\vec{\eta}}&(\sigma |\rho) 
  := \int_{{\cal C}(\sigma)}\dd z_2\, \dd z_3 \ldots \dd z_n \,
                       \exp \Big(  \sum_{1\leq i<j}^n s_{ij} G_{A}(z_{i}{-}z_{j},\tau) \Big)
                       \label{again2} \\
                     &\times  \rho\big[
                       \Omega(z_{12} , \eta_{23\ldots n}, \tau) \,\Omega(z_{23},\eta_{34\ldots n}, \tau)\cdots \Omega(z_{n-2,n-1}, \eta_{n-1,n},\tau)\, \Omega(z_{n-1,n} ,\eta_n,\tau ) \big] \, , \notag
\end{align}
with $\sigma,\rho \in {\cal S}_{n-1}$, and the planar open-string Green function $G_{A}$ on the $A$-cycle reads
\begin{align}
  G_{A}(z,\tau) = -\log \left( \frac{\theta(z,\tau)}{\eta(\tau)}\right)  + \frac{i \pi \tau}{6} + \frac{ i \pi }{2} \, .
  \label{defoGF}
\end{align}
The integration domain ${\cal C}(\sigma)$ prescribes the cyclic ordering $0=z_1{<}z_{\sigma(2)}
{<}z_{\sigma(3)}{<}\ldots{<}z_{\sigma(n)}{<}1$ of the punctures on the $A$-cycle of a torus, that
is why the $Z^\tau_{\vec{\eta}}(\sigma |\rho) $ will be referred to as $A$-cycle integrals henceforth. 
With this restriction to $z \in \RR$, the open-string Green function (\ref{defoGF}) shares the holomorphic 
derivative $\partial_z G_{A}(z,\tau)=\partial_z G(z,\tau)=- f^{(1)}(z,\tau)$ 
of its closed-string counterparts (\ref{defcGF}), and the addition of $\frac{i \pi \tau}{6} {+} \frac{ i \pi }{2}$
enforces that $\int^1_0 G_{A}(z,\tau) \, \dd z =0$ \cite{Broedel:2018izr, Zerbini:2018hgs}.
The above choice of ${\cal C}(\sigma)$ allows to generate cylinder- and M\"obius-strip contributions 
to planar one-loop open-string amplitudes from (\ref{again2}) by restricting $\tau \in i \RR_+$ and 
$\tau \in \frac{1}{2}+ i\RR_+$, respectively \cite{Green:1984ed}. Generalization to non-planar $A$-cycle integrals 
can be found in \cite{Mafra:2019ddf, Mafra:2019xms}.


\subsubsection{Component integrals and string amplitudes}
\label{sec:cptints}

The $W$-integrals in (\ref{eq:Wdef}) are engineered to generate the integrals over torus punctures
in closed-string one-loop amplitudes upon expansion in the $\eta_j$ and $\bar \eta_j$ variables.
The expansion (\ref{eq:Ofexp}) of the doubly-periodic Kronecker--Eisenstein integrands introduces
component integrals
\begin{align}
  W_{(A|B)}^\tau(\sigma|\rho) &:= W_{(a_2,a_3,\ldots,a_n|b_2,b_3,\ldots,b_n)}^\tau(\sigma|\rho) \notag \\
                              &:=\int  \dd\mu_{n-1}
                              \KN^{\tau}_{n} \rho\big[ f_{12}^{(a_2)}  \, f_{23}^{(a_3)}\ldots  f_{n-1,n}^{(a_n)} \big]  
                                \, \sigma \big[ \overline{ f_{12}^{(b_2)} } \, \overline{ f_{23}^{(b_3)}}\ldots  \overline{ f_{n-1,n}^{(b_n)} } \big]  \label{cptnpt1}
\end{align}
with $a_i,b_i\geq 0$ and where we have introduced the short-hand
\begin{align}
\label{eq:dmu}
\int \dd\mu_{n-1} = \prod_{k=2}^n \int \frac{\dd^2z_k}{\Im\tau}
\end{align}
for the integral over the $n-1$ punctures. Note that the $z_{ij}$ arguments of the $f_{ij}^{(a_{k})}$ with weights from the first index set $A=a_2,a_3,\ldots,a_n$ are permuted with the permutation $\rho$ in the second slot of the argument of $W_{(A|B)}^\tau$ and vice versa. This is to ensure consistency with the notation \eqref{again2} of open-string integrals and also to have $W^\tau_{(A|B)}$ carry modular weight $(|A|,|B|)$, where
\begin{align}
  |A| = \sum_{i=2}^{n}a_{i} \,, \ \ \ \ \ \  |B| = \sum_{i=2}^{n}b_{i}\ .\label{eq:2}
\end{align}
The component integral $W_{(A|B)}^\tau(\sigma|\rho)$ can be extracted from its generating series $W_{\vec{\eta}}^\tau(\sigma|\rho)$ by isolating the coefficients of the parameters (\ref{eq:etasum})
\begin{align}
  W_{\vec{\eta}}^\tau(\sigma|\rho) &=  \sum_{A,B}  
                                     W_{(A|B)}^\tau(\sigma|\rho)\, \rho\big[ \eta_{234\ldots n}^{a_2-1} \eta_{34\ldots n}^{a_3-1} \ldots \eta_{n}^{a_n-1} \big]    
                                     \, \sigma\big[ \bar \eta_{234\ldots n}^{b_2-1} \bar \eta_{34\ldots n}^{b_3-1} \ldots \bar \eta_{n}^{b_n-1}  \big]  \, . \label{cptnp2} 
\end{align}
Here and in the rest of this work, we use the abbreviating notation
\begin{align}
  \sum_{A,B}=\sum_{a_{2},a_{3},\ldots,a_{n}=0}^{\infty}\ \sum_{b_{2},b_{3},\ldots,b_{n}=0}^{\infty}\ .\label{eq:4}
\end{align}
Note that the permutations $\rho,\sigma$ in (\ref{cptnpt1}) and (\ref{cptnp2})
only act on the subscripts of $z_{ij}$ and $\eta_j$ but not on the superscripts $a_i$ and $b_j$.
The component integrals satisfy the reality condition
\begin{align}
\overline{W^\tau_{(A|B)}(\sigma|\rho)} = W^\tau_{(B|A)}(\rho|\sigma)\,.
\end{align}

Component integrals of the type in (\ref{cptnpt1}) arise from the conformal-field-theory correlators
underlying one-loop amplitudes of closed bosonic strings, heterotic strings and 
type-II superstrings \cite{Gerken:2018jrq}. More specifically, the $f^{(a)}_{ij}$ were found to appear 
naturally from the spin sums of the RNS formalism
\cite{Broedel:2014vla} and the current algebra of heterotic strings \cite{Dolan:2007eh}.\footnote{Also see e.g.\ \cite{Tsuchiya:1988va, Stieberger:2002wk, Bianchi:2006nf} for earlier work on RNS spin 
  sums, \cite{Mafra:2016nwr, Mafra:2017ioj, Mafra:2018nla, Mafra:2018pll, Mafra:2018qqe} for $g^{(a)}_{ij}$ and $f^{(a)}_{ij}$
  in one-loop amplitudes in the pure-spinor formalism and \cite{Bianchi:2015vsa, Berg:2016wux} 
  for applications to RNS one-loop amplitudes with reduced supersymmetry.} For these theories,
the $(n {-}1)!\times (n{-}1)!$ matrix in (\ref{cptnpt1}) is in fact claimed to contain a basis of the integrals that arise in string theory\footnote{String-theory integrals can also involve integrals over $\partial_{z_i} f^{(a)}_{ij}$ or $f^{(a_1)}_{ij} f^{(a_2)}_{ij}$ that 
are not in the form of (\ref{cptnpt1}) but can be
reduced to the conjectural basis by means of Fay identities and integration by parts 
w.r.t.\ the punctures. Similar reductions should be possible for products of $\partial_{z_i} f^{(a)}_{ij}$ or
cycles $f_{i_1i_2}^{(a_1)}f_{i_2i_3}^{(a_2)} \ldots f_{i_ki_1}^{(a_k)}$ by adapting the recursive 
techniques of \cite{Schlotterer:2016cxa, He:2018pol, He:2019drm} to a genus-one setup.}
for any massless one-loop amplitude. Moreover, massive-state amplitudes are likely to fall into the same basis.

The massless four-point one-loop amplitude of type-II superstrings \cite{Green:1982sw} for instance is 
proportional to the four-point component integrals $W^\tau_{(0,0,0|0,0,0)}$. Similarly, the five-point type-II 
amplitude involves $W^\tau_{(0,0,0,0|0,0,0,0)}$ and various permutations of $W^\tau_{(1,0,0,0|1,0,0,0)}$ 
and $W^\tau_{(0,1,0,0|1,0,0,0)}$ \cite{Richards:2008jg, Mafra:2009wi, Green:2013bza}. The $n$-point
systematics and the role of $W^\tau_{(A|B)}$ at higher $A, B$ in the context
of reduced supersymmetry are detailed in Appendix~\ref{app:string}.

The main motivation of this work is to study the $\ap$-expansion of component integrals (\ref{cptnpt1})
via generating-function methods. As detailed in Section~\ref{sec:alpha-expand-comp-ints} below, by the dimensionless Mandelstam variables (\ref{intr3}) 
in the Koba--Nielsen exponent (\ref{eq:KNn}), the coefficients in such $\ap$-expansions are
torus integrals over Green functions as well as products of $f_{ij}^{(a)}$ and $\overline{f_{kl}^{{(b)}}}$.
At each order in $\ap$, these integrals fall into the framework of modular graph forms to be reviewed below. As will be
demonstrated in later sections, the $W$-integrals (\ref{eq:Wdef}) allow for streamlined derivations
of differential equations for infinite families of modular graph forms.


\subsubsection{Relations between component integrals}
\label{sec:cptints2}

It is important to stress that the component integrals $W^\tau_{(A|B)}(\sigma|\rho)$ are not all linearly independent. There are two simple mechanisms that lead to relations between certain special cases of component integrals. Still, component integrals
$W_{(A|B)}$ with generic weights $A,B$ are not affected by the subsequent relations, that is why they do not propagate
to relations between the $(n-1)!\times (n-1)!$ generating series in (\ref{eq:Wdef}).

Firstly, there can be relations between different $W^\tau_{(A|B)}(\sigma|\rho)$ stemming from the fact that the functions $f^{(a)}_{ij}$ entering in~\eqref{cptnpt1} satisfy $f^{(a)}_{ij} = (-1)^a f^{(a)}_{ji}$ and similarly for the $\overline{f^{(b)}_{kl}}$. Since these parity properties interchange points they intertwine with the permutations $\rho$ and $\sigma$. For instance, if the last two entries of $A$ and $B$ are $A=(a_2,\ldots, a_{n-2}, 0, a_n)$ and $B=(b_2,\ldots,b_{n-2},0,b_n)$, respectively, then the only places where the points $z_{n-1}$ and $z_n$ appear are $f_{n-1,n}^{(a_n)}$, $\overline{f^{(b_n)}_{n-1,n}}$ and in the permutation invariant Koba--Nielsen factor. Applying the parity transformation to the these factors of $f^{(a_n)}$ and $\overline{f^{(b_n)}}$ therefore can be absorbed by composing the permutations $\rho$ and $\sigma$ with the transposition $n-1\leftrightarrow n$ and an overall sign $(-1)^{a_n+b_n}$. This yields a simple instance of an algebraic relation between the component integrals and we shall see an explicit instance of this for three points in Section~\ref{sec:CR3} below. 

The second mechanism is integration by parts -- integrals of total $z$-derivatives (or $\bar{z}$-derivatives) 
vanish due to the presence of the Koba--Nielsen factor. Such derivatives produce sums over $s_{ij}f^{(1)}_{ij}$
from the Koba--Nielsen factor (see~\eqref{eq:dzKN}) and may also involve $\partial_{z_i} f^{(a)}_{ij}$.
Integration by parts in combination with the Fay identities in Appendix~\ref{app:Om} can first of all be 
used to eliminate derivatives of $ f^{(a)}_{ij}$ and conjecturally any integrand that does not line up
with the form of $W^\tau_{(A|B)}(\sigma|\rho)$. Moreover, $W^\tau_{(A|B)}(\sigma|\rho)$
with $a_i,b_j=1$ for some of the weights can be related by the $s_{ij}f^{(1)}_{ij}$ from
the Koba--Nielsen derivatives. We note that these integration-by-parts relations can mix component integrals of different modular weight as they can also contain explicit instances of $\Im\tau$. A two-point instance of such an integration-by-parts identity
among component integrals can be found in~\eqref{ibp2a} below.

\subsection{Modular graph forms}
\label{sec:MGF}

Modular graph forms are a compact way of denoting certain classes of torus world-sheet integrals
that will be shown in Section~\ref{sec:alpha-expand-comp-ints} to arise also in the low-energy expansion of the component $W$-integrals (\ref{cptnpt1}).
The starting point of a modular graph form is a decorated graph $\Gamma$ on $n$ vertices, corresponding to insertion points $z_i$ ($1\leq i\leq n$), and with directed edges of loop momentum $p_e$, where $e$ runs over the set of edges $E_{\Gamma}$. The decoration corresponds to a pair of integers $(a_e,b_e)$ for each edge $e$.

The modular graph form associated with the decorated graph $\Gamma$ is then the following function of~$\tau$~\cite{DHoker:2016mwo}
\begin{align}
  \label{eq:MGF}
  \mC_\Gamma(\tau) :=  \left[\prod_{e\in E_{\Gamma}}\sum_{p_e\neq 0}  \frac1{p_e^{a_e} \bar{p}^{b_e}_e} \right]\prod_{i=1}^n \delta\!\left(\sum_{e' \rspoonarrowin i} p_{e'} -\sum_{e' \rspoonarrowout i}  p_{e'}\right)\,,
\end{align}
where our normalization conventions differ from similar definitions in the literature\footnote{\label{fn:norm}More 
specifically, the right-hand side of our definition (\ref{ourdef}) does not include the factor 
of $\prod_{e\in E_{\Gamma}} (\frac{\Im \tau}{\pi})^{\frac{1}{2}(a_e + b_e)}$ from \cite{DHoker:2016mwo, DHoker:2016quv, Gerken:2018zcy, DHoker:2019txf} and the factor of
$\prod_{e \in E_{\Gamma}} (\frac{\Im \tau}{\pi})^{b_e}$ from \cite{Gerken:2018jrq}.}.
From the definition one sees that the decorations $(a_e,b_e)$ on the edges label the powers of the holomorphic and anti-holomorphic momenta $p_e=m_{e}\tau+n_{e}$ and $\bar{p}_e=m_{e}\bar{\tau}+n_{e}$ (with $m_{e},n_{e}\in\ZZ$) propagating through the edge. At each vertex there is a momentum conserving delta function as indicated by summing over all momenta touching the vertex. The two terms with opposite signs in the momentum conserving delta function distinguish the incoming and outgoing momenta at a vertex. This definition implies in particular that if an edge connects a vertex to itself, $\mathcal{C}_{\Gamma}$ factorizes.

If for any two edges $e$ and $e'$ the sum of weights $a_{e}+b_{e}+a_{e'}+b_{e'}>2$, \eqref{eq:MGF} is absolutely convergent. Furthermore, due to the symmetry under $p_{e}\rightarrow - p_{e}$ (for all $e\in E_\Gamma$ simultaneously), it follows immediately from the definition \eqref{eq:MGF} that $\mathcal{C}_{\Gamma}(\tau)$ vanishes if the sum of all exponents $\sum_{e\in E_{\Gamma}}(a_{e}+b_{e})$ is odd.

Under an ${\rm SL}_{2}(\ZZ)$ transformation
a modular graph form transforms as
\begin{align}
  \label{eq:MGFtrm}
  {\cal C}_\Gamma\left(\frac{\alpha\tau+\beta}{\gamma\tau+\delta}\right) = (\gamma\tau+\delta)^{|A|} (\gamma\bar\tau+\delta)^{|B|} {\cal C}_\Gamma(\tau)\,,
\end{align}
where the integers $|A|$ and $|B|$ defined by\footnote{From the context it will always be clear if we are referring with $|A|$ and $|B|$ to a modular graph form as in \eqref{eq:MGFtrm2} or to a component integral as in \eqref{eq:2}.}
\begin{align}
  |A|=\sum_{e\in E_{\Gamma}} a_e \, , \ \ \ \ \ \ |B|= \sum_{e\in E_{\Gamma}} b_e
  \label{eq:MGFtrm2}
\end{align}
are commonly referred to as holomorphic and anti-holomorphic weights, respectively, and often written as a pair $(|A|,|B|)$. Similarly, $\Im\tau$ transforms as \mbox{$\Im\tau\rightarrow (\gamma\tau+\delta)^{-1}(\gamma\bar{\tau}+\delta)^{-1}\Im\tau$} and hence carries modular weight $(-1,-1)$.

We note that the modular graph form~\eqref{eq:MGF} vanishes when there are vertices with a single edge ending on them
and more generally when the graph $\Gamma$ is one-particle reducible. As a consequence, one 
of the delta functions in~\eqref{eq:MGF} is redundant due to overall momentum conservation, i.e.\ there is 
always one vertex whose in- and outgoing momenta are already fixed by the assignments of the momenta at all other vertices.

The modular graph form \eqref{eq:MGF} can also be written in terms of the lattice sums~\eqref{eq:Cabfunc} as
\begin{align}
  \label{eq:MGFint}
  \mC_\Gamma(\tau) = \int \Big( \prod_{i=1}^n \frac{\dd^2 z_i}{\Im \tau} \Big) \prod_{e\in E_{\Gamma}} C^{(a_e,b_e)}(z_{e},\tau) \,,
\end{align}
where we have denoted by $z_e$ the difference between the starting and final points of an edge $e$ and use \eqref{eq:30} for $C^{(0,0)}$. The delta function in~\eqref{eq:MGF} originates from the integral over the punctures $z_i$ on the torus and the phase factors $e^{2\pi i \langle p_e,z_i\rangle}$ for all  edges touching the vertex $z_i$. By translation invariance on the torus we could also set $z_1=0$ and integrate only over $n{-}1$ points as we have done for the $W$-integrals~\eqref{eq:Wdef}. This translation invariance corresponds to overall momentum conservation.

Modular graph forms with symmetric decorations $a_e=b_e$ for all edges are known as
modular graph functions \cite{DHoker:2015wxz}. They arise if the integrand in (\ref{cptnpt1}) is solely made
of Green functions and can be rendered modular invariant upon multiplication by a suitable power of $\Im \tau$.
More generally, modular invariant completions of this type can be attained under the weaker condition 
$|A|=|B|$ on (\ref{eq:MGFtrm2}) with $a_e \neq b_e$ for some edges.

\subsubsection{Dihedral examples}

\begin{figure}
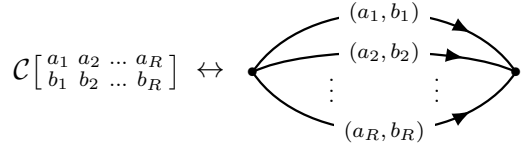

  \begin{center}
    \tikzpicture[scale=0.7,line width=0.3mm,decoration={markings,mark=at position 0.8 with {\arrow[scale=1.3]{latex}}}]
    \draw(-2,0)node{$\cform{a_1&a_2&\ldots &a_R \protect\\b_1&b_2&\ldots &b_R}$};
    \draw(0.2,0)node{$\displaystyle \leftrightarrow$};
    \draw[postaction={decorate}] (1,0)node{\scriptsize $\bullet$} .. controls (2,1.5) and (5,1.5) .. node[fill=white]{\scriptsize $(a_1,b_1)$} (6,0)node{\scriptsize $\bullet$};
    \draw[postaction={decorate}] (1,0) .. controls (2,0.5) and (5,0.5) .. node[fill=white]{\scriptsize $(a_2,b_2)$} (6,0);
    \draw (2.5,-0.2) node{\scriptsize $\vdots$};
    \draw (4.5,-0.2) node{\scriptsize $\vdots$};
    \draw[postaction={decorate}] (1,0) .. controls (2,-1.5) and (5,-1.5) .. node[fill=white]{\scriptsize $(a_R,b_R)$} (6,0);
    \endtikzpicture
    \caption{\label{fig:dihed}\it Decorated dihedral graph with associated notation for modular graph form.}
  \end{center}
\end{figure}

As an example, we consider \textit{dihedral graphs} shown in Figure~\ref{fig:dihed} that consist of two vertices connected by $R$ lines. For the associated modular graph forms we use the notation
\begin{align}
  \cform{a_1&a_2&\ldots &a_R \protect\\b_1&b_2&\ldots &b_R}\! (\tau) := \sum_{p_1,\ldots,p_R\neq 0} \frac{\delta(p_1+\ldots +p_R)}{p_1^{a_1} \bar{p}_1^{b_1} \cdots p_R^{a_R}\bar{p}_R^{b_R}}\,,
                                                        \label{ourdef}
\end{align}
where our normalization conventions differ from~\cite{DHoker:2016mwo, DHoker:2016quv, Gerken:2018zcy, Gerken:2018jrq, DHoker:2019txf}, see Footnote~\ref{fn:norm}. In the above expression we have suppressed the redundant delta function from overall momentum conservation. Following~\eqref{eq:MGFint} the integral representation of this is
\begin{align}
  \cform{a_1&a_2&\ldots &a_R \protect\\b_1&b_2&\ldots &b_R}\! (\tau) = \int \frac{\dd^2z}{\Im\tau} \prod_{e=1}^R C^{(a_e,b_e)}(z,\tau)\,.
                                                        \label{ourdef2pt}
\end{align}

There are two special cases of dihedral modular graph forms that will play an important role in our examples. These correspond to cases with only two edges,
\begin{subequations}
\label{eq:GkEk}
  \begin{align}
    \cform{k&0\protect\\ 0&0}\! (\tau)
                            &= {\rm G}_k(\tau) \hspace{-5mm} &&\textrm{(for $k>2$)}\phantom{\, ,}  \\
    \cform{k&0\protect\\ k&0}\! (\tau)
                            &= \left(\frac{\pi}{\Im\tau}\right)^k{\rm E}_k(\tau) \hspace{-5mm} &&\textrm{(for $k>1$)}\, ,
  \end{align}
\end{subequations}
where the zeros in the second column stem from $\cform{a_1&a_2\protect\\ b_1&b_2} 
= (-1)^{a_2+b_2} \cform{a_1+a_2&0\protect\\ b_1+b_2&0}$. The holomorphic Eisenstein
series ${\rm G}_k$ were defined in (\ref{eq:Gk}), 
and ${\rm E}_k$ denotes the non-holomorphic Eisenstein series in the normalization convention
\begin{align}
  {\rm E}_k(\tau) = \Big( \frac{\Im \tau}{\pi} \Big)^k \sum_{p\neq 0} \frac{1}{ |p|^{2k}} 
  \label{eq:33x}
\end{align}
that converges absolutely for $\Re(k)>1$. The non-holomorphic Eisenstein series is invariant under modular transformations
by virtue of $(\Im \tau)^k$ in the prefactor. Note that in the above formulas the column with two zeroes
is essential in order to have a non-vanishing modular graph form.

From the convergence conditions in~\eqref{eq:GkEk} we see that there are two special cases that can be considered separately. We formally define
\begin{align}
  \cform{2&0\protect\\ 0&0}\! (\tau) = \widehat{\rm G}_2(\tau)\,,\hspace{10mm}
                          \cform{1&0\protect\\ 1&0}\! (\tau) = \frac{\Im \tau}{\pi} {\rm E}_1(\tau)\,,
\end{align}
where the $\widehat{\rm G}_2$ denotes the non-holomorphic but modular function in~\eqref{eq:G2hat} while ${\rm E}_1$ is strictly divergent but can be regularized using the (first) Kronecker limit formula~\cite{Fleig:2015vky}. 

For later reference, we furthermore define the combinations
\begin{align}
  \mathrm{E}_{2,2}(\tau)&=\Big(\frac{\Im\tau}{\pi}\Big)^{4}\cform{1&1&2\\1&1&2}\!(\tau)-\frac{9}{10}\mathrm{E}_{4}(\tau)\label{eq:33}\\
  \mathrm{E}_{2,3}(\tau)&=\Big(\frac{\Im\tau}{\pi}\Big)^{5}\cform{1&1&3\\1&1&3}\!(\tau)-\frac{43}{35}\mathrm{E}_{5}(\tau)\notag
\end{align}
which capture the independent two-loop modular graph functions at weight four and five \cite{DHoker:2015gmr} and were tailored to satisfy particularly simple differential equations \cite{DHoker:2016mwo, Broedel:2018izr}. They correspond to so-called depth-two iterated integrals over holomorphic Eisenstein series~\cite{Broedel:2018izr}.

\subsubsection{Differential equations of modular graph forms}

We will frequently express modular graph forms at the leading $s_{ij}$-orders
of the $W$-integrals through $\mathrm{E}_{k}$, $\mathrm{E}_{2,2}$, $\mathrm{E}_{2,3}$ and their derivatives
w.r.t.\ the operator\footnote{The subscript of $\dgcr$
aims to avoid confusion with the differential operators of Sections~\ref{sec:maass-ops} and \ref{sec:CRgen}
and refers to the authors D'Hoker and Green of \cite{DHoker:2016mwo} which initiated the systematic 
study of relations between modular graph forms via repeated action of (\ref{eq:31}).}
\begin{align}
  \dgcr:=2i (\Im\tau)^{2}\partial_{\tau} \,,\quad\quad \dgcrbar:= - 2i (\Im\tau)^{2}\partial_{\bar\tau} \label{eq:31}
\end{align}
seen in earlier literature~\cite{DHoker:2016mwo, DHoker:2016quv, Broedel:2018izr, Gerken:2018jrq}. 
Non-holomorphic Eisenstein series give rise to the closed formula for two-column modular graph forms
\begin{align}
  \dgcr^{n}\mathrm{E}_{k}&=\frac{ (\Im \tau)^{k+n}}{\pi^{k}}\frac{(k+n-1)!}{(k-1)!}\cform{k+n&0\\k-n&0} \, ,
  \label{eq:32}
\end{align}
whereas the derivatives of their simplest generalizations in (\ref{eq:33})
are determined by
\begin{align}
   \dgcr \mathrm{E}_{2,2}&=\frac{(\Im\tau)^{5}}{\pi^{4}}\Big({-}\frac{8}{5}\cform{5&0\\3&0}+\cform{1&1&3\\1&1&1}\Big)\notag \\
  \dgcr \mathrm{E}_{2,3}&= \frac{(\Im\tau)^6}{\pi ^5} \Big( {-}\frac{50}{7}\cform{ 6 & 0 \\ 4 & 0 }- \cform{ 0 & 2 & 4 \\ 2 & 2 & 0 } \Big)  +2  \frac{(\Im\tau)^4}{\pi ^3} {\rm E}_2 \cform{ 4 &0 \\ 2 &0 } \label{eq:32b} \\
  \dgcr^2 \mathrm{E}_{2,3} &= \frac{(\Im\tau)^{7}}{\pi^{5}}\Big({-}\frac{62}{7}\cform{7&0\\3&0}-4 \cform{3&0\\1&0}\cform{4&0\\2&0}+4 \cform{0&2&5\\1&0&2}\Big)  \, .\notag
\end{align}
Higher iterations of $\dgcr$ give rise to holomorphic Eisenstein series \cite{DHoker:2016mwo}, e.g.
\begin{align}
\dgcr^k \mathrm{E}_k &= \frac{ (2k-1)! }{\pi^k (k-1)!} (\Im \tau)^{2k} {\rm G}_{2k}
\notag \\
\dgcr^3 \mathrm{E}_{2,2} &= - \frac{6}{\pi^2} (\Im \tau)^4 {\rm G}_4 \dgcr \mathrm{E}_{2} 
  \label{eq:32c}  \\
  \dgcr^3 \mathrm{E}_{2,3} &= - 2 (\dgcr \mathrm{E}_{2}) \dgcr^2 \mathrm{E}_{3} - \frac{4}{ \pi^{2} } (\Im \tau)^4 {\rm G}_4 \dgcr \mathrm{E}_{3} \, .
\notag
 \end{align}
Note, however, that the Cauchy--Riemann and Laplace equations of the $W$-integrals
take a more convenient form when trading $\dgcr$ for the differential operator that will be introduced in Sections~\ref{sec:maass-ops} and \ref{sec:CRgen}. 


\begin{figure}
  \begin{center}
    \tikzpicture[scale=1,line width=0.30mm]
    \scope[shift={(-1,0)},decoration={markings,mark=at position 0.8 with {\arrow[scale=1.3]{latex}}}]
    \fill (0,0) circle [radius=0.07]node[left]{$z_2$};
    \fill (5,0) circle [radius=0.07]node[right]{$z_1$};
    \fill (60:5) circle [radius=0.07]node[above]{$z_3$};
    \draw[postaction={decorate}] (5,0) .. controls (4,0.6) and (1,0.6) ..node[fill=white,transform shape]{$(a_{12}^1,b_{12}^1)$} (0,0);
    \draw (1.5,0.15) node{\scriptsize $\vdots$};
    \draw (3.5,0.15) node{\scriptsize $\vdots$};
    \draw[postaction={decorate}] (5,0) .. controls (4,-0.6) and (1,-0.6) ..node[fill=white,transform shape]{$(a_{12}^{R_{12}},b_{12}^{R_{12}})$} (0,0);

    \scope[rotate=60]
    \draw[postaction={decorate}] (0,0) .. controls (1,0.6) and (4,0.6) ..node[fill=white,transform shape,sloped]{$(a_{23}^1,b_{23}^1)$} (5,0);
    \draw (1.5,0.15) node[transform shape]{\scriptsize $\vdots$};
    \draw (3.5,0.15) node[transform shape]{\scriptsize $\vdots$};
    \draw[postaction={decorate}] (0,0) .. controls (1,-0.6) and (4,-0.6) ..node[transform shape,fill=white]{$(a_{23}^{R_{23}},b_{23}^{R_{23}})$} (5,0);
    \endscope

    \scope[shift={(60:5)},rotate=-60]
    \draw[postaction={decorate}] (0,0) .. controls (1,0.6) and (4,0.6) ..node[fill=white,transform shape]{$(a_{31}^1,b_{31}^1)$} (5,0);
    \draw (1.5,0.15) node[transform shape]{\scriptsize $\vdots$};
    \draw (3.5,0.15) node[transform shape]{\scriptsize $\vdots$};
    \draw[postaction={decorate}] (0,0) .. controls (1,-0.6) and (4,-0.6) ..node[transform shape,fill=white]{$(a_{31}^{R_{31}},b_{31}^{R_{31}})$} (5,0);
    \endscope
    \endscope
    \endtikzpicture
    \caption{The graph associated to the trihedral modular graph form 
      $\cformtri{A_{12}\\B_{12}}{A_{23}\\B_{23}}{A_{31}\\B_{31}}$.}
    \label{fig:trihed}
  \end{center}
\end{figure}

\subsubsection{More general graph topologies}

While the dihedral modular graph forms (\ref{ourdef2pt}) accommodate any order in the $\ap$-expansion of
two-point component integrals (\ref{cptnpt1}), higher multiplicities $n\geq 3$ introduce more general graph topologies.
Their three-point instances additionally involve the trihedral topology depicted in Figure~\ref{fig:trihed} and that we shall encounter in examples in this paper. 
For trihedral modular graph forms, the integral representation (\ref{eq:MGFint}) in terms of the building 
blocks~\eqref{eq:Cabfunc} takes the form 
\begin{align}
  \cformtri{A_{12}\\B_{12}}{A_{23}\\B_{23}}{A_{31}\\B_{31}} = \int\frac{ \dd^2 z_2}{\Im \tau} \frac{ \dd^2 z_3}{\Im \tau} 
  \Big( \prod_{i=1}^{R_{12}} C^{(a_{12}^{i},b_{12}^i)}_{12} \Big)
  \Big( \prod_{j=1}^{R_{23}} C^{(a_{23}^{j},b_{23}^j)}_{23} \Big)
  \Big( \prod_{k=1}^{R_{31}} C^{(a_{31}^{k},b_{31}^k)}_{31} \Big)\, ,\label{eq:29}
\end{align}
where $z_1=0$, and the collective labels $A_{ij},B_{ij}$ have length $R_{ij}$, e.g.\ $A_{ij}=(a_{ij}^1,a_{ij}^2,\ldots,a_{ij}^{R_{ij}})$. In special cases, this simplifies to dihedral modular graph forms, as detailed in Appendix~\ref{sec:top-simp}.

More generally, $\ap$-expansions of $n$-point component integrals generate modular graph forms associated with $n$-vertex graphs due to $\prod_{i<j}^n ( \prod_{k=1}^{R_{ij}} C_{ij}^{(a_{ij}^k, b_{ij}^k )})$.

\subsection{Low-energy expansion of component integrals}
\label{sec:alpha-expand-comp-ints}

The component integrals $W^{\tau}_{(A|B)}(\sigma|\rho)$ introduced in \eqref{cptnpt1} depend on the Mandelstam variables $s_{ij}$ through the Koba--Nielsen factor \eqref{eq:KNn}. In this section, we will study the expansion of the $W^{\tau}_{(A|B)}(\sigma|\rho)$ in the Mandelstam variables. Since the Mandelstams as defined in \eqref{intr3} carry a factor of $\alpha'$, this expansion is also an expansion in $\alpha'$ and therefore corresponds to the low-energy expansion of the corresponding amplitude in the string-theory context.

Expanding the integrands of the component integrals $W^{\tau}_{(A|B)}(\sigma|\rho)$ in the $s_{ij}$ leads to
\begin{align}
  W^{\tau}_{(A|B)}(\sigma|\rho)
  &=\int \dd \mu_{n-1} 
 \left[ \prod_{1\leq i<j}^n \sum_{k_{ij}=0}^{\infty} \frac{1}{k_{ij}!}  \big[ s_{ij}G(z_{ij},\tau) \big]^{k_{ij}}\right]
\notag\\
&\qquad\times  \rho[f^{(a_{2})}_{12}\,f^{(a_{3})}_{23}\dots f^{(a_{n})}_{n-1,n}] \, 
  \sigma[
\overline{f^{(b_{2})}_{12}}\, \overline{f^{(b_{3})}_{23}}\dots \overline{f^{(b_{n})}_{n-1,n}}]
\ ,\label{eq:20}
\end{align}
where the contributions to the $(\ap)^w$-order are characterized by $\sum_{1\leq i<j}^n k_{ij}=w$.
Once the $f^{(a)}_{ij}$, $\overline{f^{(b)}_{ij}}$ and $G(z_{ij},\tau)$ in the integrand
are identified with the doubly-periodic functions $C^{(a,b)}_{ij}$ via
\eqref{eq:21} and \eqref{eq:22}, each term in the $\ap$-expansion of \eqref{eq:20} is lined up with 
the integral representation \eqref{eq:MGFint} of modular graph forms. This shows that the coefficients at 
the $(\ap)^w$-order of $W^{\tau}_{(A|B)}(\sigma|\rho)$ are modular graph forms of weight $(|A|{+}w,|B|{+}w)$, see
Sections~\ref{sec:2ptCmptInts} and \ref{sec:3ptexmpl} for $(n\leq 3)$-point examples.
Since modular graph forms vanish if the sum of their holomorphic and anti-holomorphic modular 
weight is odd, also $W^{\tau}_{(A|B)}(\sigma|\rho)=0$ if $|A|+|B|$ is odd. 

Conversely, any convergent modular graph form can be realized through the $\ap$-expansion (\ref{eq:20})
of suitably chosen component integrals. The topology of the defining graph determines the minimal multiplicity $n$ that admits such a realization. For instance, the spanning set $\cform{a&b&0\\c&0&d}$ of two-loop 
modular graph forms \cite{DHoker:2019txf}\footnote{In particular, the odd two-loop modular graph forms
${\cal A}_{u,v;w}=( \frac{ \Im \tau }{\pi})^w(\cform{w-u&u&0\\w-v&0&v} - \cform{w-v&v&0\\w-u&0&u} ) $
studied in \cite{DHoker:2019txf} arise at the $\ap \rightarrow 0$ limit
of the component integrals $\Im W^\tau_{(w-u,u|w-v,v)}(3{,}2|2{,}3)$. The simplest odd modular graph form
${\cal A}_{1,2;5}$ which is not expressible in terms of $\dgcr^{n}\mathrm{E}_{k}$ in (\ref{eq:32}) 
and their complex conjugates is generated by the component integral $\Im W^\tau_{(4,1|3,2)}(3{,}2|2{,}3)$ over
$\Im (f^{(4)}_{12} \, f^{(1)}_{23}\, \overline{ f^{(3)}_{13} }\, \overline{ f^{(2)}_{32} } )$.} with $a,b,c,d\neq 0$ and $(b,d)\neq (1,1)$ 
arises at the $\ap^0$-order of the three-point component integral $W^\tau_{(a,b|c,d)}(3{,}2|2{,}3)$ over $f^{(a)}_{12} \, f^{(b)}_{23} \, \overline{ f^{(c)}_{13} } \, \overline{ f^{(d)}_{32}}$ since the intermediate point $1$ is two-valent and can be contracted using~\eqref{eq:35}. Further examples of dihedral
and trihedral topology can be found in Section \ref{extrastuff}.

However, the $\ap$-expansion (\ref{eq:20}) at the level of the integrand is not applicable in presence of
singularities $|z_{ij}|^{-2}$, i.e.\ in case of real combinations $|f_{ij}^{(1)}|^2=f_{ij}^{(1)} \overline{f_{ij}^{(1)}}$. By the local behavior
$\KN^{\tau}_{n} \sim |z_{ij}|^{-2s_{ij}}$ of the Koba--Nielsen factor as $z_i \rightarrow z_j$, the integration region
over $|z_{ij}| \ll 1$ yields kinematic poles $\sim s_{ij}^{-1}$. Still, component integrals $W^{\tau}_{(A|B)}(\sigma|\rho)$
with integrands $\sim |f_{ij}^{(1)}|^2$ can be Laurent-expanded via suitable 
subtraction schemes as reviewed in Appendix~\ref{app:poleA}. Since the residues of the kinematic poles\footnote{In the same way as $|f_{ij}^{(1)}|^2
\KN^{\tau}_{n}$ integrates to kinematic poles $\sim s_{ij}^{-1}$, integrands with $(p{-}1)$ factors of
$f_{i_ai_b}^{(1)}$ and $\overline{f_{i_ai_b}^{(1)}}$ each in the range $1\leq a<b\leq p$ give rise to poles 
$\sim s_{i_1 i_2\ldots i_p}^{-1}$ in multiparticle channels. Pole structures of this type can still be
accounted for via analogous subtraction schemes, and the residues are again
expressible in terms of lower-multiplicity integrals.} 
are expressible in terms of Koba--Nielsen integrals at lower multiplicity, any contribution to such 
subtraction schemes can be integrated in the framework of modular graph forms.

Note that the differential-equation approach of the next sections does not require any tracking of
kinematic poles, and our results  do not rely on any subtraction scheme.

\subsubsection{Two-point examples}\label{sec:2ptCmptInts}
At two points, the generating function and the component integrals do not depend on permutations and are given by
\begin{align}
  W^\tau_\eta &= \sum_{a,b=0}^\infty \eta^{a-1} \bar \eta^{b-1} W^\tau_{(a|b)}   \,, \ \ \ \ \ \
                W^\tau_{(a|b)} = \int \frac{ \dd^2 z_2}{\Im \tau} f^{(a)}_{12} \overline{f^{(b)}_{12} } 
                \, \KN^{\tau}_{2} \, ,\label{lapsec15}
\end{align}
where we have denoted $\eta=\eta_2$ for simplicity. By identifying the Green function as
$\frac{\Im \tau}{\pi} C^{(1,1)}(z,\tau)$ as in \eqref{eq:20}, one can easily arrive at
closed formulae for the $\ap$-expansion of $W^\tau_{(a|b)}$ in terms of dihedral 
modular graph forms (\ref{ourdef})
\begin{align}
  W^{\tau}_{(0|0)}&=1+\sum_{k=2}^{\infty}\frac{s_{12}^{k}}{k!}
 \Big(\frac{\Im \tau}{\pi}\Big)^{k}
  \cform{1_{k}\\1_{k}}\notag\\
  W^{\tau}_{(a|0)}&=-\sum_{k=1}^{\infty}\frac{s_{12}^{k}}{k!}
   \Big(\frac{\Im \tau}{\pi}\Big)^{k}
  \cform{a&1_{k}\\0&1_{k}} \, , & a&>0\label{eq:7}\\
  W^{\tau}_{(a|b)}&=(-1)^{a}\sum_{k=0}^{\infty}\frac{s_{12}^{k}}{k!}
   \Big(\frac{\Im \tau}{\pi}\Big)^{k}
  \cform{a&0&1_{k}\\0&b&1_{k}} \, , & a,b&>0\, ,\quad(a,b)\neq(1,1)\notag\,,
\end{align}
where $1_{k}$ denotes the row vector with $k$ entries of 1. The expansion of $W^{\tau}_{(0|b)}$ can be obtained by complex conjugating the expansion of $W^{\tau}_{(a|0)}$.

The $\ap$-expansion of $W^{\tau}_{(1|1)}$ requires extra care since the singularity
$f^{(1)}_{12} \overline{ f^{(1)}_{12}}  \sim \frac{1}{|z_2|^2}$ of the integrand leads to
a kinematic pole in $s_{12}$ as mentioned above. We can bypass the need for subtraction schemes through
the integration-by-parts-identities
\begin{align}
  s_{12} W^\tau_{(a|1)} + \frac{ 2\pi i }{\tau- \bar \tau} W^\tau_{(a-1|0)} = 0 \, , \ \ \ \ \ \
  s_{12} W^\tau_{(1|b)} + \frac{ 2\pi i }{\tau- \bar \tau} W^\tau_{(0|b-1)} = 0\ ,
  \label{ibp2a}
\end{align}
which can be checked by evaluating the total derivative $\partial_{ \bar z_2}( f^{(a)}_{12}  \KN^{\tau}_{2})$ 
via (\ref{antipar}) and (\ref{eq:dzKN}) and result in
\begin{align}
  W^{\tau}_{(1|1)}  &= -\frac{\pi}{s_{12}\Im\tau} W^{\tau}_{(0|0)}=-\frac{\pi}{s_{12}\Im\tau}
  -\sum_{k=2}^{\infty}\frac{s_{12}^{k-1}}{k!}\Big(\frac{\Im\tau}{\pi}\Big)^{k-1}\cformS{1_{k}\\1_{k}}\,.  \label{ibp2b}
\end{align}
Note that similar integrations by parts should suffice to rewrite higher-multiplicity 
component integrals $W^{\tau}_{(A|B)}(\sigma|\rho)$ with kinematic poles
in terms of regular representatives with a Taylor expansion in $s_{ij}$. 
The kinematic poles will
then appear as the expansion coefficients such as the factor of $s_{12}^{-1}$ in the first
step of (\ref{ibp2b}), see Appendix~\ref{app:poleB} for a three-point example.

Although the $\ap$-expansion in terms of lattice sums could be quickly generated from \eqref{eq:20}, the representation
in (\ref{eq:7}) is not optimal since many non-trivial identities between modular graph forms exist \cite{Green:2008uj, DHoker:2015gmr,DHoker:2015wxz, DHoker:2016mwo, DHoker:2016quv}. The types of identities in the references
are reviewed in Ap\-pendix~\ref{sec:idMGF} and can be used to reduce the above expansions into a putative 
basis of lattice sums.
At the lowest orders, these conjectured bases are given by the non-holomorphic Eisenstein series 
$\mathrm{E}_{k}$ in (\ref{eq:33x}), their higher-depth analogue $\mathrm{E}_{2,2}$ in \eqref{eq:33} and derivatives \eqref{eq:32} or \eqref{eq:32b} and we have for example
\begin{align}
  W^{\tau}_{(0|0)}  &= 1+\frac{1}{2}s_{12}^{2}\mathrm{E}_{2}+\frac{1}{6}s_{12}^{3}(\mathrm{E}_{3}+\zeta_{3})+s_{12}^{4}\left(\frac{\mathrm{E}_{2}^{2}}{8}+\frac{3 \mathrm{E}_{4}}{20}+\mathrm{E}_{2,2}\right)  \notag \\
  &\quad+s_{12}^5 \left( \frac{ {\rm E}_{2,3} }{2} + \frac{ {\rm E}_2 }{12} ( {\rm E}_3+ \zeta_3) 
+ \frac{ 3 {\rm E}_5 }{14} + \frac{ 2 \zeta_5 }{15} \right)+ O(s_{12}^6) \notag \\
  W^{\tau}_{(2|0)}  &= \frac{ \pi }{(\Im \tau)^2} \bigg[ {-}\frac{1}{2}s_{12} \dgcr \mathrm{E}_{2}-\frac{1}{6}s_{12}^{2}
  \dgcr \mathrm{E}_{3} \notag\\
  &\quad+s_{12}^{3}\left({-}\frac{1}{4} \mathrm{E}_{2} \dgcr \mathrm{E}_{2} -\frac{3}{20} \dgcr \mathrm{E}_{4} 
  - \dgcr \mathrm{E}_{2,2} \right)+ O(s_{12}^4) \bigg]\notag\\
  W^{\tau}_{(4|0)}  &=  \frac{ \pi^2 }{(\Im \tau)^4} \bigg[ {-}\frac{1}{12}s_{12} \dgcr^{2}\mathrm{E}_{3}
  +s_{12}^{2}\left(\frac{1}{8} (\dgcr \mathrm{E}_{2})^{2} -\frac{3}{40} \dgcr^{2}\mathrm{E}_{4} \right)+ O(s_{12}^3) \bigg]
  \label{eq:7simp}\\
  W^{\tau}_{(3|1)}  &=  \frac{ \pi^2 }{(\Im \tau)^3} \bigg[  \frac{1}{2} \dgcr \mathrm{E}_{2}
  +\frac{1}{6}s_{12} \dgcr \mathrm{E}_{3} \notag\\
                    &\quad+s_{12}^{2}\left(\frac{1}{4} \mathrm{E}_{2}\dgcr \mathrm{E}_{2}
                    +\frac{3}{20}  \dgcr \mathrm{E}_{4} + \dgcr \mathrm{E}_{2,2} \right)+ O(s_{12}^3) \bigg]\notag\\
  W^{\tau}_{(2|2)}  &= \Big(\frac{\pi}{\Im\tau}\Big)^{2} \bigg[ \mathrm{E}_{2}+s_{12} \mathrm{E}_{3}\notag\\
                    &\quad+s_{12}^{2}\left(\frac{1}{4}\frac{(\dgcrbar \mathrm{E}_{2})(\dgcr \mathrm{E}_{2})}{(\Im\tau)^{2}}
                 -\frac{1}{2}\mathrm{E}_{2}^{2}+\frac{9}{5}\mathrm{E}_{4}+2 \mathrm{E}_{2,2} \right)+ O(s_{12}^3) \bigg]\notag\\
 W^{\tau}_{(4|2)}  &= \frac{ \pi^3 }{(\Im \tau)^4} \bigg[ \frac{1}{3} \dgcr \mathrm{E}_{3}+s_{12}\Big(\frac{3}{4}
 \dgcr \mathrm{E}_{4} -\frac{1}{2}  \mathrm{E}_{2}\dgcr \mathrm{E}_{2} \Big)\notag\\
  &\quad+s_{12}^{2}\Big(\frac{27}{14} \dgcr \mathrm{E}_{5} {+}  \dgcr \mathrm{E}_{2,3}
  {+}\frac{1}{24}\frac{ (\dgcrbar\mathrm{E}_{2})\dgcr^{2}\mathrm{E}_{3}}{(\Im\tau)^{2}}{-}\frac{1}{2}  \mathrm{E}_{3}\dgcr \mathrm{E}_{2}{-}\frac{2}{3}  \mathrm{E}_{2}\dgcr \mathrm{E}_{3} \Big)+ O(s_{12}^3) \bigg]\,,\notag
\end{align}
where the simplified $\ap$-expansion of $W^{\tau}_{(1|1)}$ follows from inserting the first line
of \eqref{eq:7simp} into \eqref{ibp2b}.
Note that none of the modular graph forms on the right-hand sides of (\ref{eq:7}) is amenable to
holomorphic subgraph reduction reviewed in Appendix~\ref{sec:HSR}. In fact, since the $z_{ij}$
arguments of the chiral or anti-chiral Kronecker--Eisenstein integrands of the $W$-integrals 
do not form any cycles, none of the modular graph forms in the $n$-point $\ap$-expansions (\ref{eq:20}) will allow for
holomorphic subgraph reduction.

\subsubsection{Three-point examples}\label{sec:3ptexmpl}
From three points onward, the generating functions and component integrals start depending on a chiral and an anti-chiral permutation. Following  \eqref{cptnpt1}, we introduce three-point component integrals by
\begin{align}
  W_{\eta_{2},\eta_{3}}^\tau(\sigma|\rho)&=\sum_{a_{2},a_{3}=0}^{\infty}\sum_{b_{2},b_{3}=0}^{\infty}\rho[\eta_{23}^{a_{2}-1}\eta_{3}^{a_{3}-1}]\sigma[\bar{\eta}_{23}^{b_{2}-1}\bar{\eta}_{3}^{b_{3}-1}]W^{\tau}_{(a_{2},a_{3}|b_{2},b_{3})}(\sigma|\rho)\label{eq:12}\\
  W^{\tau}_{(a_{2},a_{3}|b_{2},b_{3})}(\sigma|\rho)&=\int\!\frac{\dd^{2}z_{2}}{\Im\tau}\frac{\dd^{2}z_{3}}{\Im\tau}\,\rho[f_{12}^{(a_{2})}f_{23}^{(a_{3})}]\sigma[\overline{f_{12}^{(b_{2})}}\,\overline{f_{23}^{(b_{3})}}]\KN^{\tau}_{3}\ ,\label{eq:11}
\end{align}
where $\rho,\sigma \in {\cal S}_2$ act on the subscripts $i,j \in \{2,3\}$ of the $\eta$ and $\bar\eta$ in \eqref{eq:12} and of $f^{(n)}$ and $\overline{f^{(n)}}$ in \eqref{eq:11} but not on those of $a_i$ and $b_j$.

As in the two-point case, kinematic poles arise if the integrand develops a $\frac{1}{|z|^{2}}$ singularity in (some combination of) the punctures. The details of how to treat these poles using integration-by-parts-identities are spelled out in Appendix~\ref{app:poleB}.

In contrast to the two-point case, the three-point $\ap$-expansions also contain trihedral modular graph forms as defined in \eqref{eq:29}. Nevertheless, using the identities from Appendix~\ref{sec:idMGF}, the leading orders displayed below can also be brought into the basis spanned by the $\mathrm{E}_{k}$, their higher-depth generalizations and derivatives:
\begin{align}
  W^{\tau}_{(0,0|0,0)}(2,3|2,3)&=1+\frac{1}{2}(s_{12}^{2}{+}s_{13}^{2}{+}s_{23}^{2})\mathrm{E}_{2}+\frac{1}{6}(s_{12}^{3}{+}s_{13}^{3}{+}s_{23}^{3})(\mathrm{E}_{3}{+}\zeta_{3})+s_{12}s_{13}s_{23}\mathrm{E}_{3}+O(s_{12}^4)\notag\\
  W^{\tau}_{(2,0|2,0)}(2,3|2,3)&=\Big(\frac{\pi}{\Im\tau}\Big)^{2} \bigg[ \mathrm{E}_{2}+s_{12} \mathrm{E}_{3}
  +  \frac{1}{2}(s_{13}^{2}+s_{23}^{2}-s_{12}^{2})\mathrm{E}_{2}^{2}\notag\\
                               &\hspace{-2.6em}+s_{13}s_{23}\Big(\frac{3}{2}\mathrm{E}_{2}^{2}-\frac{33}{10}\mathrm{E}_{4}-2 \mathrm{E}_{2,2}\Big)+s_{12}^{2}\Big(\frac{1}{4}\frac{\dgcr \mathrm{E}_{2}\dgcrbar\mathrm{E}_{2}}{(\Im\tau)^{2}}+\frac{9}{5}\mathrm{E}_{4}+2 \mathrm{E}_{2,2}\Big) +O(s_{12}^3) \bigg]\notag\\
  W^{\tau}_{(2,0|2,0)}(2,3|3,2)&= \Big(\frac{\pi}{\Im\tau}\Big)^{2} \bigg[ s_{23}\mathrm{E}_{3}
  +   (s_{12}+s_{13})s_{23}\Big({-}\frac{1}{2}\mathrm{E}_{2}^{2}+\frac{39}{20}\mathrm{E}_{4}+\frac{1}{2}\mathrm{E}_{2,2}\Big)\notag\\
                               &\hspace{-2.6em}+s_{23}^{2}\Big(\frac{9}{20}\mathrm{E}_{4}+\frac{1}{2}\mathrm{E}_{2,2}\Big)+s_{12}s_{13}\frac{1}{4}\frac{\dgcr \mathrm{E}_{2}\dgcrbar\mathrm{E}_{2}}{(\Im\tau)^{2}} +O(s_{12}^3) \bigg]\\
  W^{\tau}_{(1,2|1,2)}(2,3|3,2)&=\Big(\frac{\pi}{\Im\tau}\Big)^{3} \bigg[ \mathrm{E}_{3}-   (s_{12}+s_{13}+s_{23})\mathrm{E}_{2}^{2}\notag\\
                               &\hspace{-2.6em}+(s_{12}+s_{13}+2s_{23})\Big(\frac{9}{5}\mathrm{E}_{4}+2 \mathrm{E}_{2,2}+\frac{1}{4}\frac{(\dgcr \mathrm{E}_{2})(\dgcrbar \mathrm{E}_{2})}{(\Im\tau)^{2}}\Big) +O(s_{12}^2) \bigg]\ .\notag
\end{align}
Note in particular that although $W^{\tau}_{(2,0|2,0)}(2,3|2,3)$ and $W^{\tau}_{(2,0|2,0)}(2,3|3,2)$ differ just in their chiral permutations, their $\ap$-expansions are very different.


\subsubsection{From modular graph forms to component integrals}
\label{extrastuff}

\begin{table}
  \begin{center}
    \begin{tabular}{ccc}  
      \toprule
       Modular graph form & $\dd \mu_{n-1}$ integrand & Component integral \\
      \midrule
      $\bigg.\cform{a&1_k \\b&1_k&}\bigg.$ & $f^{(a)}_{12} \, \overline{ f^{(b)}_{13} } \, G(z_{23})^k$ & $W^\tau_{(a,0|b,0)}(3{,}2|2{,}3)$ \\
      $\bigg.\cform{a&b &0 &1_k \\c &0 &d &1_k}\bigg.$ & $f^{(a)}_{12} \,f^{(b)}_{23}\, \overline{ f^{(c)}_{13} }
\, \overline{ f^{(d)}_{32} } \, G(z_{23})^k$ & $W^\tau_{(a,b|c,d)}(3{,}2|2{,}3)$ \\
      $\bigg.\cform{a&b &1_k \\c  &d &1_k}\bigg.$ & $f^{(a)}_{12} \,f^{(b)}_{24}\, \overline{ f^{(c)}_{13} }
\, \overline{ f^{(d)}_{34} } \, G(z_{23})^k$ & $W^\tau_{(a,b,0|c,d,0)}(3{,}4{,}2|2{,}4{,}3)$ \\
      $\bigg.\cformtri{a &1_k\\d &1_k}{ b &0 \\ 0 &v }{c &0 \\0 &u} \bigg.$ & $f^{(a)}_{12} \,f^{(b)}_{23}\, f^{(c)}_{34} \, \overline{ f^{(d)}_{14} }
\, \overline{ f^{(u)}_{43} } \, \overline{ f^{(v)}_{32} } \, G(z_{24})^k$ & $W^\tau_{(a,b,c|d,u,v)}(4{,}3{,}2|2{,}3{,}4)$ \\
      \bottomrule
    \end{tabular}
    \caption{Component integrals giving rise to different modular graph forms.}
    \label{tab:MGFCInt}
  \end{center}
\end{table}

The closed formul\ae{} (\ref{eq:7}) for two-point component integrals allow
to identify infinite families of modular graph forms within their $\ap$-expansion.
Similarly, we shall now list possible realizations of more general modular graph forms 
in $(n\geq 3)$-point component integrals. Like this, the differential equations of the
modular graph forms in Table~\ref{tab:MGFCInt} can be extracted from the
differential equations of $W$-integrals in later sections. It is straightforward to extend the list to
arbitrary graph topologies, where the multiplicity 
of the associated component integrals will grow with the complexity of the graph. 
Note that the $k$ powers of Green functions $G(z_{ij})$ instruct to extract the coefficients
of $s_{ij}^{k}$ from the component integrals.

\section{Modular differential operators, Cauchy--Riemann- and Laplace equations}
\label{sec:CR}
In this section, we introduce the Cauchy--Riemann and Laplace operators that we will use 
to derive the differential equations for generating functions of world-sheet integrals.

\subsection{Maa\ss{} raising and lowering operators}\label{sec:maass-ops}

We define the following standard Maa\ss{} differential operators~\cite{Maass}:
\begin{align}
  \label{eq:MRL}
  \nabla^{(a)} := (\tau-\bar\tau)  \frac{\partial}{\partial \tau} + a \,,\hspace{15mm}
  \overline\nabla^{(b)}  :=  (\bar\tau-\tau)  \frac{\partial}{\partial \bar\tau} + b\,.
\end{align}
These have the property that they act on functions transforming with modular weights $(a,b)$ as\footnote{Note that the differential operator $\dgcr$, defined in \eqref{eq:31} to compactly write modular graph forms, is related to the Maa\ss{} operators by $\dgcr=\Im\tau \nabla^{(0)}$. Therefore its image has only nice modular properties if $\dgcr$ acts on a modular invariant. For this reason, we will not use $\dgcr$ to derive Cauchy--Riemann or Laplace equations in the following sections. We note also that, acting on modular invariant functions $f$, one has $\dgcr^n f = (\Im\tau)^n \nabla^{(n-1)} \cdots \nabla^{(1)} \nabla^{(0)} f$, a version of Bol's identity~\cite{Bol:1949}.}
\begin{align}
  \nabla^{(a)}: (a,b) \mapsto (a+1,b-1)\,,\hspace{15mm}
  \overline\nabla^{(b)}: (a,b) \mapsto (a-1,b+1)\
\end{align}
when using that the derivatives transform as
$\partial_\tau \to (\gamma\tau+\delta)^2 \partial_\tau$ and $\partial_{\bar\tau} \to (\gamma\bar\tau+\delta)^2 \partial_{\bar\tau}$
under modular transformations. The Maa\ss{} raising and lowering operators satisfy the product rule
\begin{align}
  \label{eq:RLprod}
  \nabla^{(a+a')} (fg) = (\nabla^{(a)} f) g +f (\nabla^{(a')}  g)
\end{align}
for any $a$ and $a'$ and similarly for $\overline\nabla^{(b)}$. Here, $f$ and $g$ can be any real analytic functions and do not need to have definite modular transformation properties.

The differential operators~\eqref{eq:MRL} should be thought of as raising and lowering operators of the action of ${\rm SL}_{2}(\RR)$ on the space spanned by functions transforming with modular weights $(a,b)$ for some $a,b$. The diagonal element of this action obtained by commutation is given by the scalar
\begin{align}
  \label{eq:MRLcomm}
  h^{(a,b)}:=\nabla^{(a-1)} \overline\nabla^{(b)} - \overline\nabla^{(b-1)} \nabla^{(a)} = a-b
\end{align}
acting on any function of modular weight $(a,b)$.\footnote{The connection to ${\rm SL}_{2}(\RR)$ representations can also be seen by noting that ${\rm E}_k$ and ${\rm G}_k$ are trivially eigenfunctions of the corresponding $h^{(a,b)}$, namely $h^{(0,0)}$ for ${\rm E}_k$ and $h^{(k,0)}$ for ${\rm G}_k$. The function ${\rm G}_k$ satisfies moreover $\overline{\nabla}^{(k)} {\rm G}_k=0$ and is thus a lowest-weight vector, namely that of a discrete series unitary representation. ${\rm E}_k$ is the so-called spherical vector in a principal-series representation of ${\rm SL}_{2}(\RR)$; see~\cite{Fleig:2015vky} for more information on this connection.}
We define a Laplace operator on functions of modular weight $(a,b)$ by\footnote{This differs by an overall sign from the one in~\cite{Brown:2017qwo, Brown:2017qwo2} and from the standard second-order ${\rm SL}_{2}(\RR)$-invariant Casimir $C_2^{(a,b)} := \nabla^{(a-1)}\overline{\nabla}^{(b)} + \frac14 h^{(a,b)}(h^{(a,b)}-2)$ by the constant $-\frac14(a+b)(a+b-2)$. Both operators reduce to the standard scalar Laplacian~\eqref{eq:hypLap} and satisfy $\overline{\Delta^{(a,b)} f}= \Delta^{(b,a)} \overline{f}$ since their difference is symmetric under the interchange $a\leftrightarrow b$.}
\begin{align}
  \label{eq:Daa}
  \Delta^{(a,b)} &:= \nabla^{(a-1)} \overline\nabla^{(b)} -b(a-1)=  \overline\nabla^{(b-1)} \nabla^{(a)} - a(b-1) \\
                 &\phantom{:}= - (\tau-\bar\tau)^2 \partial_\tau\partial_{\bar\tau} + b(\tau-\bar\tau)\partial_\tau + a (\bar\tau-\tau) \partial_{\bar\tau} \,,\nn
\end{align}
where we have chosen the overall normalization to be such that $\Delta^{(a,b)}$ reduces to the ordinary hyperbolic Laplacian 
\begin{align}
  \label{eq:hypLap}
  \Delta = - (\tau-\bar\tau)^2 \partial_\tau\partial_{\bar\tau} =
  (\Im\tau)^2 \left(\frac{\partial^2}{\partial (\Re\tau)^2} + \frac{\partial^2}{\partial (\Im\tau)^2}\right)
\end{align}
on modular invariant functions with $(a,b)=(0,0)$.

Let us consider the action of the Maa\ss{} operators on a modular graph form~\eqref{eq:MGF} under the assumption of absolute convergence so that differentiation and summation can be interchanged freely. Recall from~\eqref{eq:MGFtrm} that the total modular weight of a general modular graph function ${\cal C}_\Gamma(\tau)$ is given by $(|A|,|B|)$. Hence, on $\mathcal{C}_{\Gamma}(\tau)$, the action of the raising and lowering operators \eqref{eq:MRL} is given by
\begin{align}
  \label{eq:MFGRL}
  \nabla^{(|A|)} {\cal C}_\Gamma(\tau) = \sum_{e\in E_{\Gamma}} a_e \, {\cal C}_{\Gamma+s_e} (\tau)\,, \hspace{15mm}
  \overline\nabla^{(|B|)} {\cal C}_\Gamma(\tau) = \sum_{e\in E_{\Gamma}} b_e \, {\cal C}_{\Gamma+\bar{s}_e} (\tau)\,,
\end{align}
where $\Gamma+s_e$ is the same graph as $\Gamma$ but with the decoration on edge $e$ shifted to $(a_e+1,b_e-1)$. Similarly $\Gamma+\bar{s}_e$ is the same graph $\Gamma$ but with the decoration on edge $e$ shifted to $(a_e-1,b_e+1)$. In deriving~\eqref{eq:MFGRL} we have used the product rule~\eqref{eq:RLprod} and that
\begin{align}
  \nabla^{(a)} p^{-a} = a\, p^{-a-1} \,\bar{p}\,,\hspace{15mm} \overline\nabla^{(b)} \bar{p}^{-b} =b \, p\, \bar{p}^{-b-1} \,.
\end{align}

\subsection{Differential operators on generating series}
\label{sec:CRgen}

Equipped with the differential operators introduced in the previous section, we will derive and study differential equations satisfied by 
the generating integrals $W_{\vec\eta}^\tau$ defined in~\eqref{eq:Wdef} in the remainder of this work. 

These differential equations describe the dependence of $W_{\vec\eta}^\tau$ on $\tau$ at all orders in $\ap$ and in the series parameters $\vec\eta=(\eta_2,\eta_3,\ldots,\eta_n)$. As $W_{\vec\eta}^\tau$ is defined as an integral over the world-sheet torus with complex structure parameter $\tau$ we first have to clarify how the $\tau$-derivative acts on such integrals. Our convention will always be to treat such torus integrals as
\begin{align}
  \int \frac{\dd^2z}{\Im \tau} = \int_{[0,1]^2}\! \dd u  \, \dd v\,,
\end{align}
such that they are taken to \textit{not} depend on $\tau$ when the torus coordinate $z=u\tau+v$ is written in terms of two real variables along the unit square. The $\tau$-derivative is then always taken \textit{at constant $u$ and $v$} and will only act on the integrand.

While the definitions (\ref{eq:MRL}) and (\ref{eq:Daa}) of the Maa\ss{} operators and the Laplace operator
are tailored to functions of definite modular weights $(a,b)$, the $W$-integrands in (\ref{eq:Wdef}) mix different
modular weights in their expansion w.r.t.\ $\eta_j$ and $\bar \eta_j$. More precisely,
the component integrals $W_{(A|B)}^\tau(\sigma|\rho)$
in (\ref{cptnpt1}) have modular weights $(|A|,|B|)$, using the notation \eqref{eq:2}.

Hence, it remains to find a representation of the Maa\ss{} operator \eqref{eq:MRL}
such that its action on the expansion (\ref{cptnp2}) of $n$-point $W$-integrals is compatible
with the modular weights of the component integrals. The modular weights
of the $W_{(A|B)}^\tau(\sigma|\rho)$ correlate with the homogeneity degrees in the $\eta_j$ and $\bar\eta_j$ that is measured by the differential operators $\sum_{j=2}^n \eta_j \partial_{\eta_j}$ and $\sum_{j=2}^n \bar \eta_j \partial_{\bar \eta_j}$, respectively. We therefore define the following operators on functions depending on $\tau$ and $\vec\eta$
\begin{subequations}
  \label{eq:MRLn}
  \begin{align}
    \label{eq:MRn}
    \nabla_{\vec\eta}^{(k)} &:= 
    (\tau-\bar\tau) \partial_\tau + k + \sum_{j=2}^n \eta_j \partial_{\eta_j} \\
    \label{eq:MLn}
    \overline{\nabla}_{\vec{\eta}}^{(k)} &:= 
     (\bar\tau-\tau) \partial_{\bar\tau} + k + \sum_{j=2}^n \bar{\eta}_j \partial_{\bar{\eta}_j}\ .
  \end{align}
\end{subequations}
Due to the shift in the expansion of the component integrals~\eqref{cptnp2}, there is an offset between the eigenvalues of  $(\sum_{j=2}^n \eta_j \partial_{\eta_j} ,\sum_{j=2}^n \bar \eta_j \partial_{\bar \eta_j} )$ and the weights $(|A|,|B|)$ according to
\begin{subequations}
  \label{cptboth}
  \begin{align}
 \sum_{j=2}^n \eta_j \partial_{\eta_j}  \,
    \rho\big[ \eta_{234\ldots n}^{a_2-1} \eta_{34\ldots n}^{a_3-1} \ldots \eta_{n}^{a_n-1} \big]    
    &= ( |A| -(n-1))\,
      \rho\big[ \eta_{234\ldots n}^{a_2-1} \eta_{34\ldots n}^{a_3-1} \ldots \eta_{n}^{a_n-1} \big]
      \label{cptnp12}
    \\
 \sum_{j=2}^n \bar \eta_j \partial_{\bar \eta_j}  \,
    \sigma \big[ \bar \eta_{234\ldots n}^{b_2-1} \bar \eta_{34\ldots n}^{b_3-1} \ldots \bar \eta_{n}^{b_n-1}  \big] 
    &= ( |B| -(n-1))\,
      \sigma \big[ \bar \eta_{234\ldots n}^{b_2-1} \bar \eta_{34\ldots n}^{b_3-1} \ldots \bar \eta_{n}^{b_n-1}  \big] \, .
      \label{cptnp13}
  \end{align}
\end{subequations}
Thus we have to set $k=n-1$ in (\ref{eq:MRn}) in order to obtain
\begin{align}
  \nabla^{(n-1)}_{\vec{\eta}} W_{\vec{\eta}}^\tau(\sigma|\rho) &= \sum_{A,B} 
                                                         \nabla^{(|A|)} W_{(A|B)}^\tau(\sigma|\rho)\rho\big[ \eta_{234\ldots n}^{a_2-1} \eta_{34\ldots n}^{a_3-1} \ldots \eta_{n}^{a_n-1} \big]    
                                                         \, \sigma \big[ \bar \eta_{234\ldots n}^{b_2-1} \bar \eta_{34\ldots n}^{b_3-1} \ldots \bar \eta_{n}^{b_n-1}  \big] \label{cptnp8}
\end{align}
that relates the raising operator on the generating series $W^\tau_{\vec\eta}$ correctly to the raising operator on the component integrals $W^\tau_{(A|B)}$ of definite modular weight $(|A|,|B|)$. In~\eqref{cptboth}, we have also used the $\mathcal{S}_{n-1}$ permutation invariance of the raising and lowering operators~\eqref{eq:MRLn} that descends to the specific sums (\ref{eq:etasum}) of the $\eta_j$-variables in the expansion
of the $W$-integrals via
\begin{align}
  \label{eq:etasum2}
  \sum_{j=2}^n \eta_j \partial_{\eta_j} = \sum_{j=2}^n \eta_{j,j+1\ldots n} \partial_{\eta_{j,j+1\ldots n}}
  = \sum_{j=2}^n \rho \big[ \eta_{j,j+1\ldots n} \partial_{\eta_{j,j+1\ldots n}} \big] \, .
\end{align}

The Laplace operator can be defined in a similar fashion to (\ref{eq:MRLn}) as
\begin{align}
\Delta_{\vec{\eta}} := \overline{\nabla}^{(n-2)}_{\vec{\eta}} \nabla^{(n-1)}_{\vec{\eta}} - \Big( n-1 + \sum_{j=2}^n \eta_j \partial_{\eta_j} \Big)  
\Big( n-2 + \sum_{j=2}^n \bar{\eta}_j \partial_{\bar{\eta}_j} \Big) \label{eq:lapW}
\end{align}
such that it acts on component integrals $W_{(A|B)}^\tau(\sigma|\rho)$ 
via $\Delta^{(|A|,|B|)}$ in \eqref{eq:Daa} with appropriate weights $(|A|,|B|)$:
\begin{align}
  \Delta_{\vec{\eta}} W_{\vec{\eta}}^\tau(\sigma|\rho) &= \sum_{A,B} 
                                                         \Delta^{(|A|,|B|)} W_{(A|B)}^\tau(\sigma|\rho)   \label{cptnp9} \\
                                                       &\ \ \ \  \times \rho\big[ \eta_{234\ldots n}^{a_2-1} \eta_{34\ldots n}^{a_3-1} \ldots \eta_{n}^{a_n-1} \big]    
                                                         \, \sigma \big[ \bar \eta_{234\ldots n}^{b_2-1} \bar \eta_{34\ldots n}^{b_3-1} \ldots \bar \eta_{n}^{b_n-1}  \big]  \, .\notag 
\end{align}
These expressions are invariant under permutation of $\eta_2,\eta_3,\ldots,\eta_n$, i.e.\ 
$\rho \big[\nabla_{\vec\eta}^{(k)}\big] = \nabla_{\vec\eta}^{(k)}$ for any $\sigma,\rho \in {\cal S}_{n-1}$, and
valid at any order in the $(\eta_j,\bar \eta_j)$- and $\ap$-expansions of $W^\tau_{\vec{\eta}}$ at $n$ points.


In the remainder of this section we work out the first-order Cauchy--Riemann equation and the second-order Laplace equation satisfied by $W^\tau_{\vec\eta}$ for two points in order to illustrate the basic manipulations.  We will dedicate Sections~\ref{sec:bigCRn} and \ref{sec:Lap} to the
Cauchy--Riemann equations and Laplace equations of $n$-point $W$-integrals.

\subsection{Two-point warm-up for differential equations}
\label{sec:CR2}

For the simplest case of $n=2$, there are no permutations to consider, and the $(n{-}1)!\times (n{-}1)!$ matrix
in (\ref{eq:Wdef}) reduces to the real scalar
\begin{align}
  \label{eq:W2pt}
  W^\tau_{\eta} = \int \frac{\dd^2z_2}{\Im \tau}\, \Omega(z_{12},\eta,\tau) \, \overline{\Omega(z_{12},\eta,\tau)} \, \KN^{\tau}_{2} \,,
\end{align}
where we have denoted $\eta=\eta_2$ for simplicity. 


\subsubsection{Cauchy--Riemann equation}
\label{sec:cr2pts}

Under the two-point instance $\nabla_\eta^{(1)} = (\tau-\bar \tau) \partial_\tau + 1 + \eta \partial_\eta$ of the operator~\eqref{eq:MRn}, the two-point $W$-integral (\ref{eq:W2pt}) satisfies
\begin{align}
  \label{eq:N1W2}
  2\pi i \nabla_\eta^{(1)} W^\tau_\eta  &= \int \frac{\dd^2z_2}{\Im\tau} \, \bigg[-(\tau-\bar\tau)\big( \partial_{z_2}\partial_\eta \Omega (z_{12},\eta,\tau)\big)  \, \overline{\Omega(z_{12},\eta,\tau)} \, \KN^{\tau}_{2} \nn\\
                                        &\hspace{40mm}- s_{12} (\tau-\bar\tau)  f^{(2)}_{12} \Omega (z_{12},\eta,\tau)  \, \overline{\Omega(z_{12},\eta,\tau)} \, \KN^{\tau}_{2} 
                                          \bigg]\nn\\
                                        &= \int \frac{\dd^2z_2}{\Im\tau} \, \bigg[2\pi i \bar{\eta} \partial_\eta +s_{12}  (\tau-\bar\tau) \left( f^{(1)}_{12} \partial_\eta -f^{(2)}_{12} \right)\bigg]\Omega (z_{12},\eta,\tau)  \, \overline{\Omega(z_{12},\eta,\tau)} \, \KN^{\tau}_{2}\nn\\
                                        &= \int \frac{\dd^2z_2}{\Im\tau} \, \bigg[2\pi i \bar{\eta} \partial_\eta +s_{12}  (\tau-\bar\tau) \left( \frac12\partial_\eta^2 - \wp(\eta,\tau) \right)\bigg]\Omega (z_{12},\eta,\tau)  \, \overline{\Omega(z_{12},\eta,\tau)} \, \KN^{\tau}_{2}\nn\\
                                        &= \bigg[ 2\pi i \bar{\eta}\partial_\eta + (\tau-\bar\tau) \sv D^\tau_\eta \bigg] W_\eta^\tau\, ,
\end{align}
where the first line on the right-hand side stems from~\eqref{eq:Omixh2} (for fixed coordinates $u$ and $v$ in the torus integral) and the second one from the Koba--Nielsen derivative~\eqref{eq:dtauKN}. In passing to the third line, we have integrated $\partial_{z_2}$ by parts in the first term\footnote{There are no boundary terms arising in this process since they are suppressed by the Koba--Nielsen factor.} and used~\eqref{eq:dzbO}, \eqref{eq:dzKN} and the fact that $\partial_\eta$ only acts on the $\Omega$ factor in the product. For the next equality, one simplifies
$(f^{(1)}_{12} \partial_\eta -f^{(2)}_{12} ) \Omega (z_{12},\eta,\tau)$ via (\ref{eq:25})
that produces a Weierstra\ss{} function $\wp(\eta,\tau)$. Since the differential operator in $\eta$ does not depend on $z_{2}$, 
we have moved it out of the integral and defined
\begin{align}
  \sv D_\eta^\tau := s_{12} \left( \frac12 \partial_\eta^2 -\wp(\eta,\tau)\right)
  \label{svD2pt}
\end{align}
in passing to the last line of (\ref{eq:N1W2}).
Note that this operator is meromorphic in $\eta$ and $\tau$. 

The notation $\sv D_\eta^\tau$ in (\ref{svD2pt}) is motivated by the analogous differential equations
for $A$-cycle integrals in (\ref{again2}). Their two-point instance $Z^\tau_\eta = \int^1_0 \dd z_2 \, \Omega(z_{12},\eta,\tau)e^{ s_{12}G_A(z_{12},\tau)}$ was shown in~\cite{Mafra:2019ddf, Mafra:2019xms} to
obey the differential equation
\begin{align}
  2\pi i \partial_\tau Z^\tau_\eta = D^\tau_\eta Z^\tau_\eta \, , \ \ \ \ \ \ 
  D^\tau_\eta = s_{12} \left( \frac12 \partial_\eta^2 -\wp(\eta,\tau)-2 \zeta_2\right) \, .
  \label{dsv1}
\end{align}
The open-string differential operator $D^\tau_\eta$ only differs from its closed-string counterpart 
(\ref{svD2pt}) through the additional term $-2\zeta_2 s_{12}$. As a formal prescription to drop 
the $\zeta_2$-contribution to $D^\tau_\eta$, we refer to the single-valued map for (motivic) 
MZVs $\sv: \, \zeta_2 \rightarrow 0$ \cite{Schnetz:2013hqa, Brown:2013gia} in 
the notation for $\sv D_\eta^\tau$ in (\ref{dsv1}).


\subsubsection{Laplace equation}
\label{sec:Lap2}

According to~\eqref{eq:lapW}, the representation of the Laplacian on the two-point
$W$-integral (\ref{eq:W2pt}) is given by $\Delta_\eta =  \overline{\nabla}_\eta^{(0)} \nabla_\eta^{(1)}
-(1+ \eta \partial_\eta) \bar \eta \partial_{\bar \eta}$.
In order to evaluate the action of the Maa\ss{} operators, we introduce the short-hand
\begin{align}
  Q^\tau_\eta := 2\pi i \bar{\eta}\partial_\eta + (\tau-\bar\tau) \sv D^\tau_\eta
\end{align}
for the operator in the Cauchy--Riemann equation $2\pi i \nabla_\eta^{(1)} W^\tau_\eta  = Q^\tau_\eta  W^\tau_\eta$ derived in (\ref{eq:N1W2}). The action of $\overline{\nabla}^{(0)}_{\eta}= \overline{\nabla}_\eta^{(1)}-1$ on $Q^\tau_\eta  W^\tau_\eta$ can be conveniently inferred by means of the commutation relation
\begin{align}
  [\overline{\nabla}_\eta^{(1)},Q^\tau_\eta ] 
    & = Q^\tau_\eta  \label{nabQ2}
\end{align}
along with the complex conjugate $-2\pi i \overline{\nabla}_\eta^{(1)} W_\eta^\tau =
\overline{Q^\tau_{\eta}}   W_\eta^\tau $ of (\ref{eq:N1W2}),
\begin{align}
  (2\pi i)^2 \overline{\nabla}_\eta^{(0)} \nabla_\eta^{(1)} W^\tau_\eta &= 2\pi i  ( \overline{\nabla}_\eta^{(1)} - 1) Q^\tau_{\eta} W_\eta^\tau = 2\pi i \Big(   Q^\tau_{\eta} \overline{\nabla}_\eta^{(1)} + [\overline{\nabla}_\eta^{(1)} , Q^\tau_{\eta}] - Q^\tau_{\eta} \Big) W^\tau_\eta \notag \\
                                                                        &=   2\pi i  Q^\tau_{\eta} (  \overline{\nabla}_\eta^{(1)} W_\eta^\tau )
                                                                          = - Q^\tau_{\eta} \overline{Q^\tau_{\eta}}   W_\eta^\tau \label{eq:Nb0N1W2}\\
 &= \left(2\pi i \bar\eta \partial_\eta +(\tau-\bar\tau) \sv D_\eta^\tau\right)\left( 2\pi i \eta\partial_{\bar\eta} +(\tau-\bar\tau) \overline{\sv D_\eta^\tau}\right) W_\eta^\tau \, . \notag
\end{align}
While the operator $\overline{\sv D^\tau_\eta} = s_{12} \left( \frac12 \partial_{\bar\eta}^2 -\wp(\bar\eta,\bar\tau)\right)$
commutes with $\sv D_\eta^\tau$, two extra contributions arise when reordering $\sv D_\eta^\tau \eta\partial_{\bar\eta} = \eta\partial_{\bar\eta}\sv D_\eta^\tau +   s_{12} \partial_{\eta}  \partial_{\bar\eta} $
and $ \bar\eta \partial_\eta  \eta\partial_{\bar\eta} = \bar\eta \partial_{\bar\eta}  
+ \bar\eta \eta \partial_\eta \partial_{\bar\eta} $, cf.\ (\ref{svD2pt}). Hence, we are led 
to the following alternative form of (\ref{eq:Nb0N1W2}),
\begin{align}
  \label{eq:Nb0N1W23}
  (2\pi i)^2 \overline{\nabla}_\eta^{(0)} \nabla_\eta^{(1)} W^\tau_\eta &= 
                                                                          \bigg[ (2\pi i)^2 \bar\eta\partial_{\bar\eta} +(2\pi i)^2 \eta\bar\eta\partial_\eta\partial_{\bar\eta}+2\pi i s_{12}(\tau-\bar\tau)\partial_\eta \partial_{\bar\eta}\\
                                                                        &\hspace{9mm} + 2\pi i (\tau-\bar\tau) (\bar\eta \partial_\eta \overline{\sv D_\eta^\tau} + \eta \partial_{\bar\eta} \sv D_\eta^\tau) +(\tau-\bar\tau)^2 \sv D_\eta^\tau \overline{\sv D_\eta^\tau}\bigg] W_\eta^\tau \, .\nn
\end{align}
Finally, the first two terms in the square bracket cancel when completing the 
Laplacian:
\begin{align}
  \label{eq:Lap2}
  (2\pi i)^2 &\Delta_\eta W^\tau_\eta =  (2\pi i)^2 \left[ \overline{\nabla}_\eta^{(0)} \nabla_\eta^{(1)} -(1+\eta\partial_\eta)\bar\eta \partial_{\bar\eta} \right] W^\tau_\eta \\
             &= \bigg[ 2\pi i s_{12}(\tau-\bar\tau)\partial_\eta \partial_{\bar\eta} +2\pi i (\tau-\bar\tau) (\bar\eta \partial_\eta \overline{\sv D_\eta^\tau} + \eta \partial_{\bar\eta} \sv D_\eta^\tau) +(\tau-\bar\tau)^2 \sv D_\eta^\tau \overline{\sv D_\eta^\tau}\bigg] W_\eta^\tau\nn\,.
\end{align}


\subsection{Two-point warm-up for component integrals}
\label{sec:comp2}

We shall now translate the Cauchy--Riemann- and Laplace equations
(\ref{eq:N1W2}) and (\ref{eq:Lap2}) of the generating integral $W^\tau_\eta$ to the equations satisfied by its component integrals $W_{(a|b)}^\tau$ defined in~\eqref{lapsec15}.


\subsubsection{Cauchy--Riemann equation}
\label{sec:comp2.a}

At the level of component integrals (\ref{lapsec15}), the Cauchy--Riemann equations (\ref{eq:N1W2}) 
are equivalent to
\begin{align}
  2\pi i \nabla^{(a)}W^{\tau}_{(a|b)} &= s_{12}(\tau-\bar\tau) \Big[
                                        \frac{1}{2}(a{+}2)(a{-}1) W^{\tau}_{(a+2|b)} - \sum_{k=4}^{a+2} (k{-}1) {\rm G}_k W^{\tau}_{(a+2-k |b)}
                                        \Big] \notag \\
                                      &\ \ \ \ +  2\pi i a  W^{\tau}_{(a+1|b-1)} 
                                        \label{cr2ptcp}
\end{align}
with the understanding that $W^{\tau}_{(a|-1)}=W^{\tau}_{(-1|b)}=0$ for all $a ,b\geq0$. The simplest 
examples for low weights $a,b$ include
\begin{align}
  2\pi i  \nabla^{(0)}W^{\tau}_{(0|0)}&=-s_{12}(\tau-\bar\tau) W^{\tau}_{(2|0)}  \notag \\
  2\pi i  \nabla^{(2)}W^{\tau}_{(2|0)}&=-s_{12}(\tau-\bar\tau) (3 \mathrm{G}_{4}W^{\tau}_{(0|0)}-2W^{\tau}_{(4|0)}) \notag \\
  2\pi i  \nabla^{(1)}W^{\tau}_{(1|1)}&=2\pi i W^{\tau}_{(2|0)}  \notag\\
  2\pi i  \nabla^{(0)}W^{\tau}_{(0|2)}&=-s_{12}(\tau-\bar\tau) W^{\tau}_{(2|2)}  \notag\\
  2\pi i  \nabla^{(4)}W^{\tau}_{(4|0)}&=-s_{12}(\tau-\bar\tau) (5 \mathrm{G}_{6}W^{\tau}_{(0|0)}+3 \mathrm{G}_{4}W^{\tau}_{(2|0)}-9W^{\tau}_{(6|0)}) \label{cr2ptcpa} \\
  2\pi i  \nabla^{(3)}W^{\tau}_{(3|1)}&=-s_{12}(\tau-\bar\tau) (3 \mathrm{G}_{4}W^{\tau}_{(1|1)}-5W^{\tau}_{(5|1)}) + 3(2\pi i )W^{\tau}_{(4|0)}  \notag\\
  2\pi i  \nabla^{(2)}W^{\tau}_{(2|2)}&=-s_{12}(\tau-\bar\tau) (3 \mathrm{G}_{4}W^{\tau}_{(0|2)}-2W^{\tau}_{(4|2)}) + 2(2\pi i )W^{\tau}_{(3|1)}  \notag\\
  2\pi i  \nabla^{(1)}W^{\tau}_{(1|3)}&=2\pi i  W^{\tau}_{(2|2)}  \notag\\
  2\pi i  \nabla^{(0)}W^{\tau}_{(0|4)}&=-s_{12}(\tau-\bar\tau) W^{\tau}_{(2|4)}\ . \notag
\end{align}
These equations are obtained by direct evaluation of~\eqref{eq:N1W2} and are valid at all orders in $\ap$. In Section~\ref{sec:2pt-lessons-for-MGF}, we shall analyze their $\ap$-expansion and relate them to equations for modular graph forms.


\subsubsection{Laplace equation}
\label{sec:Lap2cpt}

The Laplace equation (\ref{eq:Lap2}) of the generating integrals $W^\tau_{\eta}$ 
implies the following component relations for the $W^\tau_{(a|b)}$ of modular weight $(a,b)$ 
in (\ref{lapsec15}):
\begin{align}
  (2\pi i)^2 \Delta^{(a,b)} W^\tau_{(a|b)} &= (\tau{-}\bar \tau)^2 s_{12}^2 \Big\{ \frac{1}{4}(a{+}2)(a{-}1)(b{+}2)(b{-}1) W^\tau_{(a+2|b+2)}  \notag \\
                                           & \ \ \ \ - \frac{1}{2}(a{+}2)(a{-}1)  \sum_{k=4}^{b+2}(k{-}1) \overline{\rm G}_k W^\tau_{(a+2|b+2-k)} \notag \\
                                           & \ \ \ \  - \frac{1}{2} (b{+}2)(b{-}1) \sum_{\ell=4}^{a+2} (\ell{-}1) {\rm G}_\ell W^\tau_{(a+2-\ell|b+2)} \label{lapsec95} \\
                                           & \ \ \ \  +\sum_{k=4}^{b+2}(k{-}1)  \sum_{\ell=4}^{a+2}  (\ell{-}1) {\rm G}_\ell \overline{\rm G}_k W^\tau_{(a+2-\ell|b+2-k)} \Big\}  \notag\\
                                           &+ 2\pi i (\tau{-}\bar \tau) s_{12} \Big\{ \frac{ 1}{2}(ab{-}2) (a{+}b)W^\tau_{(a+1|b+1)}
                                             \notag \\
& \ \ \ \ - b \sum_{k=4}^{a+1} (k{-}1) {\rm G}_k W^\tau_{(a+1-k|b+1)} 
  - a \sum_{\ell=4}^{b+1} (\ell{-}1)\overline{ {\rm G}_\ell} W^\tau_{(a+1|b+1-\ell)}\Big\} \, . \notag
\end{align}
The simplest examples include
\begin{align} 
  (2\pi i)^2\Delta^{(0,0)} W^\tau_{(0|0)} &=  s_{12}^2 ( \tau{-}\bar\tau )^2 W^\tau_{(2|2)} \notag \\
  (2\pi i)^2\Delta^{(1,1)} W^\tau_{(1|1)} &=  -2\pi i s_{12} (\tau{-}\bar\tau) W^\tau_{(2| 2)} \notag \\
  (2\pi i)^2\Delta^{(2,0)} W^\tau_{(2|0)} &=  s_{12}^2 (\tau{-}\bar\tau)^2 (3 {\rm G}_4 W^\tau_{(0|2)} - 2 W^\tau_{(4| 2)})
                                            -4 \pi i s_{12} (\tau{-}\bar\tau) W^\tau_{(3| 1)}  
                                            \label{lapsec16} \\
  (2\pi i)^2\Delta^{(2,2)} W^\tau_{(2|2)} &=   
                                            s_{12}^2 (\tau{-}\bar\tau)^2 (9 \bar{\rm G}_4 {\rm G}_4 W^\tau_{(0| 0)} - 6 {\rm G}_4 W^\tau_{(0| 4)} - 
                                            6 \bar{\rm G}_4 W^\tau_{(4| 0)} + 4 W^\tau_{(4|4)}) \notag \\
                                          & \ \ \ \ +  8\pi is_{12} (\tau{-}\bar\tau) W^\tau_{(3| 3)} \notag \\ 
  (2\pi i)^2\Delta^{(3,1)} W^\tau_{(3|1)} &=  2 \pi is_{12} (\tau{-}\bar\tau) (-3 {\rm G}_4 W^\tau_{(0| 2)} + 2 W^\tau_{(4| 2)}) \notag \\
  (2\pi i)^2\Delta^{(4,0)} W^\tau_{(4|0)} &= 
                                            s_{12}^2 (\tau{-}\bar\tau)^2 (5 {\rm G}_{6} W^\tau_{(0| 2)}  +     3 {\rm G}_4 W^\tau_{(2| 2)} - 9 W^\tau_{(6|2)})
                                            -8 \pi is_{12} (\tau{-}\bar\tau) W^\tau_{(5| 1)} \, .\notag 
\end{align}
These are again valid to all orders in $\ap$ and we shall discuss their $\ap$-expansion below. 

We note that the evaluation of the differential operators in~\eqref{eq:Lap2} a priori leads to double poles
in $\eta$ or $\bar \eta$ on the right-hand side.
These do not occur on the left-hand side and therefore have to cancel. While this is not necessarily manifest, their appearance can be traced back to the integration by parts that was used in the derivation (\ref{eq:N1W2}) of the differential equation. For this reason, the residues of the putative double poles vanish by the integration-by-parts identities \eqref{ibp2a}, that is why the $\eta$-expansions on both sides of~\eqref{eq:Lap2} are identical.


\subsubsection{Lessons for modular graph forms}\label{sec:2pt-lessons-for-MGF}

In Section~\ref{sec:2ptCmptInts}, we calculated the leading orders in the $\ap$-expansions 
of two-point component integrals in terms of modular graph forms. Using these expansions, \eqref{cr2ptcp} and \eqref{lapsec95} imply Cauchy--Riemann and Laplace equations for modular graph forms.

As an example, using (\ref{eq:7}) to expand the right-hand side of the Cauchy--Riemann 
equation \eqref{cr2ptcpa} for $W^{\tau}_{(2|0)}$ leads to
\begin{align}
\nabla^{(2)}W^{\tau}_{(2|0)}&=
 -s_{12}\frac{\Im\tau}{\pi} (3 \mathrm{G}_{4}W^{\tau}_{(0|0)}-2W^{\tau}_{(4|0)})
 \label{eq:5} \\
                            &=-3s_{12}\frac{\Im\tau}{\pi}\mathrm{G}_{4}-2s_{12}^{2}\Big(\frac{\Im\tau}{\pi}\Big)^{2}\cform{5&0\\1&0}+O(s_{12}^3)\ .
 \notag
\end{align}
Similarly, the right-hand side of the Laplace equation  \eqref{lapsec16} expands to\footnote{We have not yet inserted the simplified form (\ref{eq:7simp}) of the $\ap$-expansions which are obtained after using identities between modular graph forms, since we want to illustrate that \eqref{cr2ptcp} and \eqref{lapsec95} can be used to generate these kinds of identities.}
\begin{align}
\Delta^{(2,0)} W^\tau_{(2|0)} &= s_{12}^2 \Big(\frac{\Im\tau}{\pi}\Big)^{2} (3 {\rm G}_4 W^\tau_{(0|2)} - 2 W^\tau_{(4| 2)}) -2 s_{12} \frac{\Im\tau}{\pi} W^\tau_{(3| 1)} \label{eq:8}\\
  &=-2s_{12}\frac{\Im\tau}{\pi}\cform{3&0\\1&0}+2s_{12}^{2}\Big(\frac{\Im\tau}{\pi}\Big)^{2}(\cform{0&1&3\\1&1&0}-\cform{4&0\\2&0})+O(s_{12}^3)\ .
  \notag
\end{align}
The right-hand sides of \eqref{eq:5} and \eqref{eq:8} do not manifestly match the direct action~\eqref{eq:MFGRL}
of $\nabla^{(2)}$ and $\Delta^{(2,0)}$ on the modular graph forms in the expansion (\ref{eq:7})
of $W^{\tau}_{(2|0)}$,
\begin{align}
  \nabla^{(2)}W^{\tau}_{(2|0)}=&-s_{12}\nabla^{(2)}\left(\frac{\Im\tau}{\pi}\cform{3&0\\1&0}\right)-\frac{1}{2}s_{12}^{2}\nabla^{(2)}\left(\Big(\frac{\Im\tau}{\pi}\Big)^{2}\cform{1&1&2\\1&1&0}\right)+O(s_{12}^3)\label{eq:9}\\
  =&-3s_{12}\frac{\Im\tau}{\pi}\mathrm{G}_{4}-s_{12}^{2}\Big(\frac{\Im\tau}{\pi}\Big)^{2}(\cform{1&1&3\\1&1&-1}+\cform{1&2&2\\1&0&0})+O(s_{12}^3)\notag \\
  \Delta^{(2,0)}W^{\tau}_{(2|0)}=&-s_{12}\,\Delta^{(2,0)}\!\!\left(\frac{\Im\tau}{\pi}\cform{3&0\\1&0}\right)-\frac{1}{2}s_{12}^{2}\,\Delta^{(2,0)}\!\!\left(\Big(\frac{\Im\tau}{\pi}\Big)^{2}\cform{1&1&2\\1&1&0}\right)+O(s_{12}^3)\label{eq:10}\\
  =&-2s_{12}\frac{\Im\tau}{\pi}\cform{3&0\\1&0}-s_{12}^{2}\Big(\frac{\Im\tau}{\pi}\Big)^{2}(2 \cform{0&1&3\\2&1&-1}+\cform{0&2&2\\2&0&0})+O(s_{12}^3) \, .\notag
\end{align}
Comparing \eqref{eq:5} and \eqref{eq:9} as well as \eqref{eq:8} and \eqref{eq:10} order by order in $\ap$ yields 
infinitely many identities for modular graph forms, e.g.
\begin{align}
  2\cform{5&0\\1&0}&=\cform{1&1&3\\1&1&-1}+\cform{1&2&2\\1&0&0}\label{eq:23}\\
  2(\cform{0&1&3\\1&1&0}-\cform{4&0\\2&0})&=-2 \cform{0&1&3\\2&1&-1}-\cform{0&2&2\\2&0&0}\ .\notag
\end{align}
In particular, the differential equations \eqref{cr2ptcp} and \eqref{lapsec95} of the component integrals
bypass the need to perform holomorphic subgraph reduction (see Appendix~\ref{sec:HSR}) to all orders in $\ap$.
This is exemplified by the identities for modular graph forms $\cform{A&c&d\\B&0&0}$ in (\ref{eq:23})
and becomes particularly convenient at $n\geq 3$ points, where the state-of-the-art methods
for holomorphic subgraph reduction \cite{Gerken:2018zcy} are recursive and may generate huge numbers of terms
in intermediate steps.

Since for low weights many identities between modular graph forms are known \cite{Green:2008uj, DHoker:2015gmr, DHoker:2015wxz, DHoker:2016mwo, DHoker:2016quv}, the expansions above also allow for an explicit test of the differential equations. In particular, the identities such as \eqref{eq:23} generated by $W^{\tau}_{(2|0)}$ can be confirmed by applying simple identities at low weights. In general, by applying identities for dihedral modular graph forms,~\eqref{cr2ptcp} can be verified to all orders in $\ap$, as detailed in Appendix~\ref{sec:2ptMGFVrf}. The Laplace equations (\ref{lapsec16}) have been verified to the order $(\ap)^{5}$ for $W^\tau_{(0|0)}$, to the order $(\ap)^{4}$ for $W^\tau_{(1|1)}$ and $W^\tau_{(2|0)}$ and to the orders $(\ap)^{3}$ for $W^\tau_{(2|2)},W^\tau_{(3|1)}$ and $W^\tau_{(4|0)}$.


\section{Cauchy--Riemann differential equations}
\label{sec:bigCRn}


In this section, we derive the general first-order differential equation in $\tau$ on the generating series $W_{\vec\eta}^\tau(\rho|\sigma)$ for $n$ points. The steps will generalize the two-point derivation in Section~\ref{sec:cr2pts} with some additional steps due to the permutations $\rho,\sigma\in \mathcal{S}_{n-1}$. After deriving the general $n$-point formula we exemplify it by studying in detail the cases $n=3$ and $n=4$.

\subsection{Cauchy--Riemann differential equation at $n$ points}
\label{sec:CRn}

In order to act with $\nabla_{\vec\eta}^{(n-1)}$ on the generating series $W_{\vec{\eta}}^\tau(\sigma|\rho)$ defined in~\eqref{eq:Wdef}, we observe that the Maa\ss{} raising and lowering operator distributes correctly according to~\eqref{eq:RLprod} and acts only on the product of chiral $\Omega$-series and on the Koba--Nielsen factor. 
Moreover, the differential operator and the Koba--Nielsen factor are invariant under $\rho\in {\cal S}_{n-1}$ as can be seen from the definition~\eqref{eq:KNn} and the property \eqref{eq:etasum2}. Using the mixed heat equation~\eqref{eq:Omixh} for the $\tau$-derivative of $\Omega$ as well as~\eqref{eq:dtauKN} for the $\tau$-derivative of the Koba--Nielsen factor this leads to
\begin{align}
  \label{eq:CRW1}
  2\pi i \nabla_{\vec\eta}^{(n-1)} W_{\vec{\eta}}^\tau(\sigma|\rho)
  &=\! \int \dd\mu_{n-1} \,\rho\Big[ 2\pi i \nabla_{\vec\eta}^{(n-1)}  \Big\{\! \KN^{\tau}_{n} \prod_{p=2}^n \Omega(z_{p-1,p},\xi_p,\tau) \Big\}  \Big] \sigma\!\Big[\prod_{q=2}^n \overline{\Omega(z_{q-1,q},\xi_q,\tau)}\Big]\nn\\
  &= (\tau-\bar\tau)\! \int \dd\mu_{n-1} \, \rho\bigg[-\!\sum_{i=2}^n\! \big( \partial_{z_{i}}\partial_{\xi_i}\Omega(z_{i-1,i},\xi_i,\tau)  \big)\KN^{\tau}_{n}\prod_{\substack{p=2\\p\neq i}}^n \Omega(z_{p-1,p},\xi_p,\tau)  \nn\\
  &\hspace{7mm}-    \sum_{1\leq i<j}^n   s_{ij} f^{(2)}_{ij}  \KN^{\tau}_{n} \prod_{p=2}^n \Omega(z_{p-1,p},\xi_p,\tau) \bigg]\sigma \Big[\prod_{q=2}^n \overline{\Omega(z_{q-1,q},\xi_q,\tau)}\Big]\,,  
\end{align}
where we have introduced the following 
short-hand:\footnote{Note that the permutation $\rho$ does not act on the immediate indices of $\xi_i$ but on the indices of the constituent~$\eta_i$.}
\begin{align}
     \xi_i = \eta_{i,i+1,\ldots,n} = \sum_{k=i}^n \eta_k\,.\label{eq:6}
\end{align}
In each of the terms in the $i$-sum in \eqref{eq:CRW1} one can replace $\partial_{z_i} \to \partial_{z_i} + \partial_{z_{i+1}} + \ldots + \partial_{z_n}=\sum_{j=i}^n \partial_{z_j}$ as the function it acts on does not depend on the other $z$-variables. This has the advantage that one can integrate by parts all $z$-derivatives without producing any contribution from the other chiral Kronecker--Eisenstein series since they all depend on differences such that the corresponding terms cancel. This leads to two contributions: In the first the $z$-derivatives act on the Koba--Nielsen factor, and the second contribution comes from the action on the anti-chiral Kronecker--Eisenstein series. These two contributions are of different kinds and we first focus on the one when the $z$-derivative acts on the anti-chiral $\overline{\Omega}$.

A partial $z$-derivative acting on a single anti-chiral $\overline{\Omega}$ was given in~\eqref{eq:dzbO} and generates the corresponding $\bar\eta$. It can be checked that the combination of all terms does not depend on the permutations $\rho$ and $\sigma$ that one started with and in total produces the operator $2\pi i \sum_{i=2}^{n} \bar\eta_i \partial_{\eta_i}$ acting on the whole expression. We emphasize that this is the only part that does not involve only holomorphic or anti-holomorphic $\eta$-parameters but mixes them. Carrying out the full integration by parts, we can therefore rewrite~\eqref{eq:CRW1} as
\begin{align}
  2\pi i \nabla_{\vec\eta}^{(n-1)} W_{\vec{\eta}}^\tau(\sigma|\rho) &= 2\pi i \sum_{i=2}^n \bar{\eta}_i \partial_{\eta_i} W_{\vec\eta}^\tau(\sigma|\rho)\\
                                                                      &\hspace{5mm} + (\tau-\bar\tau) \int  \dd\mu_{n-1} \, \rho\bigg[\sum_{i=2}^n \big(\sum_{j=i}^n \partial_{z_j}\KN^{\tau}_{n} \big) \partial_{\xi_i} \prod_{p=2}^n \Omega(z_{p-1,p},\xi_p,\tau) \nn\\
                                                                    &\hspace{10mm}- \sum_{1\leq i<j}^n s_{ij} f^{(2)}_{ij} \KN^{\tau}_{n}\prod_{p=2}^n \Omega(z_{p-1,p},\xi_p,\tau)  \bigg]\sigma \Big[\prod_{q=2}^n \overline{\Omega(z_{q-1,q},\xi_q,\tau)}\Big]\,.\nn
\end{align}

We next focus on analyzing the terms inside the $\rho$-permutation by using~\eqref{eq:dzKN2} for evaluating the $z$-derivative acting on the Koba--Nielsen factor:
\begin{align}
  \label{eq:CRsimple1}
  &\quad\sum_{i=2}^n  \big( \sum_{j=i}^n \partial_{z_j}\KN^{\tau}_{n} \big) \partial_{\xi_i} \prod_{p=2}^n \Omega(z_{p-1,p},\xi_p,\tau)- \sum_{1\leq i<j}^n s_{ij} f^{(2)}_{ij}  \KN^{\tau}_{n} \prod_{p=2}^n \Omega(z_{p-1,p},\xi_p,\tau)  \nn\\
  & \ \ \  =  \sum_{1\leq i<j}^n s_{ij} \left[ \sum_{k=i+1}^j f_{ij}^{(1)} \partial_{\xi_k} - f_{ij}^{(2)}\right]   \KN^{\tau}_{n}\prod_{p=2}^n \Omega(z_{p-1,p} , \xi_p ,\tau)\,.
\end{align}
As shown in Appendix~\ref{app:sij}, the cyclic product of Kronecker--Eisenstein series can be brought into a form such that the differential operator in square brackets has a simple action, see~\eqref{eq:sijdiff}, generalizing~\eqref{eq:25}. The result can be written in terms of certain shuffles explained in detail in the Appendix and illustrated here in Figure~\ref{fig:sijform}. At $n$ points for $1\leq i<j\leq n$ and $i<k\leq j$ one has to consider all sequences $(a_1,\ldots,a_n)$ in the set
\begin{align}
  \label{eq:Snijk}
  S_n(i,j,k) := \Big\{ \big(\{ 1,\ldots, i-1\}\shuffle \{k-1,\ldots,i+1\}, i, j, \{j-1,\ldots,k\}\shuffle\{j+1,\ldots,n\}\big) \Big\}\,,
\end{align}
where $i$ and $j$ are at fixed position determined by $k$ and the values to the left and right of them are given by shuffling very specific lists, half of which are reversed in order. 

\begin{figure}[t]
  \centering
  \begin{tikzpicture}
    \draw (0,0)--(8,0);
    \draw (0,-.2)--(0,.2);
    \draw (0,.5) node {$1$};
    \draw (8,-.2)--(8,.2);
    \draw (8,.5) node {$n$};
    \draw (2.5,-.2)--(2.5,.2);
    \draw (2.5,.5) node {$i$};
    \draw (7,-.2)--(7,.2);
    \draw (7,.5) node {$j$};
    \draw (4.5,-.2)--(4.5,.2);
    \draw (4.5,.5) node {$k$};
    \draw [color=darkgreen] (0,-.4) -- (0,-.5) -- (2.3,-.5) -- (2.3,-.4);
    \draw [color=darkgreen] (2.7,-.4) -- (2.7,-.5) -- (4.3,-.5) -- (4.3,-.4);
    \draw [color=darkred] (4.5,-.4) -- (4.5,-.5) -- (6.8,-.5) -- (6.8,-.4);
    \draw [color=darkred] (7.2,-.4) -- (7.2,-.5) -- (8,-.5) -- (8,-.4);
    \draw [color=darkgreen] (2.65,0.05) node[anchor=north west] {(reverse)};
    \draw [color=darkred] (4.8,0.05) node[anchor=north west] {(reverse)};
    \draw [color=darkgreen,thick,->] (1.15,-.6)--(2,-2.5);
    \draw [color=darkgreen,thick,->] (3.5,-.6)--(2.1,-2.5);
    \draw [color=darkgreen] (1.5,-1.3) node[anchor=north west] {shuffle};
    \draw [color=darkred,thick,->] (5.65,-.6)--(6,-2.5);
    \draw [color=darkred,thick,->] (7.6,-.6)--(6.1,-2.5);
    \draw [color=darkred] (5.7,-1.3) node[anchor=north west] {shuffle};
    \draw [color=darkgreen] (1.3,-2.8) -- (1.3,-2.7) -- (3.2 ,-2.7) -- (3.2,-2.8);
    \draw [color=darkred] (4.9,-2.8) -- (4.9,-2.7) -- (6.9 ,-2.7) -- (6.9,-2.8);
    \draw (1,-3) node[anchor=west] {$( {\color{darkgreen}a_1,\ldots ,a_{k-2}}, a_{k-1}, a_k, {\color{darkred}a_{k+1},\ldots, a_n})$};
    \draw [thick,->] (2.5,-0.3) .. controls (2.5,-1) and (3.8,-1.2) .. (3.8,-2.8);
    \draw [thick,->] (7,-0.3) .. controls (7,-1) and (4.5,-1.2) .. (4.5,-2.8);
  \end{tikzpicture}
  \caption{\label{fig:sijform}\it The shuffles appearing in the $s_{ij}$-form on the right-hand side of (\ref{eq:CRsimple2}) for a fixed $i<k\leq j$. The values $i$ and $j$ are not included in the indicated ranges. The middle two intervals are reversed to descending order before the shuffles. The origin of this reversal is~\eqref{eq:sijext}. All sequences obtained in this way constitute the set $S_n(i,j,k)$ defined in~\eqref{eq:Snijk}.}
\end{figure}
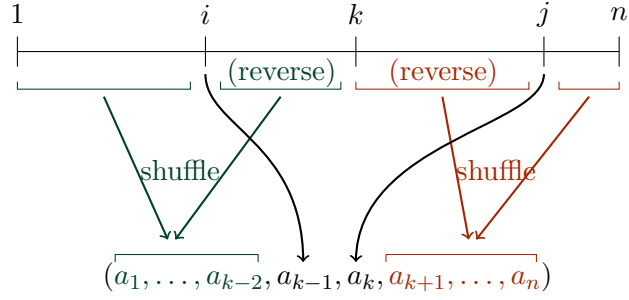

Using the result~\eqref{eq:sijdiff},  equation~\eqref{eq:CRsimple1} can then be evaluated to
\begin{align}
  &\quad  \sum_{1\leq i<j}^n s_{ij} \left[ \sum_{k=i+1}^j f_{ij}^{(1)} \partial_{\xi_k} - f_{ij}^{(2)}\right]   \KN^{\tau}_{n}\prod_{p=2}^n \Omega(z_{p-1,p} , \xi_p ,\tau) \notag\\
  & \ \ \ = \sum_{1\leq i<j}^n s_{ij} \Bigg[ \frac12 (\partial_{\eta_j}-\partial_{\eta_i})^2 \KN^{\tau}_{n} \prod_{p=2}^n \Omega(z_{p-1,p} , \xi_p ,\tau)    \label{eq:CRsimple2} \\
  &\hspace{10mm}
    -(-1)^{j-i+1} \sum_{k=i+1}^j \wp(\xi_k,\tau) \sum_{(a_1,\ldots,a_n)\in S_n(i,j,k)} \KN^{\tau}_{n}\prod_{p=2}^n \Omega\bigg(z_{a_{p-1},a_p},\sum_{\ell=p}^n \eta_{a_\ell},\tau\bigg) \Bigg]\,. \notag
\end{align}
Here, we have set $\partial_{\eta_1}=0$ in the case $i=1$ to avoid a separate bookkeeping of the
terms $\sim \sum_{j=2}^n s_{1j} \partial_{\eta_j}^2$. We see that there is a `diagonal term' containing the differential operators that goes back to the standard ordering of points. The terms including the Weierstra\ss{} functions mix the standard ordering with other orderings described by the set $S_n(i,j,k)$. Since neither the differential operators nor the Weierstra\ss{} functions depend on $z$ they can be pulled out of the world-sheet integral. 

We note that these operations are also related to the so-called $S$-map~\cite{Mafra:2014oia,Mafra:2014gsa} which enters the expressions of \cite{Mafra:2019xms} for the $\tau$-derivatives of $A$-cycle integrals (\ref{again2}). In fact, the $(n\geq 6)$-point instances of the open-string differential operator $D^\tau_{\vec{\eta}}$ in (\ref{intr4}) were conjectural in the reference, and (\ref{eq:CRsimple2}) together with Appendix~\ref{app:sij} furnish the missing proof.

Equation~\eqref{eq:CRsimple2} is expressed in an over-complete basis since a sequence $(a_1,\ldots, a_n)\in S_n(i,j,k)$ can have the index $1$ at any place. Assume that the index $1$ appears at position $m>1$, i.e. $a_m=1$ and write $(a_1,\ldots, a_n)=(A,1,B)$ with $A=(a_1,\ldots,a_{m-1})$ and $B=(a_{m+1},\ldots, a_n)$. The index $1$ can be moved to the front using the fact that, as a consequence of the Fay identity~\eqref{eq:FayO}, products of Kronecker--Eisenstein series obey the shuffle identity~\cite{Schocker} 
\begin{align}
  \label{eq:Oshuffle}
  \prod_{p=2}^n \Omega\bigg(z_{a_{p-1},a_p},\sum_{\ell=p}^n \eta_{a_\ell},\tau\bigg) = (-1)^{m-1} \sum_{(c_2,\ldots,c_n)\in A^t\shuffle B  } \prod_{p=2}^{n} \Omega\bigg(z_{c_{p-1},c_p},\sum_{\ell=p}^n \eta_{c_\ell} ,\tau\bigg)\,,
\end{align}
where $A^t=(a_{m-1},a_{m-2},\ldots,a_1)$ denotes the reversed sequence and we have set $c_1=1$ always.
Applying this identity replaces one sequence $(a_1,\ldots,a_n)\in S_n(i,j,k)$ by a sum of sequences and the resulting integrals are then all of $W$-type but with different orderings of the $n-1$ unfixed points. This replaces the second term in~\eqref{eq:CRsimple2} by a sum over all possible permutations $\alpha\in\mathcal{S}_{n-1}$ multiplying $W_{\vec\eta}^\tau(\sigma|\alpha)$ with coefficients $T_{\vec\eta}^\tau(\rho|\alpha)$ constructed out of Mandelstam invariants and Weierstra\ss{} functions. We write the total contribution of~\eqref{eq:CRsimple2} to the Cauchy--Riemann derivative, up to an overall $(\tau-\bar\tau)$, as the operator
\begin{align}
\label{eq:svCR}
  \sum_{\alpha\in {\cal S}_{n-1}} \sv D^\tau_{\vec\eta}(\rho|\alpha) W_{\vec{\eta}}^\tau(\sigma|\alpha)
  := \sum_{1\leq i<j}^n s_{ij} \Big[ \frac12 (\partial_{\eta_j}-\partial_{\eta_i})^2\Big]W_{\vec{\eta}}^\tau(\sigma|\rho) + \sum_{\alpha\in {\cal S}_{n-1}} T_{\vec\eta}^\tau(\rho|\alpha)  W_{\vec{\eta}}^\tau(\sigma|\alpha)
  \,.
\end{align}
Explicit expressions for $\sv D_{\vec\eta}^\tau(\rho|\alpha)$, detailing in particular the coefficients $T_{\vec\eta}^\tau(\rho|\alpha)$ will be given in Section~\ref{sec:CR3} and~\ref{sec:CR4} below for $3$ and $4$ points. At two points, one
can identify $T_{\eta}^\tau = -s_{12} \wp(\eta,\tau)$ from the expression (\ref{svD2pt}) for $\sv D_{\eta}^\tau$. The `single-valued' notation here again instructs to drop the diagonal term $\sim\!- 2  \zeta_2 s_{12\ldots n} \delta_{\rho,\alpha}$ in the analogous open-string differential operator $D_{\vec\eta}^\tau(\rho|\alpha)$ in (\ref{intr4}) \cite{Mafra:2019ddf, Mafra:2019xms}, i.e.\
\begin{align}
\sv D_{\eta}^\tau(\alpha|\rho) = D_{\eta}^\tau(\alpha|\rho)\Big|_{\zeta_{2}\rightarrow0} =  D_{\eta}^\tau(\alpha|\rho) + 2  \zeta_2 s_{12\ldots n} \delta_{\rho,\alpha}\, .
\end{align}
We have further separated $\sv D_{\eta}^\tau(\alpha|\rho)$ into a part that contains the holomorphic derivatives with respect to $\vec\eta$ and terms $T_{\vec\eta}^\tau(\alpha|\rho)$ that are completely meromorphic in $\vec\eta$ and $\tau$ and contain no derivatives. 

\medskip

Putting everything together we conclude that~\eqref{eq:Wdef} obeys the Cauchy--Riemann equation
\begin{align}
  \label{eq:CRn}
  2\pi i \nabla_{\vec\eta}^{(n-1)} W_{\vec{\eta}}^\tau(\sigma|\rho) &= 2\pi i \sum_{i=2}^n \bar{\eta}_i\partial_{\eta_i} W_{\vec\eta}^\tau(\sigma|\rho) + (\tau-\bar\tau) \sum_{\alpha \in {\cal S}_{n-1}} \sv D^\tau_{\vec\eta}(\rho| \alpha)W_{\vec{\eta}}^\tau(\sigma|\alpha)\nn\\
                                                                    &=: \sum_{\alpha\in {\cal S}_{n-1}} Q_{\vec\eta}^\tau(\rho| \alpha) W_{\vec\eta}^\tau(\sigma|\alpha)\,,
\end{align}
defining a short-hand for the action of the Maa\ss{} operator and with $\sv D_{\vec\eta}^\tau$ given in~\eqref{eq:svCR}. By expanding this equation in the $\eta$-parameters one can obtain systems of Cauchy--Riemann equations for the component integrals which in turn yield Cauchy--Riemann equations for modular graph forms.


\subsection{Three-point examples}
\label{sec:CR3}

The three-point analogue of~\eqref{eq:W2pt} is given by
\begin{align}
  \label{eq:W3pt}
  W_{\eta_{2},\eta_{3}}^\tau(\sigma|\rho) = \int \frac{\dd^2z_2}{\Im\tau} \frac{\dd^2z_3}{\Im\tau} \,  \,\rho\Big[ \Omega(z_{12},\eta_{23},\tau) \Omega(z_{23},\eta_3,\tau)\Big] \sigma\!\Big[ \overline{\Omega(z_{12},\eta_{23},\tau)}\, \overline{\Omega(z_{23},\eta_3,\tau)}\Big] \KN^{\tau}_{3}\,.
\end{align}
This is a $(2\times 2)$ matrix of functions parametrized by the two permutations $\rho,\sigma\in {\cal S}_2$ that act on the indices $2$ and $3$ of the $z_j$ and $\eta_j$. 

We first explain how to obtain the operators $\sv D_{\vec\eta}^\tau(\rho|\alpha)$ in~\eqref{eq:svCR}. Writing out~\eqref{eq:CRsimple2} that is obtained from the combination of the $\tau$-derivative and the $\partial_z$-derivatives acting on the Koba--Nielsen factor yields for $n=3$
\begin{align}
&\hspace{10mm}\bigg[\frac12 s_{12} \partial_{\eta_2}^2 + \frac12 s_{13} \partial_{\eta_3}^2 + \frac12 s_{23} \big( \partial_{\eta_2}^2-\partial_{\eta_3}^2\big) \bigg] \Omega(z_{12},\eta_{2}{+}\eta_3,\tau) \Omega(z_{23},\eta_3,\tau) \KN_3^\tau\\
&-\!s_{12}\wp(\xi_2,\tau) \Omega(z_{12},\eta_{2}{+}\eta_3,\tau)\Omega(z_{23},\eta_{3},\tau) \KN_3^\tau -s_{23}\wp(\xi_{3},\tau) \Omega(z_{12},\eta_{2}{+}\eta_3,\tau)\Omega(z_{23},\eta_{3},\tau) \KN_3^\tau\nn\\
&+\!s_{13} \bigg[ \wp(\xi_{2},\tau)\Omega(z_{13},\eta_{2}{+}\eta_{3},\tau)\Omega(z_{32},\eta_{2},\tau) \KN_3^\tau + \wp(\xi_{3},\tau)\Omega(z_{21},\eta_{1}{+}\eta_3,\tau)\Omega(z_{13},\eta_{3},\tau) \KN_3^\tau \bigg]\,,\nn
\end{align}
where we have used $\partial_{\eta_1}=0$ and $\xi_2=\eta_2+\eta_3$ and $\xi_3=\eta_3$. The sequence sets $S_n(i,j,k)$ that appear in this case are $S_3(1,2,2)=\{(1,2,3)\}$ (for terms proportional to $s_{12}$), $S_3(2,3,3)=\{(1,2,3)\}$ (for $s_{23}$) and $S_3(1,3,2)=\{(1,3,2)\}$ as well as $S_3(1,3,3)=\{(2,1,3)\}$ (for $s_{13}$). The very last sequence $(2,1,3)$ does not start with the index $1$ and needs to reordered using~\eqref{eq:Oshuffle}, yielding contributions to the sequences $(1,2,3)$ and $(1,3,2)$:
\begin{align}
\Omega(z_{21},\eta_{1}{+}\eta_3,\tau)\Omega(z_{13},\eta_{3},\tau) = - \Omega(z_{12},\eta_{2}{+}\eta_3,\tau)\Omega(z_{23},\eta_{3},\tau) - \Omega(z_{13},\eta_{2}{+}\eta_3,\tau)\Omega(z_{32},\eta_{2},\tau)\,.
\end{align}

Therefore we see that for $\rho(2,3) =(2,3)$ we obtain the following components of the first row of $\sv D_{\eta_2,\eta_3}^\tau(\rho|\alpha)$:
\begin{align}
  \sv D^{\tau}_{\eta_2,\eta_3}(2,3|2,3) &=
                                          s_{12} \big[ \tfrac{1}{2} \partial^2_{\eta_2}- \wp(\eta_2{+}\eta_3,\tau) \big] + s_{23}  \big[ \tfrac{1}{2} (\partial_{\eta_2}{-}\partial_{\eta_3})^2- \wp(\eta_3,\tau) \big]
                                          \notag \\
                                        & \ \ \ \ \ \ \ \ + s_{13} \big[ \tfrac{1}{2} \partial^2_{\eta_3}- \wp(\eta_3,\tau) \big]  \label{dsv3.1} \\
  \sv D^{\tau}_{\eta_2,\eta_3}(2,3|3,2) &=
                                          s_{13} \big[ \wp(\eta_2{+}\eta_3,\tau) - \wp(\eta_3,\tau) \big]  \, . \notag
\end{align}
The second row associated with $\rho(2,3) =(3,2)$ follows from relabeling $s_{12}\leftrightarrow s_{13}$ and $\eta_2\leftrightarrow \eta_3$,
\begin{align}
  \sv D^{\tau}_{\eta_2,\eta_3}(3,2|3,2)&=
                                         s_{13} \big[ \tfrac{1}{2} \partial^2_{\eta_3}- \wp(\eta_2{+}\eta_3,\tau) \big]
                                         + s_{23}  \big[ \tfrac{1}{2} (\partial_{\eta_2}{-}\partial_{\eta_3})^2- \wp(\eta_2,\tau) \big]
                                         \notag\\
                                       & \ \ \ \ \ \ \ \
                                         + s_{12} \big[ \tfrac{1}{2} \partial^2_{\eta_2}- \wp(\eta_2,\tau) \big]  \label{dsv3.2} \\
  \sv D^{\tau}_{\eta_2,\eta_3}(3,2|2,3)&=
                                         s_{12} \big[ \wp(\eta_2{+}\eta_3,\tau) - \wp(\eta_2,\tau) \big]  \, . \notag
\end{align}
By comparing these expressions with the open-string expression $D^{\tau}_{\vec{\eta}}(\sigma|\rho)$ given in~\cite{Mafra:2019ddf, Mafra:2019xms}, we see that they agree up to a term $-2\zeta_2 s_{123}$ in the diagonal entries of the open-string operators. As the standard single-valued map for zeta values implies $\sv(\zeta_2)=0$, our notation $\sv D_{\vec\eta}^\tau(\sigma|\rho)$ is consistent with the same operators in the open-string case.

One can then work out the three-point Cauchy--Riemann equations~\eqref{eq:CRn} that read for three points
\begin{align}
  2\pi i \nabla_{\eta_{2},\eta_{3}}^{(2)}W_{\eta_{2},\eta_{3}}^{\tau}(\sigma|\rho)=\sum_{\alpha\in\mathcal{S}_{2}}\left[2\pi i\delta_{\rho,\alpha} (\bar{\eta}_{2}\partial_{\eta_{2}}{+}\bar{\eta}_{3}\partial_{\eta_{3}})+(\tau-\bar{\tau})\sv D^{\tau}_{\eta_{2},\eta_{3}}(\rho|\alpha)\right]W_{\eta_{2},\eta_{3}}^{\tau}(\sigma|\alpha)\ ,\label{eq:13}
\end{align}
by substituting in the matrix elements of $\sv D^{\tau}_{\eta_{2},\eta_{3}}(\rho|\alpha)$ given in~\eqref{dsv3.1} and~\eqref{dsv3.2}. 

We carry out the derivation of the Cauchy--Riemann equations for component integrals \eqref{eq:11} in detail in Appendix~\ref{app:3pt} where we explain a subtlety in translating (\ref{eq:13}) to the component level: Both sides of~\eqref{eq:13} have to be expanded in the same $\eta$ variables (e.g. $\eta_{23}=\eta_2{+}\eta_3$ and $\eta_3$) but other permutations naturally come with different $\eta$ variables that have to be rearranged using the binomial theorem. 

A general formula for the Cauchy-Riemann equation of the components \eqref{eq:11} can be found in~\eqref{eq:16}.
It can be specialized to yield
\begin{align}
  2\pi i \nabla^{(0)} W^{\tau}_{(0,0|b_{2},b_{3})}(\sigma|2{,}3)&=-(\tau{-}\bar{\tau})s_{12}W^{\tau}_{(2,0|b_{2},b_{3})}(\sigma|2{,}3)-(\tau{-}\bar{\tau})s_{23}W^{\tau}_{(0,2|b_{2},b_{3})}(\sigma|2{,}3)\notag\\
                                                                    &\hspace{-3em}\quad-(\tau{-}\bar{\tau})s_{13}\Big(W^{\tau}_{(0,2|b_{2},b_{3})}(\sigma| 2{,}3)+W^{\tau}_{(2,0|b_{2},b_{3})}(\sigma| 3{,}2)-W^{\tau}_{(0,2|b_{2},b_{3})}(\sigma|3{,}2)\Big)\notag\\[1em]
  2\pi i \nabla^{(1)} W^{\tau}_{(1,0|b_{2},b_{3})}(\sigma|2{,}3)&=2\pi i W^{\tau}_{(2,0|b_{2}-1,b_{3})}(\sigma| 2{,}3)-(\tau{-}\bar{\tau})s_{23}W^{\tau}_{(1,2|b_{2},b_{3})}(\sigma| 2{,}3)\label{eq:17main}\\
                                                                    &\hspace{-3em}\quad+(\tau{-}\bar{\tau})s_{13}\Big(W^{\tau}_{(3,0|b_{2},b_{3})}(\sigma| 2{,}3){-}W^{\tau}_{(1,2|b_{2},b_{3})}(\sigma| 2{,}3)\notag\\
                                                                    &\hspace{-3em}\qquad+2W^{\tau}_{(0,3|b_{2},b_{3})}(\sigma| 3{,}2){+}W^{\tau}_{(1,2|b_{2},b_{3})}(\sigma| 3{,}2){-}W^{\tau}_{(3,0|b_{2},b_{3})}(\sigma|3{,}2)\Big) \notag
\end{align}
and further examples listed in (\ref{eq:17}). 

The very simplest instance of this is for $(b_2,b_3)=(0,0)$
\begin{align}
  2\pi i \nabla^{(0)} W^{\tau}_{(0,0|0,0)}(\sigma|2{,}3)&=-(\tau{-}\bar{\tau})s_{12}W^{\tau}_{(2,0|0,0)}(\sigma|2{,}3)-(\tau{-}\bar{\tau})s_{23}W^{\tau}_{(0,2|0,0)}(\sigma|2{,}3)\label{eq:CR3W0000}\\
                                                                    &\quad-(\tau{-}\bar{\tau})s_{13}\Big(W^{\tau}_{(0,2|0,0)}(\sigma| 2{,}3)+W^{\tau}_{(2,0|0,0)}(\sigma| 3{,}2)-W^{\tau}_{(0,2|0,0)}(\sigma|3{,}2)\Big)\nn\\
&=-(\tau{-}\bar{\tau}) \big[ s_{12}W^{\tau}_{(2,0|0,0)}(\sigma|2{,}3){+}s_{13}W^{\tau}_{(2,0|0,0)}(\sigma| 3{,}2) 
{+} s_{23} W^{\tau}_{(0,2|0,0)}(\sigma|2{,}3) \big]\,, \notag
\end{align}
where we have used the corollary
\begin{align}
W^{\tau}_{(0,2|0,0)}(\sigma| 2{,}3) = W^{\tau}_{(0,2|0,0)}(\sigma|3{,}2) 
\end{align}
of $f^{(2)}_{23}=f^{(2)}_{32}$. This is an example of the first type of linear dependence between component integrals mentioned in Section~\ref{sec:cptints2}.


\subsubsection{Lessons for modular graph forms}
\label{sec:CR3.2}

We now consider one instance of such a Cauchy--Riemann equation to probe its contents in the $\ap$-expansion. The example we shall look at involves the component integral
\begin{align}
\label{eq:W1221}
W_{(1,2|2,1)}^{\tau }(2,3|2,3) &= -s_{13} \frac{\Im\tau }{\pi} \cformtri{0&1\\2&0}{0&2\\1&0}{1\\1} + O(s_{ij}^2)\,,
\end{align}
of modular weight $(3,3)$. We have written out the $\alpha'$-expansion along the lines of Section~\ref{sec:3ptexmpl} to the lowest non-trivial order which here contains a trihedral function of the class defined in~\eqref{eq:29}. The component Cauchy--Riemann equation in this case is
\begin{align}
\nabla^{(3)} &W_{(1,2|2,1)}^{\tau }(2,3|2,3) =  2 W_{(1,3|2,0)}^{\tau }(2,3|2,3)  +   W_{(2,2|1,1)}^{\tau }(2,3|2,3)\nn\\
&\quad  + \frac{\Im \tau}{\pi}  \Big\{ 2 s_{13} W_{(1,4|2,1)}^{\tau }(2,3|2,3) + 2 s_{23}  W_{(1,4|2,1)}^{\tau }(2,3|2,3)- 2 s_{13}  W_{(1,4|2,1)}^{\tau }(2,3|3,2)  \nn\\
&\quad\quad \quad\quad  + 2 s_{13}   W_{(2,3|2,1)}^{\tau }(2,3|2,3) - 2 s_{13}  W_{(2,3|2,1)}^{\tau }(2,3|3,2) + s_{13}   W_{(3,2|2,1)}^{\tau }(2,3|2,3) \nn\\
&\quad\quad\quad\quad  - s_{13}   W_{(3,2|2,1)}^{\tau }(2,3|3,2) -3s_{13}   {\rm G}_4 W_{(1,0|2,1)}^{\tau }(2,3|2,3) \label{eq:CRW1221}\\
&\quad\quad\quad\quad - 3 s_{23}  {\rm G}_4W_{(1,0|2,1)}^{\tau }(2,3|2,3) + 3  s_{13}{\rm G}_4  W_{(1,0|2,1)}^{\tau }(2,3|3,2) \Big\} \nn\\
&=  \frac{\Im\tau}{\pi} s_{13} \Big\{ 2  \cform{0&1&4\\2&0&1} +   \cformtri{0&2\\1&0}{0&2\\1&0}{1\\1}
+2  \cform{0&1&4\\1&2&0} +  \cform{0&2&3\\1&0&2} 
\nn\\
&\quad  \quad  \quad\quad  -2  \cform{0&2&3\\1&2&0}   - 3  {\rm G}_4  \cform{1&0\\3&0}+2    {\rm E}_2 \cform{3&0\\1&0}  \Big\}+ O(s_{ij}^2)\,,
\notag
\end{align}
where we have also written out the leading $\alpha'$-order of the right-hand side. Alternatively, we could have applied directly the differential operator to the expansion~\eqref{eq:W1221} using~\eqref{eq:MFGRL} which yields
\begin{align}
\nabla^{(3)} W_{(1,2|2,1)}^{\tau }(2,3|2,3) &= -  \frac{\Im\tau}{\pi} s_{13} \Big\{ \cformtri{0&2\\2&-1}{0&2\\1&0}{1\\1} +2
  \cformtri{0&1\\2&0}{0&3\\1&-1}{1\\1}  \notag\\
  &\hspace{6em}+  \cformtri{0&1\\2&0}{0&2\\1&0}{2\\0}  \Big\}+O(s_{ij}^2)\,.
 \label{eq:34}
\end{align}
Equating this to~\eqref{eq:CRW1221} again leads to a non-trivial identity between modular graph forms, now mixing trihedral and dihedral type. This identity can be checked by using various identities for modular graph forms to bring both \eqref{eq:CRW1221} and \eqref{eq:34} into the form
\begin{align}
  s_{13}\left(\frac{5}{2}\frac{\pi^{3} \mathrm{E}_{2}\dgcr \mathrm{E}_{2}}{(\Im\tau)^{4}}-3 \frac{\pi^{3}\dgcr \mathrm{E}_{4}}{(\Im\tau)^{4}}-5 \frac{\pi^{3}\dgcr \mathrm{E}_{2,2}}{(\Im\tau)^{4}}-\frac{3}{2}\mathrm{G}_{4}\frac{\pi\dgcrbar\mathrm{E}_{2}}{(\Im\tau)^{2}}\right)+\mathcal{O}(s_{ij}^{2})\ .
\end{align}
Similarly to the two-point results outlined in Section~\ref{sec:2pt-lessons-for-MGF}, also the three-point Cauchy--Riemann equations imply infinitely many relations between modular graph forms, now also including trihedral topologies. In particular, these identities allow to circumvent the convoluted trihedral holomorphic subgraph reduction~\cite{Gerken:2018zcy}. In the example above, when \eqref{eq:34} is simplified by means of the factorization and momentum-conservation identities spelled out in Appendices~\ref{sec:MGFfact} and~\ref{sec:MomCons}, one obtains
\begin{align}
  \nabla^{(3)} W_{(1,2|2,1)}^{\tau }(2,3|2,3)=\frac{\Im\tau}{\pi}s_{13}\Big\{&{-}2 \cformtri{ 0 & 1 \\ 2 & 0 }{ 0 & 3 \\ 1 & 0 }{ 1 \\ 0 }-\cformtri{ 0 & 2 \\ 1 & 0 }{ 0 & 2 \\ 2 & 0 }{ 1 \\ 0 }\\
                                             &-\cformtri{ 0 & 1 \\ 2 & 0 }{ 0 & 2 \\ 1 & 0 }{ 2 \\ 0 }+\cformtri{ 0 & 2 \\ 1 & 0 }{ 0 & 2 \\ 1 & 0 }{ 1 \\ 1 }+2 \cform{ 0 & 1 & 4 \\ 2 & 0 & 1 }\Big\} + O(s_{ij}^2)\ .
\notag
\end{align}
In this expression, the first three trihedral modular graph forms have to be simplified using holomorphic subgraph reduction. In \eqref{eq:CRW1221}, by contrast, no holomorphic subgraph reduction is necessary which exemplifies a general feature of the Cauchy--Riemann equations generated by \eqref{eq:13}: They avoid a large number of iterated momentum conservations and all instances of holomorphic subgraph reductions.


\subsection{Four-point examples}
\label{sec:CR4}

At four points we restrict ourselves to providing the expression for the operators in~\eqref{eq:svCR}. The following expressions for $\sv D^{\tau}_{\vec{\eta}}(\sigma|\rho)$ can be obtained by applying the general method with the same steps as for three points:                                                                                    %
\begin{align}
  \sv D^{\tau}_{\vec{\eta}}(2,3,4|2,3,4) &=
               \frac{1}{2} \sum_{1\leq i<j}^4 s_{ij} (\partial_{\eta_i}{-}\partial_{\eta_j})^2
                                           - s_{12} \wp(\eta_2{+}\eta_3{+}\eta_4,\tau)
                                           \notag \\
                                         & \ \ \ - (s_{13}{+}s_{23}) \wp(\eta_3{+}\eta_4,\tau)
                                           -(s_{14}{+}s_{24}{+}s_{34}) \wp(\eta_4,\tau) 
                                           \notag \\
  \sv D^{\tau}_{\vec{\eta}}(2,3,4|2,4,3) &= (s_{14}{+}s_{24}) \big[ \wp(\eta_3{+}\eta_4,\tau) - \wp(\eta_4,\tau)\big] \notag \\
  \sv D^{\tau}_{\vec{\eta}}(2,3,4|3,2,4) &= s_{13} \big[ \wp(\eta_2{+}\eta_3{+}\eta_4,\tau) - \wp(\eta_3{+}\eta_4,\tau)\big] \label{dsv4.1}\\
  \sv D^{\tau}_{\vec{\eta}}(2,3,4|3,4,2) &= s_{13} \big[ \wp(\eta_2{+}\eta_3{+}\eta_4,\tau) - \wp(\eta_3{+}\eta_4,\tau)\big] \notag \\
  \sv D^{\tau}_{\vec{\eta}}(2,3,4|4,2,3) &= s_{14}\big[ \wp(\eta_3{+}\eta_4,\tau) - \wp(\eta_4,\tau)\big]  \notag \\
  \sv D^{\tau}_{\vec{\eta}}(2,3,4|4,3,2) &= s_{14} \big[ \wp(\eta_3{+}\eta_4,\tau) - \wp(\eta_2{+}\eta_3{+}\eta_4,\tau)\big] \, .\notag
\end{align}
They agree with the corresponding open-string expressions $D^{\tau}_{\vec{\eta}}(\sigma|\rho)$ in \cite{Mafra:2019ddf, Mafra:2019xms} after dropping the term $-2\zeta_2 s_{1234}$ in the diagonal entries which is annihilated by the single-valued map.


\section{Laplace equations}
\label{sec:Lap}

In this section, we extend the first-order Cauchy--Riemann equation~\eqref{eq:CRn} to a second-order Laplace equation. This is first done in general for $n$ points and then examples are worked out for a low number of points. The derivation follows the ideas of Section~\ref{sec:Lap2} on two-point Laplace equations.

\subsection{Laplace equation at $n$ points}
\label{sec:Lapn}

In order to extend the Cauchy--Riemann equation~\eqref{eq:CRn} to the Laplacian we need to act with $\overline{\nabla}_{\vec\eta}^{(n-2)}$ from~\eqref{eq:MLn} on~\eqref{eq:CRn} and subtract an appropriate combination of weight terms according to~\eqref{eq:lapW}. The action of $\overline{\nabla}_{\vec\eta}^{(n-2)}=\overline{\nabla}_{\vec\eta}^{(n-1)}-1$ on~\eqref{eq:CRn} is simple since the differential operator passes through most terms in $Q_{\vec\eta}^\tau(\rho| \alpha)$ except for the explicit $\bar{\eta}_i$ in the diagonal term and the explicit $\bar\tau$ in front of $\sv D_{\vec\eta}(\rho| \alpha)$, leading to the simple commutation relation generalizing (\ref{nabQ2})
\begin{align}
  \left[ \overline{\nabla}_{\vec\eta}^{(n-1)}, Q_{\vec{\eta}}^\tau(\rho| \alpha) \right] =  Q_{\vec{\eta}}^\tau(\rho| \alpha)\,.
  \label{nabQn}
\end{align}
Taking the complex conjugate of \eqref{eq:CRn} leads to\footnote{Note that one has $\overline{W_{\vec\eta}^\tau(\sigma|\rho)} = W_{\vec\eta}^\tau(\rho|\sigma)$, leading to the summation over the first permutation labeling $W^\tau_{\vec{\eta}}$
in the complex conjugate equation.}
\begin{align}
2\pi i \overline{\nabla}_{\vec\eta}^{(n-1)} W_{\vec\eta}^\tau (\sigma|\alpha) &
= -\sum_{\beta\in {\cal S}_{n-1}} \overline{Q_{\vec\eta}^\tau}(\sigma|\beta) W_{\vec\eta}^\tau(\beta|\alpha) \\
&= 2\pi i \sum_{i=2}^n \eta_i \partial_{\bar{\eta}_i} W_{\vec\eta}^\tau(\sigma|\alpha) +(\tau - \bar\tau) \sum_{\beta\in {\cal S}_{n-1}}  \overline{\sv D_{\vec\eta}}(\sigma|\beta) W_{\vec\eta}^\tau(\beta|\alpha)\ , \notag
\end{align}
which implies (see (\ref{eq:Nb0N1W2}) for the analogous two-point calculation)
\begin{align}
  (2\pi i )^2 \overline{\nabla}_{\vec\eta}^{(n-2)}\nabla_{\vec\eta}^{(n-1)} W_{\vec\eta}^\tau (\sigma|\rho) 
  &= 2\pi i \! \!  \sum_{\alpha \in {\cal S}_{n-1} } \! \!\big( Q_{\vec\eta}^\tau(\rho| \alpha) \overline{\nabla}_{\vec\eta}^{(n-1)}
  {+} [\overline{\nabla}_{\vec\eta}^{(n-1)} \! , Q_{\vec\eta}^\tau(\rho| \alpha) ] {-} Q_{\vec\eta}^\tau(\rho| \alpha) \big) W_{\vec\eta}^\tau(\sigma|\alpha)
  \notag \\
  &= \sum_{\alpha \in {\cal S}_{n-1}} 
   Q_{\vec\eta}^\tau(\rho| \alpha)  \big( 2\pi i  \overline{\nabla}_{\vec\eta}^{(n-1)}  W_{\vec\eta}^\tau(\sigma|\alpha) \big)\nn\\
                                                                                                            &= - \sum_{\alpha,\beta\in {\cal S}_{n-1}} Q_{\vec\eta}^\tau(\rho|\alpha) \overline{Q_{\vec\eta}^\tau}(\sigma|\beta) W_{\vec\eta}^\tau(\beta|\alpha)\,. \label{nabQn22}
\end{align}
This expression can be expanded further by moving all $\eta$-differential operators in $Q_{\vec\eta}^\tau$ to the right to act directly on $W_{\vec\eta}^\tau$ since most terms commute. The only extra contributions come from $\bar \eta_i \partial_{\eta_i}$ and $s_{ij}(\partial_{\eta_i}-\partial_{\eta_j})^2$ in $Q_{\vec\eta}^\tau$ acting on the $\eta_k\partial_{\bar\eta_k}$ in $\overline{Q_{\vec\eta}^\tau}$. The result is
\begin{align}
  & \qquad \Big(2\pi i \sum_{i=2}^n \bar{\eta}_i \partial_{\eta_i} + \frac12 (\tau-\bar\tau) \sum_{1\leq i<j}^n s_{ij} (\partial_{\eta_i}-\partial_{\eta_j})^2 \Big)  2\pi i \sum_{k=2}^n \eta_k \partial_{\bar{\eta}_k}  \nn\\
  &= (2\pi i)^2 \sum_{i=2}^n \bar\eta_i \partial_{\bar\eta_i} 
    +(2\pi i)^2 \sum_{i,j=2}^n \eta_i\bar\eta_j \partial_{\eta_j}\partial_{\bar\eta_i}  + 2\pi i (\tau-\bar\tau) \sum_{1\leq i<j}^n s_{ij} (\partial_{\eta_j}-\partial_{\eta_i}) (\partial_{\bar\eta_j}-\partial_{\bar\eta_i})\nn\\
  &\hspace{10mm} + (2\pi i ) \sum_{k=2}^n \eta_k \partial_{\bar\eta_k} \,  \frac12  (\tau-\bar\tau) \sum_{1\leq i<j}^n s_{ij} (\partial_{\eta_j}-\partial_{\eta_i})^2 \, ,
\end{align}
where the last line is part of $2\pi i (\tau-\bar\tau) \sum_{k=2}^n \eta_k \partial_{\bar\eta_k} \sv D_{\vec\eta}^\tau(\rho|\alpha)$,
and therefore
\begin{align}
  (2\pi i )^2 \overline{\nabla}_{\vec\eta}^{(n-2)}\nabla_{\vec\eta}^{(n-1)} W_{\vec\eta}^\tau (\sigma|\rho) &= \sum_{\alpha,\beta\in {\cal S}_{n-1}} \bigg\{\delta_{\alpha,\rho}\delta_{\beta,\sigma}\bigg[ (2\pi i)^2  \Big(\sum_{i=2}^n \bar\eta_i \partial_{\bar\eta_i}  + \sum_{i,j=2}^n \eta_i\bar\eta_j \partial_{\eta_j}\partial_{\bar\eta_i} \Big)\nn\\
                                                                                                            &\hspace{20mm} + 2\pi i (\tau-\bar\tau) \sum_{1\leq i<j}^n s_{ij} (\partial_{\eta_j}-\partial_{\eta_i}) (\partial_{\bar\eta_j}-\partial_{\bar\eta_i})\bigg]\nn\\
                                                                                                            &\hspace{5mm} + 2\pi i (\tau-\bar\tau) \Big[ \delta_{\beta,\sigma} \sum_{i=2}^n \eta_i \partial_{\bar\eta_i} \sv D_{\vec\eta}^\tau(\rho|\alpha) + \delta_{\alpha,\rho} \sum_{i=2}^n \bar\eta_i \partial_{\eta_i} \overline{\sv D_{\vec\eta}^\tau}(\sigma|\beta)\Big]\nn\\
                                                                                                            &\hspace{5mm} + (\tau-\bar\tau)^2 \sv D_{\vec\eta}^\tau(\rho| \alpha) \, \overline{\sv D_{\vec\eta}^\tau}(\sigma|\beta)\bigg\} W_{\vec\eta}^\tau (\beta|\alpha)\,.
                                                                                                              \label{almostlapn}
\end{align}
According to~\eqref{eq:lapW}, the Laplacian differs from this by terms proportional to the weights that are also given by differential operators in $\eta$. The final result for the general Laplace equation is then
\begin{align}
  \label{eq:Lapn}
  (2\pi i)^2 \Delta_{\vec\eta} W_{\vec\eta}^\tau (\sigma|\rho)  &=\sum_{\alpha,\beta\in {\cal S}_{n-1}} \bigg\{\delta_{\alpha,\rho}\delta_{\beta,\sigma}\bigg[ (2\pi i)^2  (2-n) \Big(n-1+ \sum_{i=2}^n (\eta_i \partial_{\eta_i}+ \bar\eta_i \partial_{\bar\eta_i} ) \Big)\nn\\
                                                                &\hspace{20mm}+ (2\pi i)^2 \sum_{2\leq i<j}^n (\eta_i\bar\eta_j - \eta_j\bar\eta_i)(  \partial_{\eta_j}\partial_{\bar\eta_i} - \partial_{\eta_i}\partial_{\bar\eta_j})
                                                                  \nn\\
                                                                &\hspace{20mm} + 2\pi i (\tau-\bar\tau) \sum_{1\leq i<j}^n s_{ij} (\partial_{\eta_j}-\partial_{\eta_i}) (\partial_{\bar\eta_j}-\partial_{\bar\eta_i})\bigg]\nn\\
                                                                &\hspace{5mm}+ 2\pi i (\tau-\bar\tau) \Big[ \delta_{\beta,\sigma} \sum_{i=2}^n \eta_i \partial_{\bar\eta_i} \sv D_{\vec\eta}^\tau(\rho| \alpha) + \delta_{\alpha,\rho} \sum_{i=2}^n \bar\eta_i \partial_{\eta_i} \overline{\sv D_{\vec\eta}^\tau}(\sigma|\beta)\Big]\nn\\
                                                                &\hspace{5mm} +(\tau-\bar\tau)^2 \sv D_{\vec\eta}^\tau(\rho| \alpha) \, \overline{\sv D_{\vec\eta}^\tau}(\sigma|\beta)\bigg\} W_{\vec\eta}^\tau (\beta|\alpha)\,.
\end{align}
The term in the second line is due to the fact that the second-derivative terms of (\ref{almostlapn}) and 
$\Delta_{\vec\eta}-\overline{\nabla}_{\vec\eta}^{(n-2)}\nabla_{\vec\eta}^{(n-1)}$ comprise different 
contractions of the summation variables $i,j=2,3,\ldots,n$: 
One is $(\eta\partial_{\bar\eta}) (\bar\eta\partial_{\eta})$ while the other is $(\eta\partial_{\eta}) (\bar\eta\partial_{\bar\eta})$, so that only the diagonal terms cancel and one is left with a rotation-type term that contributes for $n>2$, as does the first line. These terms were not visible in the two-point example~\eqref{eq:Lap2}. Above we still set $\partial_{\eta_1}=\partial_{\bar\eta_1}=0$. We note that the equation~\eqref{eq:Lapn} has the correct reality property under complex conjugation associated with a real Laplacian at $n$ points.

Similar to the discussion at the end of Section~\ref{sec:Lap2cpt} the consistency of the $\eta$-expansions of the left-hand side and right-hand side of~\eqref{eq:Lapn} follows from integration-by-parts-identities for the component integrals. The $\eta$-expansion around different variables is also analyzed in Appendix~\ref{app:3pt} in a three-point example.

The general formula~\eqref{eq:Lapn} can be evaluated for any number of points $n$, any permutations $\rho,\sigma\in \mathcal{S}_{n-1}$ and for any component integral $W_{(A|B)}^\tau(\sigma|\rho)$. The complexity of doing so grows very rapidly, therefore we restrict ourselves here to giving only a few low-weight examples.

\subsection{Three-point examples}
\label{sec:Lap3}

This section is dedicated to the three-point instance of the Laplace equation
\eqref{eq:Lapn}
\begin{align}
  \label{eq:Lapn3}
  (2\pi i)^2 &\Delta_{\eta_2,\eta_3} W_{\eta_2,\eta_3}^\tau (\sigma|\rho)  =\sum_{\alpha,\beta\in {\cal S}_{2}} \bigg\{\delta_{\alpha,\rho}\delta_{\beta,\sigma}\bigg[
  (2\pi i)^2  (\eta_2\bar\eta_3 - \eta_3\bar\eta_2)(  \partial_{\eta_3}\partial_{\bar\eta_2} - \partial_{\eta_2}\partial_{\bar\eta_3}) 
   \nn\\
 &\hspace{28mm}- (2\pi i)^2  \Big(2+  \eta_2 \partial_{\eta_2}+ \bar\eta_2 \partial_{\bar\eta_2}
  +  \eta_3 \partial_{\eta_3}+ \bar\eta_3 \partial_{\bar\eta_3}  \Big) \nn\\
 &\hspace{28mm}+  2\pi i (\tau-\bar\tau)
 \big( s_{12} \partial_{\eta_2} \partial_{\bar\eta_2}+s_{13}  \partial_{\eta_3} \partial_{\bar\eta_3}
 +s_{23} (\partial_{\eta_2}-\partial_{\eta_3}) (\partial_{\bar\eta_2}-\partial_{\bar\eta_3}) \big) \bigg]\nn\\
&\hspace{6mm}+ 2\pi i (\tau-\bar\tau) \Big[ \delta_{\beta,\sigma}
    ( \eta_2 \partial_{\bar\eta_2}+\eta_3 \partial_{\bar\eta_3} ) \sv D_{\eta_2,\eta_3}^\tau(\rho| \alpha) 
    + \delta_{\alpha,\rho}   (\bar\eta_2 \partial_{\eta_2}+\bar\eta_3 \partial_{\eta_3}) \overline{\sv D_{\eta_2,\eta_3}^\tau}(\sigma|\beta)\Big]\nn\\
&\hspace{6mm} +(\tau-\bar\tau)^2 \sv D_{\eta_2,\eta_3}^\tau(\rho| \alpha) \, \overline{\sv D_{\eta_2,\eta_3}^\tau}(\sigma|\beta)\bigg\} W_{\eta_2,\eta_3}^\tau (\beta|\alpha)
\end{align}
and its implications for component integrals $W_{(a_2,a_3|b_2,b_3)}^{\tau }(\sigma|\rho)$ defined in~\eqref{eq:12}.
The matrix entries of $\sv D_{\eta_2,\eta_3}^\tau$ can be found in (\ref{dsv3.1}) and (\ref{dsv3.2}). In 
the simplest case with weights $(A|B)=(0,0|0,0)$ and $\rho(2,3)=\sigma(2,3)=(2,3)$, one obtains the following equation from~\eqref{eq:Lapn3}:
\begin{align}
(2\pi i)^2 \Delta  W_{(0,0|0,0)}^{\tau }&(2{,}3|2{,}3) = (\tau-\bar\tau)^2 \big[ s_{13}^2  W_{(0,2|0,2)}^{\tau }(2{,}3|2{,}3) + 2 s_{13} s_{23}   W_{(0,2|0,2)}^{\tau }(2{,}3|2{,}3) \nn\\
&\hspace{-5mm} + s_{23}^2   W_{(0,2|0,2)}^{\tau }(2{,}3|2{,}3) - s_{13}^2   W_{(0,2|0,2)}^{\tau }(2{,}3|3{,}2)  - s_{13} s_{23}   W_{(0,2|0,2)}^{\tau }(2{,}3|3{,}2)  \nn\\
&\hspace{-5mm} - s_{13}^2   W_{(0,2|0,2)}^{\tau }(3{,}2|2{,}3)  - s_{13} s_{23}   W_{(0,2|0,2)}^{\tau }(3{,}2|2{,}3) + s_{13}^2   W_{(0,2|0,2)}^{\tau }(3{,}2|3{,}2) \nn\\
&\hspace{-5mm} + s_{12} s_{13}   W_{(0,2|2,0)}^{\tau }(2{,}3|2{,}3) + s_{12} s_{23}   W_{(0,2|2,0)}^{\tau }(2{,}3|2{,}3) - s_{12} s_{13}   W_{(0,2|2,0)}^{\tau }(2{,}3|3{,}2)  \nn\\
&\hspace{-5mm} + s_{13}^2   W_{(0,2|2,0)}^{\tau }(3{,}2|2{,}3)+ s_{13} s_{23}   W_{(0,2|2,0)}^{\tau }(3{,}2|2{,}3) - s_{13}^2   W_{(0,2|2,0)}^{\tau }(3{,}2|3{,}2) \nn\\
&\hspace{-5mm} + s_{12} s_{13}   W_{(2,0|0,2)}^{\tau }(2{,}3|2{,}3) + s_{12} s_{23}   W_{(2,0|0,2)}^{\tau }(2{,}3|2{,}3)  + s_{13}^2   W_{(2,0|0,2)}^{\tau }(2{,}3|3{,}2)\nn\\
&\hspace{-5mm} - s_{12} s_{13}   W_{(2,0|0,2)}^{\tau }(3{,}2|2{,}3) - s_{13}^2   W_{(2,0|0,2)}^{\tau }(3{,}2|3{,}2) + s_{13} s_{23}   W_{(2,0|0,2)}^{\tau }(2{,}3|3{,}2) \nn\\
&\hspace{-5mm} + s_{12}^2   W_{(2,0|2,0)}^{\tau }(2{,}3|2{,}3) + s_{12} s_{13}   W_{(2,0|2,0)}^{\tau }(2{,}3|3{,}2)\nn\\
&\hspace{-5mm}  + s_{12} s_{13}   W_{(2,0|2,0)}^{\tau }(3{,}2|2{,}3) + s_{13}^2 W_{(2,0|2,0)}^{\tau }(3{,}2|3{,}2) \big]\,.
\end{align}
This equation exhibits the non-trivial mixing of the permutations due to the sum over $\alpha$ and $\beta$ in~\eqref{eq:Lapn3}. As was the case for the Cauchy--Riemann equation~\eqref{eq:CR3W0000} at three points, there are relations between the various component integrals. For instance one has the relation
\begin{align}
W^\tau_{(0,2|0,2)}(2{,}3|3{,}2) = W^\tau_{(0,2|0,2)}(3{,}2|2{,}3)
\end{align}
that can be deduced by looking at the world-sheet integral they represent and using that $f^{(2)}$ is an even function of its $z$-argument. Substituting in this and similar relations between the component integrals one arrives at
\begin{align}
\label{eq:LapW3}
(2\pi i)^2\Delta  W_{(0,0|0,0)}^{\tau }(2{,}3|2{,}3) &= (\tau-\bar\tau)^2
\Big\{ s_{12} s_{13}  \big[W_{(2,0|2,0)}(3{,}2|2{,}3) +   W_{(2,0|2,0)}(2{,}3|3{,}2)\big] \nn\\
&\hspace{-5mm}  +  s_{12}s_{23}  \big[   W_{(2,0|0,2)}(2{,}3|2{,}3) + W_{(0,2|2,0)}(2{,}3|2{,}3) \big]   \\
&\hspace{-5mm} + s_{13} s_{23} \big[  W_{(2,0|0,2)}(2{,}3|3{,}2)  + W_{(0,2|2,0)}(3{,}2|2{,}3)  \big]  \nn\\
&\hspace{-5mm}  + s_{12}^2   W_{(2,0|2,0)}(2{,}3|2{,}3)
+ s_{13}^2   W_{(2,0|2,0)}(3{,}2|3{,}2) + s_{23}^2  W_{(0,2|0,2)}(2{,}3|2{,}3)
  \Big\} \notag
\end{align}
that can be verified by explicitly acting with $\Delta= -(\tau-\bar\tau)^2\partial_\tau\partial_{\bar\tau}$ on the
modular invariant pure Koba--Nielsen integral with
\begin{align}
(2\pi i)^2 \Delta \KN^{\tau}_{3} = (\tau-\bar \tau)^2  (s_{12}f^{(2)}_{12} + s_{13}f^{(2)}_{13} + s_{23}f^{(2)}_{23} )
(s_{12} \overline{f^{(2)}_{12} }+ s_{13} \overline{ f^{(2)}_{13} } + s_{23} \overline{ f^{(2)}_{23} } ) \KN^{\tau}_{3}\,.
\label{sampleslap0}
\end{align}
The low-energy expansions of the above component integrals again translate into modular graph forms. The right-hand side of~\eqref{eq:LapW3} to third order in $\ap$ expands as
\begin{align}
\Delta  W_{(0,0|0,0)}^{\tau }(2{,}3|2{,}3) &=  (s_{12}^2+s_{23}^2+s_{13}^2) \left(\frac{\Im\tau}{\pi} \right)^2 \cform{2&0\\2&0} + 6 s_{12}s_{23}s_{13}  \left(\frac{\Im\tau}{\pi}\right)^3 \cform{3&0\\3&0} \nn\\
&\quad + (s_{12}^3+ s_{23}^3+s_{13}^3) \left(\frac{\Im\tau}{\pi}\right)^3 \cform{0&1&2\\2&1&0} + O(s_{ij}^4)\label{sampleslap1} \\
&= (s_{12}^2+s_{23}^2+s_{13}^2) {\rm E}_2 + (s_{12}^3+ s_{23}^3+s_{13}^3 + 6 s_{12}s_{23}s_{13}) {\rm E}_3 +O(s_{ij}^4)\,,\nn
\end{align}
where in the second step we have substituted in simplifications of modular graph forms of the type discussed in Appendix~\ref{sec:idMGF}. The expansion of the component integral $W_{(0,0|0,0)}^{\tau }(2{,}3|2{,}3)$ itself is
\begin{align}
W_{(0,0|0,0)}^{\tau }(2{,}3|2{,}3)  &= 1 + \frac12 (s_{12}^2+s_{23}^2+s_{13}^2)\left(\frac{\Im\tau}{\pi} \right)^2 \cform{2&0\\2&0} +  s_{12}s_{23}s_{13}  \left(\frac{\Im\tau}{\pi}\right)^3 \cform{3&0\\3&0} \nn\\
&\quad +  \frac16 (s_{12}^3+ s_{23}^3+s_{13}^3) \left(\frac{\Im\tau}{\pi}\right)^3 \cform{1&1&1\\1&1&1} + O(s_{ij}^4)\,.
\end{align}
Acting on this expression with the Laplacian using~\eqref{eq:MFGRL} leads again to non-trivial relations between modular graph forms including $\Delta \big(\frac{\Im\tau}{\pi}\big)^3  \cform{1&1&1\\1&1&1} = 6 {\rm E}_3$.
Higher orders in $\ap$ reproduce Laplace equations such as \cite{DHoker:2015gmr}
\begin{align}
(\Delta - 2) {\rm E}_{2,2} = - {\rm E}_2^2 \, , \ \ \ \ \ \ 
(\Delta - 6) {\rm E}_{2,3} =  \frac{ \zeta_5}{10} - 4 {\rm E}_2 {\rm E}_3
\label{sampleslap2}
\end{align}
from generating-function methods.

Moreover, we have extracted the Laplace equations of various further three-point component integrals
from (\ref{eq:Lapn3}) and verified consistency with the leading four or more orders in the $\ap$-expansion of
\begin{align}
&W_{(1,0|1,0)}^{\tau }(2{,}3|2{,}3) \,,&&W_{(1,0|0,1)}^{\tau }(2{,}3|2{,}3)\,, &&W_{(2,0|0,0)}^{\tau }(2{,}3|2{,}3) \,, \label{sampleslap3}\\
&W_{(2,0|2,0)}^{\tau }(2{,}3|2{,}3)\,, &&W_{(2,0|0,2)}^{\tau }(2{,}3|2{,}3)\,,&&W_{(1,1|2,0)}^{\tau }(2{,}3|2{,}3) \, . \notag
\end{align}
Expressions for $\Delta^{(1,1)}  W_{(1,0|0,1)}^{\tau }(2{,}3|2{,}3)$ and $\Delta^{(2,0)}   W_{(2,0|0,0)}^{\tau }(2{,}3|2{,}3)$
in terms of component integrals similar to (\ref{eq:LapW3}) can be found in Appendix~\ref{sec:3ptLaExpls}. We 
note that the general Laplace equation~\eqref{eq:Lapn} does not produce any modular graph 
forms with negative entries on the edge labels and never requires using holomorphic subgraph reduction.

\subsection{$n$-point examples}
\label{sec:Lapnex}

We have seen for the simplest three-point component integral $W^\tau_{(0,0|0,0)}(2{,}3|2{,}3)$
that the Laplace equation (\ref{eq:LapW3}) derived from the generating function
(\ref{eq:Lapn}) can be alternatively obtained from the Koba--Nielsen derivative (\ref{sampleslap0}). Similarly, the 
Laplacian of the $n$-point Koba--Nielsen factor
\begin{align}
(2\pi i)^2 \Delta \KN^{\tau}_{n} = (\tau-\bar \tau)^2 \bigg( \sum_{1\leq i < j}^n  s_{ij}f^{(2)}_{ij}  \bigg)
\bigg( \sum_{1\leq p < q}^n  s_{pq} \overline{f^{(2)}_{pq}}  \bigg) \KN^{\tau}_{n}
\label{sampleslap9}
\end{align}
allows for a shortcut derivation of
\begin{align}
(2\pi i)^2  \Delta &W^\tau_{(0,0,\ldots,0|0,0,\ldots,0)}(2{,}3{,}\ldots{,}n|2{,}3{,}\ldots{,}n) = (2\pi i)^2  \Delta \int \dd \mu_{n-1} \, \KN^{\tau}_{n}
\notag \\
&= (\tau-\bar \tau)^2  \int \dd \mu_{n-1}  \bigg( \sum_{1\leq i < j}^n  s_{ij}f^{(2)}_{ij}  \bigg)
\bigg( \sum_{1\leq p < q}^n  s_{pq} \overline{f^{(2)}_{pq}}  \bigg) \KN^{\tau}_{n}\, .
\label{sampleslap8}
\end{align}
The component integral on the left-hand side generates the modular graph functions with
only Green functions in the integrand. Hence, (\ref{sampleslap8}) reduces the Laplacian of arbitrary 
modular graph functions to the $\ap$-expansion of component integrals over $f^{(2)}_{ij} \overline{f^{(2)}_{pq}}$.
The latter can be straightforwardly lined up with $W^\tau_{(2,0,\ldots,0|2,0,\ldots,0)}(\sigma|\rho)$
and permutations of the subscripts $2$ and $0$.

In principle, this kind of direct computation involving the Koba--Nielsen derivative (\ref{sampleslap9}) can also be used beyond the simplest cases, e.g.
\begin{align}
(2\pi i)^2  \Delta^{(2,0)} &W^\tau_{(2,0,\ldots,0|0,0,\ldots,0)}(2{,}3{,}\ldots{,}n|2{,}3{,}\ldots{,}n) = (2\pi i)^2  \Delta^{(2,0)} \int \dd \mu_{n-1} \,f^{(2)}_{12} \, \KN^{\tau}_{n}
\notag \\
&= (\tau-\bar \tau)  \int \dd \mu_{n-1}  \KN^{\tau}_{n} \bigg\{  (\tau-\bar \tau)  f^{(2)}_{12} \bigg( \sum_{1\leq i < j \atop{(i,j) \neq (1,2)}}^n  s_{ij}f^{(2)}_{ij}  \bigg)
\bigg( \sum_{1\leq p < q}^n  s_{pq} \overline{f^{(2)}_{pq}}  \bigg) \notag \\
&\hspace{1cm} + (\tau-\bar \tau) \Big[  s_{12}(3 {\rm G}_4 - 2 f_{12}^{(4)})
+2 f^{(3)}_{12} \sum_{j=3}^n s_{2j} f^{(1)}_{2j}
\Big] \bigg( \sum_{1\leq p < q}^n  s_{pq} \overline{f^{(2)}_{pq}}  \bigg) \notag \\
&\hspace{1cm} + 4\pi i \, f^{(3)}_{12} \Big[ {-}s_{12} \overline{f^{(1)}_{12} } +   \sum_{j=3}^n s_{2j}  \overline{ f^{(1)}_{2j}} \Big]
\bigg\} \, .
\label{sampleslap7}
\end{align}
This generalizes the expression~\eqref{eq:42} for $\Delta^{(2,0)} W_{(2,0|0,0)}^{\tau }(2{,}3|2{,}3)$ to the $n$-point component integral over $f^{(2)}_{12}$. However, we needed
to use component identities such as 
\begin{align}
2\pi i ( (\tau{-}\bar \tau) \partial_\tau f^{(2)}_{12} + 2  f^{(2)}_{12} ) 
= 2(\tau{-}\bar \tau)  \partial_{z_1}  f^{(3)}_{12}  \, , \ \ \ \ \ \ 
f^{(2)}_{12}f^{(2)}_{12}-2 f^{(3)}_{12}f^{(1)}_{12} = -2 f^{(4)}_{12} + 3 {\rm G}_4
\end{align}
in intermediate steps to express the right-hand
side in terms of the basis of $W^\tau_{(A|B)}(\sigma|\rho)$ (possibly after use of the Fay identity (\ref{eq:Fay})). Manipulations of this type become increasingly complicated
with additional factors of $f^{(a)}_{ij}$ and $\overline{f^{(b)}_{ij}}$ in the integrand while the generating-function methods
underlying (\ref{eq:Lapn}) are insensitive to the choice of component integral under investigation. In summary,
this section exemplifies the Laplacian action at the level of $n$-point component integrals and
illustrates the kind of laborious manipulations that are bypassed in the generating-function approach.


\section{Towards all-order $\ap$-expansions of closed-string one-loop amplitudes}
\label{outlook}

As mentioned in the introduction, one main reason for deriving the differential equations for the generating series is to obtain a direct relation to iterated Eisenstein integrals that is valid at all orders in $\alpha'$. In order to explain this connection we briefly review the corresponding connection for the open string, in particular the $A$-cycle integrals $Z^\tau_{\vec{\eta}}(\sigma |\rho) $ defined in (\ref{again2}), and outline a strategy to perform analogous $\ap$-expansions for closed-string integrals.

\subsection{The open-string analogues}
\label{outlook.1}

We reiterate that the open-string integral is over the boundary of the cylinder with a certain ordering $\sigma$ 
of the punctures and we restrict to the planar case of all punctures on the same boundary for simplicity. 
As shown in~\cite{Mafra:2019ddf, Mafra:2019xms}, these integrals satisfy the differential equation (\ref{intr4})
with the differential operator $D^\tau_{\vec\eta}$ that is linear in the Mandelstam variables and whose single-valued version appears in~\eqref{eq:svCR}. This homogeneous first-order differential equation can be solved formally by Picard iteration (with $q=e^{2\pi i \tau}$)
\begin{align}
  Z^\tau_{\vec{\eta}}(\sigma|\rho) &=Z^{i\infty}_{\vec{\eta}}(\sigma|\rho)
                                     + \frac{1}{2\pi i } \int^\tau_{i\infty} \dd \tau_1 \sum_{\alpha \in {\cal S}_{n-1}}   D^{\tau_1}_{\vec{\eta}}(\rho|\alpha) Z^{i\infty}_{\vec{\eta}}(\sigma|\alpha)
                                     \notag
  \\
                                   & \ \ \ \ + \frac{1}{(2\pi i )^2} \int^\tau_{i\infty} \dd \tau_1 \int^{\tau_1}_{i\infty} \dd \tau_2 \sum_{\alpha,\beta \in {\cal S}_{n-1}}   D^{\tau_1}_{\vec{\eta}}(\rho|\alpha) D^{\tau_2}_{\vec{\eta}}(\alpha|\beta) Z^{i\infty}_{\vec{\eta}}(\sigma|\beta)  + \ldots
                                     \label{eq:pic.1}  \\
                                   &= \sum_{\ell=0}^{\infty} \frac{1}{(2\pi i)^{2\ell}}
                                     \! \! \! \! \! \!  \int \limits_{0<q_1<q_2<\ldots<q_\ell<q}    \! \! \! \! \! \!
                                     \frac{\dd q_1}{q_1} \frac{\dd q_2}{q_2}\ldots \frac{\dd q_\ell}{q_\ell}
                                     \sum_{\alpha \in {\cal S}_{n-1}} (D^{\tau_\ell}_{\vec{\eta}}\cdot \ldots \cdot D^{\tau_2}_{\vec{\eta}} \cdot D^{\tau_1}_{\vec{\eta}})(\rho|\alpha)
                                     Z^{i\infty}_{\vec{\eta}}(\sigma|\alpha) \, .
                                     \notag
\end{align}
The summation variable $\ell$ in the last line tracks the orders of $\ap$ carried by
the $D^{\tau_j}_{\vec{\eta}}$ matrices.
The important point here is that the initial values $Z^{i\infty}_{\vec{\eta}}(\sigma|\alpha)$ are by themselves
series in $\ap$ that have been identified with \textit{disk} integrals of Parke--Taylor type at $n+2$ points~\cite{Mafra:2019ddf, Mafra:2019xms}. Their $\ap$-expansion is expressible in terms of MZVs 
\cite{Aomoto, Terasoma, Brown:2009qja, Stieberger:2009rr, Schlotterer:2012ny}, 
and the dependence on $s_{ij}$ can for instance be imported from
the all-multiplicity methods of \cite{Broedel:2013aza, Mafra:2016mcc}. Hence, any
given $\ap$-order of the $A$-cycle integrals is accessible from finitely many terms
in the sum over $\ell$ in (\ref{eq:pic.1}), i.e.\ after finitely many steps of Picard iteration.

By expanding the Weierstra\ss{} functions in the matrix entries of
$D^{\tau}_{\vec{\eta}}$ in terms of holomorphic Eisenstein series using~\eqref{eq:28}, 
one can uniquely decompose
\begin{align}
  D^{\tau}_{\vec{\eta}} = \sum_{k=0}^{\infty} (1-k) {\rm G}_k(\tau) r_{\vec{\eta}}(\epsilon_k)
  \label{eq:pic.2}
\end{align}
with ${\rm G}_0=-1$. All the reference to $\eta_j$ resides in the differential operators 
$r_{\vec{\eta}}(\epsilon_k)$, where $\epsilon_k$ is a formal letter with $k=0,4,6,8,\ldots$.  Explicit expression for the $r_{\vec\eta}(\epsilon_k)$ can be found in~\cite{Mafra:2019ddf, Mafra:2019xms} and they are believed to form matrix representations
of Tsunogai's derivations dual to Eisenstein series~\cite{Tsunogai}. They inherit linearity in $s_{ij}$ from $D^\tau_{\vec{\eta}}$, and we have $r_{\vec{\eta}}(\epsilon_2)=0$ at all multiplicities
by the absence of ${\rm G}_2$ in the Laurent expansion (\ref{eq:28}) of $\wp(\eta,\tau)$.

Substituting~\eqref{eq:pic.2} into~\eqref{eq:pic.1}, the entire $\tau$-dependence in
the $\ap$-expansion of $A$-cycle integrals $Z^\tau_{\vec{\eta}}$ is carried by 
iterated Eisenstein integrals
\begin{align}
  \gamma(k_1,k_2,\ldots,k_\ell |\tau) := \frac{(-1)^\ell}{(2\pi i)^{2\ell} } 
  \! \! \! \! \! \!  \int \limits_{0<q_1<q_2<\ldots<q_\ell<q}    \! \! \! \! \! \!
  \frac{\dd q_1}{q_1} \frac{\dd q_2}{q_2}\ldots \frac{\dd q_\ell}{q_\ell} {\rm G}_{k_1}(\tau_1)
  {\rm G}_{k_2}(\tau_2)\ldots {\rm G}_{k_\ell}(\tau_\ell)
  \label{eq:pic.3}
\end{align}
subject to tangential-base-point regularization \cite{Brown:mmv} such that $\gamma(0|\tau) = \frac{\tau }{2\pi i}$.
We arrive at a sum over words $k_1,k_2,\ldots,k_\ell$ composed from the alphabet $k_j \in \{0,4,6,8,\ldots\}$
\cite{Mafra:2019ddf, Mafra:2019xms},                                                                                                                                                                                                                                                       %
\begin{align}
  Z^\tau_{\vec{\eta}}(\sigma|\rho) &=
                                     \sum_{\ell=0}^{\infty} \sum_{k_1,k_2,\ldots,k_\ell \atop{=0,4,6,8,\ldots }}\Big[ \prod_{j=1}^\ell (k_j{-}1) \Big] \gamma(k_1,k_2,\ldots,k_\ell|\tau) \sum_{\alpha \in {\cal S}_{n-1}}r_{\vec{\eta}}(\epsilon_{k_\ell} \ldots \epsilon_{k_2} \epsilon_{k_1})_{\rho}{}^\alpha
                                     Z^{i\infty}_{\vec{\eta}}(\sigma|\alpha)  \, ,
                                     \label{eq:pic.4}
\end{align}
where we write $r_{\vec{\eta}}(\epsilon_{k_\ell} \ldots \epsilon_{k_2} \epsilon_{k_1})=
r_{\vec{\eta}}(\epsilon_{k_\ell} )\ldots r_{\vec{\eta}}(\epsilon_{k_2} )r_{\vec{\eta}}(\epsilon_{k_1})$ for ease of notation.

The formula~\eqref{eq:pic.4} provides an explicit evaluation of the one-loop $A$-cycle integrals in terms of iterated Eisenstein integrals and initial values that can be traced back to tree-level amplitudes. As all $\tau$-dependence is carried by iterated integrals $\gamma(k_1,k_2,\ldots,k_\ell|\tau)$, the differential equation~\eqref{intr4} is satisfied by the defining property
\begin{align}
  2\pi i  \partial_\tau\gamma(k_1,k_2,\ldots,k_\ell|\tau) &= - {\rm G}_{k_\ell}(\tau)
                                                            \gamma(k_1,k_2,\ldots,k_{\ell-1}|\tau)\, 
\end{align}
of iterated Eisenstein integrals, cf.\ \eqref{eq:pic.3}. Moreover, (\ref{eq:pic.4}) yields the
correct initial condition for $Z^\tau_{\vec{\eta}}(\sigma|\rho)$ since 
$\lim_{\tau\to i\infty} \gamma(k_1,k_2,\ldots,k_\ell|\tau) = 0$.

\subsection{An improved form of closed-string differential equations}
\label{outlook.2}

We shall now outline a starting point for the corresponding procedure to expand closed-string generating series $W_{\vec\eta}^\tau(\sigma|\rho)$. One can first rewrite the Cauchy--Riemann equation~\eqref{eq:CRn} as
\begin{align}
2\pi i(\tau-\bar\tau) \partial_\tau    W_{\vec{\eta}}^\tau(\sigma|\rho) &= 2\pi i  \Big[ 1-n+\sum_{i=2}^n (\bar{\eta}_i- \eta_i)\partial_{\eta_i}  \Big]W_{\vec\eta}^\tau(\sigma|\rho) \notag \\
&\hspace{10mm}  + (\tau-\bar\tau) \sum_{\alpha \in {\cal S}_{n-1}} \sv D^\tau_{\vec\eta}(\rho| \alpha)W_{\vec{\eta}}^\tau(\sigma|\alpha) \, .
\label{eq:pic.21}
\end{align}
The terms in the first line are independent of the Mandelstam variables and also mix the holomorphic and anti-holomorphic orders in the variables of the generating series. This obstructs a direct link between Picard iteration and the $\alpha'$-expansion in analogy with the open-string construction. In the following, we will present a redefinition of the 
$W_{\vec{\eta}}^\tau$ integrals such that one can still obtain each order in the $\ap$-expansion of
the component integrals through a finite number of elementary operations.

The contributions $ 1-n-\sum_{i=2}^n  {\eta}_i \partial_{\eta_i}$ to the $\ap$-independent square-bracket
in \eqref{eq:pic.21} can be traced back to the connection term in the Maa\ss{} operator (\ref{eq:MRn}) which
simply adjusts the modular weights. In general, one can suppress the connection term in \eqref{eq:MRL}
by enforcing vanishing holomorphic modular weight on the functions it acts on, which is always
possible by multiplication with suitable powers of $(\tau - \bar \tau)$. Hence, we will consider a modified
version of the $W$-integrals, where each component integral in (\ref{cptnp2}) of modular weight $(|A|,|B|)$
is multiplied by $(\tau-\bar \tau)^{|A|}$ such as to attain the shifted modular weights $(0,|B|-|A|)$.

Since the component integrals $W_{(A|B)}^\tau$ defined in (\ref{cptnpt1})
have modular weights $(|A|,|B|)$, the desired modification of (\ref{cptnp2})
is given by
\begin{align}
  \WAS_{\vec{\eta}}^\tau(\sigma|\rho) &:= \sum_{A,B}
                                        (\tau-\bar \tau)^{|A|}W_{(A|B)}^\tau(\sigma|\rho)    
\, \sigma \big[ \bar \eta_{234\ldots n}^{b_2-1} \bar \eta_{34\ldots n}^{b_3-1} \ldots \bar \eta_{n}^{b_n-1}  \big]
 \, \rho\big[ \eta_{234\ldots n}^{a_2-1} \eta_{34\ldots n}^{a_3-1} \ldots \eta_{n}^{a_n-1} \big] \notag \\
&\phantom{:}= \sum_{A,B} W_{(A|B)}^\tau(\sigma|\rho) 
  \, \sigma \big[ \bar \eta_{234\ldots n}^{b_2-1} \bar \eta_{34\ldots n}^{b_3-1} \ldots \bar \eta_{n}^{b_n-1}  \big]     \label{eq:pic.22} \\
& \ \ \ \ \times  (\tau-\bar \tau)^{n-1}\, \rho\Big[  \big( (\tau - \bar \tau)\eta_{234\ldots n} \big)^{a_2-1} 
 \big( (\tau - \bar \tau)\eta_{34\ldots n} \big)^{a_3-1} \ldots    \big( (\tau - \bar \tau) \eta_{n} \big)^{a_n-1} \Big]     \notag \\
 &\phantom{:}= (\tau-\bar\tau)^{n-1} W_{\vec\eta}^\tau(\sigma|\rho)\Big|_{\substack{\eta \to (\tau-\bar\tau)\eta\\\bar\eta \to \bar\eta\phantom{(\tau-\bar\tau)}}}  \, . \notag
\end{align}
Given that the entire $Y$-integral has holomorphic modular weight zero, the action of the
Maa\ss{} operator (\ref{eq:MRn}) reduces to $(\tau-\bar \tau)\partial_\tau$, and the Cauchy--Riemann
equation (\ref{eq:pic.21}) simplifies to 
\begin{align}
  2\pi i(\tau{-}\bar \tau)^2 \partial_\tau \WAS^\tau_{\vec{\eta}}(\sigma|\rho) &= 2\pi i \sum_{j=2}^n \bar \eta_j \partial_{\eta_j} \WAS^\tau_{\vec{\eta}}(\sigma|\rho) + (\tau{-}\bar \tau)^2 \sum_{\alpha \in {\cal S}_{n-1}}
  \sv D^\tau_{(\tau - \bar \tau) \vec{\eta}}(\rho|\alpha) \WAS^\tau_{\vec{\eta}}(\sigma|\alpha) \nn\\
&  = \sum_{k=0}^{\infty} (1-k)(\tau {-} \bar \tau)^k {\rm G}_k(\tau) \sum_{\alpha \in {\cal S}_{n-1}} 
  R_{\vec{\eta}}(\epsilon_k)_\rho{}^\alpha
  \WAS^\tau_{\vec{\eta}}(\sigma|\alpha) \,,
    \label{eq:pic.24} 
\end{align}
where we have expanded the closed-string differential operator in terms of Eisenstein series in analogy with~\eqref{eq:pic.2}. The operator $R_{\vec\eta}(\epsilon_0)$ contains also the $s_{ij}$-independent term $\sim \bar \eta_j \partial_{\eta_j}$ in its diagonal components
\begin{align}
  R_{\vec{\eta}}(\epsilon_0)_\rho{}^\rho  &= \rho\Big[ \sum_{1\leq i<j}^n \frac{s_{ij} }{\eta_{j,j+1\ldots n}^2} \Big]
 - \frac{1}{2} \sum_{1\leq i<j}^n s_{ij} (\partial_{\eta_i}-\partial_{\eta_j})^2  - 2\pi i \sum_{j=2}^n \bar \eta_j \partial_{\eta_j}\, ,
 \label{pic.27A}
\end{align}
whereas  $R_{\vec{\eta}}(\epsilon_0)_\rho{}^\alpha = r_{\vec{\eta}}(\epsilon_0)_\rho{}^\alpha$ for $\rho \neq \alpha$.
For $k\geq 4$, we have agreement with the open-string expression $R_{\vec\eta}(\epsilon_k)=r_{\vec\eta}(\epsilon_k)$ and these matrices should again form a matrix representation of Tsunogai's derivations \cite{Tsunogai}. 
We have checked that they preserve the commutation relations of the 
$\epsilon_k$~\cite{LNT, Pollack, Broedel:2015hia} to the
same orders as done for their open-string analogues, see Section~4.5 of \cite{Mafra:2019xms}. 
The appearance of the term $\sim \bar \eta_j \partial_{\eta_j}$ in $R_{\vec\eta}(\epsilon_0)$ does not obstruct the $\ap$-expansion of component integrals from finitely many steps\footnote{This follows from the fact that none of the $R_{\vec\eta}(\epsilon_k)$ has a contribution that lowers the powers of $\bar \eta_j$. Hence, any component integral at a given order of $\bar \eta_j$ in (\ref{eq:pic.22}) can only be affected by finitely many instances
of the term $\bar \eta_j \partial_{\eta_j}$ in $R_{\vec\eta}(\epsilon_0)$ which is the only contribution to the
$R_{\vec\eta}(\epsilon_k)$ without any factors of $s_{ij}$.} in a formal solution with the structure of (\ref{eq:pic.4}).

Note that the asymmetric role of the $\eta_j$ and $\bar \eta_j$ variables in the definition (\ref{eq:pic.22})
of the $Y$-integrals will modify the Laplace equation (\ref{eq:Lapn}) and obscure its reality properties.
For this reason we have worked with the generating series $W_{\vec\eta}^\tau(\sigma| \rho)$ in earlier sections of this work.


\subsection{A formal all-order solution to closed-string $\ap$-expansion}
\label{outlook.3}

In analogy with the open-string $\ap$-expansion (\ref{eq:pic.4}),
one can construct a formal solution to (\ref{eq:pic.24}) from the infinite sum over
words in the alphabet $k_j\in \{0,4,6,8,\ldots\}$:
\begin{align}
  \WAS^\tau_{\vec{\eta}}(\sigma|\rho) &=
                                        \sum_{\ell=0}^{\infty} \sum_{k_1,k_2,\ldots,k_\ell \atop{=0,4,6,8,\ldots }}\Big[ \prod_{j=1}^\ell (k_j{-}1) \Big] \beta^{\rm sv}(k_1,k_2,\ldots,k_\ell|\tau) \sum_{\alpha \in {\cal S}_{n-1}}R_{\vec{\eta}}(\epsilon_{k_\ell} \ldots \epsilon_{k_2} \epsilon_{k_1})_{\rho}{}^\alpha
                                        \WAS^{i\infty}_{\vec{\eta}}(\sigma|\alpha)
                                        \label{eq:pic.28}
\end{align}
This ansatz obeys (\ref{eq:pic.24}) with the correct initial conditions at $\tau \rightarrow i\infty$
if the quantities $\beta^{\rm sv}(k_1,k_2,\ldots,k_\ell|\tau)$ satisfy the following initial-value problem
\begin{subequations}
  \label{eq:pic.30A}
  \begin{align}
    2\pi i (\tau - \bar \tau)^2 \partial_\tau\beta^{\rm sv}(k_1,k_2,\ldots,k_\ell|\tau) &=  - (\tau - \bar \tau)^{k_\ell} 
    {\rm G}_{k_\ell}(\tau)  \beta^{\rm sv}(k_1,k_2,\ldots,k_{\ell-1}|\tau)  \label{eq:pic.29}\,,\\
    \lim_{\tau \rightarrow i\infty}\beta^{\rm sv}(k_1,k_2,\ldots,k_\ell| \tau )&= 0 \, .   \label{eq:pic.30}
  \end{align}
\end{subequations}
However, an iterative solution of (\ref{eq:pic.30A}) does not completely determine them since one
can still add anti-holomorphic functions of $\tau$ that vanish at the cusp at each step. We 
expect these anti-holomorphic `integration constants' to be fixed by the reality condition $\overline{W_{\vec\eta}^\tau(\sigma|\rho)} = W_{\vec\eta}^\tau(\rho|\sigma)$ and the modularity properties of the component integrals. In particular, the anti-holomorphic differential equation of the generating series (\ref{eq:pic.22})
\begin{align}
-2\pi i \partial_{\bar \tau} Y^\tau_{\vec{\eta}}(\sigma|\rho) &= 2\pi i \sum_{j=2}^n \Big[  \eta_j \partial_{\bar \eta_j} + \frac{  \eta_j \partial_{\eta_j} - \bar \eta_j \partial_{\bar \eta_j}  }{\tau - \bar \tau}
\Big] Y^\tau_{\vec{\eta}}(\sigma|\rho) + \! \! \! \sum_{\beta \in {\cal S}_{n-1} } \! \! \overline{\sv D_{\vec\eta}^\tau}(\sigma|\beta) Y^\tau_{\vec{\eta}}(\beta |\rho) 
 \label{antiY} 
 \end{align}
might be a convenient starting point. Still, the ansatz (\ref{eq:pic.28}) may not yet be the optimal formulation of closed-string $\ap$-expansions since the initial value~\eqref{eq:pic.30} is in general incompatible with the usual shuffle 
relations for the $\beta^{\rm sv}(k_1,k_2,\ldots,k_\ell|\tau)$.

Another important point in this context is the determination of the initial values $\WAS^{i\infty}_{\vec\eta}$ from its $\alpha'$-expanded Laurent polynomial and the connection with sphere integrals. The recent all-order results for the Laurent-polynomial part of the two-point component integral $W^\tau_{(0|0)}$ \cite{DHoker:2019xef, Zagier:2019eus} 
and the results on the $\tau \rightarrow i \infty$ degeneration of the $A$-cycle integrals 
\cite{Mafra:2019ddf, Mafra:2019xms} should harbor valuable guidance.
The correct prescription to iteratively determine the $\beta^{\rm sv}(k_1,\ldots |\tau)$ is expected to involve a notion of `single-valued' integration 
along the lines of~\cite{Brown:2018omk, Brown:2019wna} and may be related to the non-holomorphic modular forms 
of~\cite{Brown:2017qwo, Brown:2017qwo2}. We plan to report on this in the near future.

\section{Summary and outlook}
\label{summary}

In this paper, we have derived first- and second-order differential equations for generating functions $W_{\vec\eta}^\tau(\sigma|\rho)$ of general torus integrals over any number $n$ of world-sheet punctures. The torus integrands involve products of doubly-periodic Kronecker--Eisenstein series and Koba--Nielsen factors that are tailored to $n$-point correlation functions of massless one-loop closed-string amplitudes in bosonic, heterotic and type-II theories. These differential equations, given in~\eqref{eq:CRn} and~\eqref{eq:Lapn}, are exact in $\alpha'$. The differential operators appearing in the differential equations coincide with the single-valued versions of similar operators appearing in the open string~\cite{Mafra:2019ddf,Mafra:2019xms}.

The flavor of open-closed string relations motivating the present work is different from
the Kawai--Lewellen--Tye (KLT) relations \cite{Kawai:1985xq} that express closed-string tree-level
amplitudes via squares of open-string trees. Tree-level KLT relations do not manifest 
the disappearance of MZVs outside the single-valued subspace, and loop-level KLT relations
in string theory are unknown at the time of writing\footnote{See Section 5.3 of \cite{Casali:2019ihm} 
for a recent account on the key challenges in setting up loop-level KLT relations among string 
amplitudes from the viewpoint of intersection theory \cite{Mizera:2017cqs}. 
A field-theory version of one-loop KLT relations among loop integrands in gauge theories and supergravity 
has been derived from ambitwistor strings \cite{He:2016mzd, He:2017spx}. 
A connection with the present 
results may be investigated by adapting the generating functions of this work to the chiral-splitting formalism \cite{Verlinde:1986kw, DHoker:1988pdl, DHoker:1989cxq},
where a string-theory analogue of the loop momentum is introduced at the expense of obscuring modular 
invariance. Still, it is unclear whether this approach yields an alternative form of the field-theory KLT relations in \cite{He:2016mzd, He:2017spx}.}. The long-term
goal of this work is to directly relate closed- and single-valued open-string amplitudes at one loop, 
generalizing the tree-level connection of \cite{Schlotterer:2012ny, Stieberger:2013wea, Stieberger:2014hba, Schlotterer:2018abc, Brown:2018omk,Vanhove:2018elu, Brown:2019wna}. Hence, it remains to be seen to which extent 
the present results and follow-up work \cite{toappsoon} relate to a tentative one-loop KLT construction.

The generating series $W_{\vec\eta}^\tau(\sigma|\rho)$ can be expanded in powers of the bookkeeping variables $\eta_2,\eta_3,\ldots,\eta_n$ to yield equations for the component integrals that are modular forms of fixed weights. These in turn can be expanded in powers of Mandelstam variables, corresponding to a low-energy expansion, to generate systematically differential equations for modular graph forms that arise in explicit string scattering calculations. 
Our generating-function approach yields Cauchy--Riemann and Laplace equations for
modular graph forms associated with an arbitrary number of vertices and edges.
The equations generated in this way can be used to deduce non-trivial identities between modular graph forms. Our method in particular never generates modular graph forms with negative entries and does not necessitate holomorphic subgraph reduction to expose holomorphic Eisenstein series. 

It would be interesting to generalize the methods of this work such as to
generate differential equations for higher-genus modular graph forms 
\cite{DHoker:2013fcx, DHoker:2014oxd, DHoker:2017pvk,DHoker:2018mys,Basu:2018bde}.
As a first step, one would need to identify a suitable Kronecker--Eisenstein series with $2g$
periodicities and an analogue of the $\eta$-expansion of this work, where the building blocks
of multiloop string amplitudes are recovered.
Moreover, in view of the recent advances in integrating modular graph forms
over $\tau$ \cite{Green:2008uj, DHoker:2015gmr, Basu:2017nhs,DHoker:2019mib, DHoker:2019blr}, 
our results are hoped to yield useful input for the flat-space
limit of string amplitudes in $AdS_5\times S^5$
\cite{Okuda:2010ym, Penedones:2010ue, Alday:2018pdi, Alday:2018kkw, Drummond:2019odu}.


\section*{Acknowledgments}

We are grateful to Johannes Broedel, Daniele Dorigoni, Carlos Mafra, Nils Matthes, Erik Panzer and Federico Zerbini 
for inspiring discussions and collaboration on related topics. Moreover, we thank Johannes Broedel for valuable comments on a draft version. JG and OS are
grateful to the organizers of the programme ``Modular forms, periods and
scattering amplitudes'' at the ETH Institute for Theoretical Studies for
providing a stimulating atmosphere and financial support. We gratefully acknowledge support from the Simons Center for Geometry and Physics in Stony Brook during the workshop ``Automorphic Structures in String Theory''. JG and AK thank the University of Uppsala for hospitality and JG furthermore thanks Chalmers University of Technology for hospitality during the final stage of the project. OS is 
supported by the European Research Council under
ERC-STG-804286 UNISCAMP. JG is supported by the International Max Planck Research School for Mathematical
and Physical Aspects of Gravitation, Cosmology and Quantum Field Theory.


\appendix

\section{Identities for $\Omega(z,\eta,\tau)$}
\label{app:Om}

In this appendix, we collect a number of useful identities satisfied by the doubly-periodic version of the Kronecker--Eisenstein series and in particular Fay identities for coincident $z$-arguments.

As was mentioned in Section~\ref{sec:fay-ids}, the Fay identity \eqref{eq:24} of the meromorphic Kronecker--Eisenstein series $F(z,\eta,\tau)$ implies a similar identity for the doubly-periodic version $\Omega(z,\eta,\tau)$ which can also be written as
  \begin{align}
    \Omega(z_1, \eta_1, \tau) \Omega(z_2,\eta_2,\tau) &= \Omega(z_1+z_2,\eta_1,\tau)\Omega(z_2,\eta_2-\eta_1,\tau)  + \Omega(z_1+z_2, \eta_2,\tau)\Omega(z_1,\eta_1-\eta_2,\tau)\ ,
  \end{align}
where we have used $\Omega(-z,-\eta,\tau) = - \Omega(z,\eta,\tau)$ to rewrite \eqref{eq:FayO}. The Fay identity implies an infinity of bilinear identities between the $f^{(a)}$ functions by expanding in $\eta_1$ and $\eta_2$. These can be written more succinctly by using the notation $z_{ij}=z_{i}-z_{j}$ and letting $z_1 \to z_{12}$ and $z_2\to z_{13}$ since the $z_{12}$ appearing in the relation above then becomes $z_{32}$. Expanding in $\eta_{1}$ and $\eta_{2}$ yields the bilinear Fay identities for the $f^{(a)}$ (where $a_1,a_2\geq 0$) \cite{Broedel:2014vla}:
\begin{align}
  \label{eq:Fay}
  f^{(a_1)}_{12} f^{(a_2)}_{13} = (-1)^{a_1-1} f^{(a_1+a_2)}_{23} &+ \sum_{j=0}^{a_1} \binom{a_2+j-1}{j} f_{32}^{(a_1-j)} f_{13}^{(a_2+j)}\\
&+ \sum_{j=0}^{a_2} \binom{a_1+j-1}{j}f_{12}^{(a_1+j)} f_{23}^{(a_2-j)}\,.  \nn
\end{align}
In the following, we want to extend \eqref{eq:Fay} to the case where the arguments of the two $f^{(a)}$ coincide. To this end, we take the limit $z_1 \rightarrow z_2$ in the meromorphic version \eqref{eq:24} of the Fay identity and obtain \eqref{eq:26}, which we quote here for reference:\footnote{One might be tempted to take the coincident limit directly in \eqref{eq:Fay}, however, this is not possible since the limit $\lim_{\epsilon\rightarrow0}f^{(a)}(\epsilon,\tau)$ is ill-defined for $a=2$.}
\begin{align}
  \Omega(z,\eta_{1},\tau)\Omega(z,\eta_{2},\tau)&=\Omega(z,\eta_{1}+\eta_{2},\tau)\left(g^{(1)}(\eta_{1},\tau)+g^{(1)}(\eta_{2},\tau)+\frac{\pi}{\Im \tau}(\eta_{1}+\eta_{2})\right)\notag\\
  &\qquad-\partial_{z}\Omega(z,\eta_{1}+\eta_{2},\tau)\ .
\end{align}
Expanding this in $\eta_{1}$ and $\eta_{2}$ and using
\begin{align}
 g^{(1)}(\eta,\tau)&=\frac{1}{\eta}-\sum_{k=2}^{\infty}\eta^{k-1}\mathrm{G}_{k}(\tau)
\end{align}
leads to a version of \eqref{eq:Fay} with both $f^{(a)}$ evaluated at the same point:
\begin{align}
  &f^{(a_{1})}(z)f^{(a_{2})}(z)=f^{(a_{1}+a_{2})}(z)\binom{a_{1}{+}a_{2}}{a_{2}}-\binom{a_{1}{+}a_{2}{-}2}{a_{2}-1}\left(\widehat{\mathrm{G}}_{2}f^{(a_{1}+a_{2}-2)}(z)+\partial_{z}f^{(a_{1}+a_{2}-1)}(z)\right)\notag\\
  &\hspace{2.3em}-\sum_{k=4}^{a_{1}+a_{2}}\binom{a_{1}{+}a_{2}{-}1{-}k}{a_{2}-1}\mathrm{G}_{k}f^{(a_{1}+a_{2}-k)}(z)-\sum_{k=4}^{a_{2}}\binom{a_{1}{+}a_{2}{-}1{-}k}{a_{1}-1}\mathrm{G}_{k}f^{(a_{1}+a_{2}-k)}(z)  \ ,\label{eq:27}
\end{align}
where $a_{1},a_{2}\geq0$ and we set $f^{(a)}=0$ for $a<0$. Note that the manifest exchange symmetry $a_{1}\leftrightarrow a_{2}$ is broken just by the different summation ranges in the second line. However, the terms with \mbox{$k=a_{1}{+}1,\dots, a_{1}{+}a_{2}{-}1$} in the first sum all vanish due to the binomial coefficient and the only remaining term is due to $k=a_{1}{+}a_{2}$ and given by $(-1)^{a_{2}}\mathrm{G}_{a_{1}+a_{2}}$, which is invariant under $a_{1}\leftrightarrow a_{2}$. We did not make this manifest since $\mathrm{G}_{a_{1}+a_{2}}$ does not appear if $a_{1}{+}a_{2}<4$. Specializing \eqref{eq:27} to $a_{1}=1,2$ yields
\begin{subequations}
\label{eq:Faydz}
  \begin{align}
    \label{eq:Fay1ad}
    f^{(1)}(z) f^{(a)}(z) &=  - \partial_z f^{(a)}(z) +  (a+1) f^{(a+1)}(z) - \widehat{\rm G}_2 f^{(a-1)}(z)- \sum_{k=4}^{a+1} {\rm G}_k f^{(a+1-k)}(z)\,,\\
    \label{eq:Fay2ad}
    f^{(2)}(z) f^{(a)}(z)  &= - a \partial_z f^{(a+1)}(z) + \frac12 (a+1)(a+2) f^{(a+2)}(z) -a \widehat{\rm G}_2 f^{(a)}(z)  \nn\\
                          &\hspace{20mm}- \sum_{k=4}^{a+2} (a+1-k){\rm G}_k f^{(a+2-k)}(z)\,.
  \end{align}
\end{subequations}
Here, $a\geq0$ and again we set $f^{(a)}=0$ for $a<0$. Using \eqref{eq:Fay1ad} and \eqref{eq:Fay2ad} we can derive \eqref{eq:25} in the main text by expanding in $\eta$:
\begin{align}
  \left( f^{(1)} \partial_\eta - f^{(2)} \right) \Omega &= -\eta^{-2} f^{(1)} + \sum_{a\geq 0} \eta^{a-1} \left( a f^{(1)} f^{(a+1)} - f^{(2)} f^{(a)} \right)\notag\\
&= -\eta^{-2} f^{(1)} + \sum_{a\geq 0} \eta^{a-1} \left( \frac12(a{+}2)(a{-}1) f^{(a+2)} - \sum_{k=4}^{a+2}(k{-}1) {\rm G}_k f^{(a+2-k)}  \right)\notag\\
&=\left( \frac12 \partial_\eta^2 - \wp(\eta,\tau)\right) \Omega(z,\eta,\tau)\,.   \label{thiseqhere}
\end{align}
We note that the terms involving $z$-derivatives of the $f^{(a)}$ cancel in this particular combination and we are left with a purely algebraic expression in these functions. In passing to the last line of (\ref{thiseqhere}), we used the expansion \eqref{eq:28} of the Weierstra\ss{} function. Note that the Weierstra\ss{} function also satisfies the identity
\begin{align}
  \wp(\eta,\tau) &= g^{(1)}(\eta,\tau) g^{(1)}(\eta,\tau) - 2 g^{(2)}(\eta,\tau) = - \partial_\eta g^{(1)}(\eta,\tau) - {\rm G}_2(\tau)\ .
\end{align}

Finally, the identity \eqref{eq:Fay1ad} can also be derived by expanding the following relations for derivatives of Kronecker-Eisenstein series:
\begin{subequations}
  \begin{align}
    \partial_z F(z,\eta,\tau) -\partial_\eta F(z,\eta,\tau) &= \left(g^{(1)}(\eta,\tau)-g^{(1)}(z,\tau)\right) F(z,\eta,\tau)\,,\\
    \label{eq:Ozeta}
    \partial_z \Omega (z,\eta,\tau) -\partial_\eta \Omega(z,\eta,\tau) &= \left(g^{(1)}(\eta,\tau) +\frac{\pi\eta}{\Im\tau} -f^{(1)}(z,\tau)\right) \Omega(z,\eta,\tau)\,.
  \end{align}
\end{subequations}


\section{Identities between modular graph forms}\label{sec:idMGF}

There are many non-trivial identities relating modular graph forms and these are crucial in the simplification of $\ap$-expansions of Koba--Nielsen integrals~\cite{Green:2008uj, DHoker:2015gmr, DHoker:2016mwo, DHoker:2016quv, Kleinschmidt:2017ege,Gerken:2018jrq,Dorigoni:2019yoq}. In this appendix, we will review the most important techniques by which these identities can be obtained.

The most straightforward approach to simplifying modular graph forms is probably to manipulate their lattice-sum representation. This was done e.g.\ in the case of $\cform{1&1&1\\1&1&1}$ by Zagier~(see~\cite{DHoker:2015gmr}), who proved that
\begin{align}
  \cform{1&1&1\\1&1&1}=\Big(\frac{\pi}{\Im\tau}\Big)^{3}(\mathrm{E}_{3}+\zeta_{3})\ .
\end{align}
However, this kind of direct manipulation is hard and only possible in a limited number of cases. In the following subsections we exhibit some more systematic ways of obtaining identities between modular graph forms.

\subsection{Topological simplifications}\label{sec:top-simp}
The simplest way to derive identities arises when the graph associated to the modular graph form has certain special topologies~\cite{DHoker:2016mwo}. For example, if the graph contains a two-valent vertex, momentum conservation at that vertex together with the form of the propagators implies that the vertex can be removed and the weights of the adjacent edges added:
\begin{align}
  \tikzpicture[baseline=-3,scale=0.7,line width=0.3mm,decoration={markings,mark=at position 0.6 with {\arrow[scale=1.3]{latex}}}]
  \fill (0,0) circle [radius=0.1];
  \fill (3,0) circle [radius=0.1];
  \fill (6,0) circle [radius=0.1];
  \draw[postaction={decorate}](0,0)node[below]{$i$}--node[above]{\scriptsize $(a_{1},b_{1})$}(3,0)node[below]{$j$};
  \draw[postaction={decorate}](3,0)--node[above]{\scriptsize $(a_{2},b_{2})$}(6,0)node[below]{$k$};
  \endtikzpicture
  \hspace{2em} = \hspace{2em}
  \tikzpicture[baseline=-3,scale=0.7,line width=0.3mm,decoration={markings,mark=at position 0.6 with {\arrow[scale=1.3]{latex}}}]
  \fill (0,0) circle [radius=0.1];
  \fill (4,0) circle [radius=0.1];
  \draw[postaction={decorate}](0,0)node[below]{$i$}--node[above]{\scriptsize $(a_{1}{+}a_{2},b_{1}{+}b_{2})$}(4,0)node[below]{$k$};
  \endtikzpicture
  \hspace{2em}.\label{eq:35}
\end{align}
A special case of this arises when the graph is dihedral. In this case, the simplification \eqref{eq:35} implies
\begin{align}
 \cform{a_{1}&a_{2}\\b_{1}&b_{2}}=(-1)^{a_{2}+b_{2}}\cform{a_{1}+a_{2}&0\\b_{1}+b_{2}&0}=(-1)^{a_{2}+b_{2}}\sum_{p\neq0}\frac{1}{p^{a_{1}+a_{2}}\bar{p}^{b_{1}+b_{2}}} \ ,
\end{align}
where the $[\begin{smallmatrix}0\\0\end{smallmatrix}]$ column is necessary to be consistent with the general notation for dihedral graphs introduced in \eqref{ourdef}. For a trihedral graph with a two-valent vertex, \eqref{eq:35} implies
the following simplification to the dihedral topology:
\begin{align}
  \cformtri{a_{12}\\b_{12}}{a_{23}\\b_{23}}{A_{31}\\B_{31}}&=(-1)^{a_{12}+b_{12}+a_{23}+b_{23}}\cform{a_{12}+a_{23}&A_{31}\\b_{12}+b_{23}&B_{31}}\ .
\end{align}

Furthermore, if removing one vertex is sufficient to make the graph disconnected, the associated modular graph form factorizes:
\begin{align}
  \tikzpicture[baseline=-3,scale=0.7,line width=0.3mm]
  \fill (2,0) circle [radius=0.1];
  \draw(0,0.75)..controls(1.2,0.75)..(2,0);
  \draw(0,-0.75)..controls(1.2,-0.75)..(2,0);
  \draw(1,0.15) node{$\vdots$};
  \filldraw[fill=black!10,draw=black](0,0) ellipse (0.5 and 1);
  \draw(0,0) node{$\Gamma_{1}$};
  \draw(2,0)..controls(2.8,0.75)..(4,0.75);
  \draw(2,0)..controls(2.8,-0.75)..(4,-0.75);
  \draw(3,0.15) node{$\vdots$};
  \filldraw[fill=black!10,draw=black](4,0) ellipse (0.5 and 1);
  \draw(4,0) node{$\Gamma_{2}$};
  \endtikzpicture
  \hspace{2em} = \hspace{2em}
  \tikzpicture[baseline=-3,scale=0.7,line width=0.3mm]
  \fill (2,0) circle [radius=0.1];
  \draw(0,0.75)..controls(1.2,0.75)..(2,0);
  \draw(0,-0.75)..controls(1.2,-0.75)..(2,0);
  \draw(1,0.15) node{$\vdots$};
  \filldraw[fill=black!10,draw=black](0,0) ellipse (0.5 and 1);
  \draw(0,0) node{$\Gamma_{1}$};
  \endtikzpicture
  \hspace{1em} \times \hspace{1em}
  \tikzpicture[baseline=-3,scale=0.7,line width=0.3mm]
  \fill (2,0) circle [radius=0.1];
  \draw(2,0)..controls(2.8,0.75)..(4,0.75);
  \draw(2,0)..controls(2.8,-0.75)..(4,-0.75);
  \draw(3,0.15) node{$\vdots$};
  \filldraw[fill=black!10,draw=black](4,0) ellipse (0.5 and 1);
  \draw(4,0) node{$\Gamma_{2}$};
  \endtikzpicture
  \hspace{2em}.\label{eq:36}
\end{align}
For a trihedral graph with a vertex pair that is not directly connected by any edges, \eqref{eq:36} 
implies the following identity:
\begin{align}
  \cformtri{}{A_{23}\\B_{23}}{A_{31}\\B_{31}}&=\cform{A_{23}\\B_{23}}\cform{A_{31}\\B_{31}}\ ,
\end{align}
where the empty first column reflects that vertices $1$ and $2$ are not directly connected.

\subsection{Factorization}\label{sec:MGFfact}
The simplifications in the previous section dealt with identities arising from special topologies, i.e.\ absent edges. If an edge is not absent but carries weight $(0,0)$, one can perform the sum over the momentum flowing through this edge explicitly~\cite{DHoker:2016mwo}. The result has two contributions according to the two-term expression (\ref{eq:30}) for $C^{(0,0)}$: One factorizes, the other one is of lower loop order. Explicitly, for the dihedral case, the resulting identity is
\begin{align}
  \cform{a_1&a_2&\ldots &a_{R} &0 \\b_1&b_2&\ldots &b_{R} &0} = \prod_{j=1}^{R} \cform{a_j &0 \\b_j &0} - \cform{a_1&a_2&\ldots &a_{R}  \\b_1&b_2&\ldots &b_{R} }  \, .
\end{align}
The corresponding identity for trihedral graphs reads
\begin{align}
  \cformtri{A_{12}&0\\B_{12}&0}{A_{23}\\B_{23}}{A_{31}\\B_{31}}=(-1)^{|A_{23}|+|B_{23}|}\mathcal{C}\!\left[\begin{smallmatrix}A_{23}&A_{31}\\B_{23}&B_{31}\end{smallmatrix}\right]\prod_{i=1}^{R_{12}} \mathcal{C}\!\left[\begin{smallmatrix}a_{12}^{i}&0\\b_{12}^{i}&0\end{smallmatrix}\right]-\cformtri{A_{12}\\B_{12}}{A_{23}\\B_{23}}{A_{31}\\B_{31}}\ .
\end{align}

\subsection{Momentum conservation}\label{sec:MomCons}
Momentum conservation implies that the sum of all momenta flowing into a vertex (with sign) is zero. Writing this vanishing momentum sum into the numerator of the modular graph form and expanding the result leads to non-trivial identities~\cite{DHoker:2016mwo}. For dihedral graphs, they are
\begin{align}
  0=\sum_{j=1}^R \cform{a_1&a_2&\ldots &a_j{-}1 &\ldots &a_R \\b_1&b_2&\ldots &b_j &\ldots &b_R}=\sum_{j=1}^R \cform{a_1&a_2&\ldots &a_j &\ldots &a_R \\b_1&b_2&\ldots &b_j{-}1 &\ldots &b_R} \, .\label{eq:39}
\end{align}
Similarly, in the trihedral case we have
\begin{align}
  \sum_{i=1}^{R_{12}}\cformtri{A_{12}-S_{i}\\B_{12}}{A_{23}\\B_{23}}{A_{31}\\B_{31}}-\sum_{j=1}^{R_{23}}\cformtri{A_{12}\\B_{12}}{A_{23}-S_{j}\\B_{23}}{A_{31}\\B_{31}}&=0\label{eq:40}\\
  \sum_{i=1}^{R_{12}}\cformtri{A_{12}\\B_{12}-S_{i}}{A_{23}\\B_{23}}{A_{31}\\B_{31}}-\sum_{j=1}^{R_{23}}\cformtri{A_{12}\\B_{12}}{A_{23}\\B_{23}-S_{j}}{A_{31}\\B_{31}}&=0\notag\ ,
\end{align}
where $S_{i}$ is the row vector whose $j$\textsuperscript{th} component is $\delta_{ij}$.

\subsection{Holomorphic subgraph reduction}\label{sec:HSR}
Finally, if a modular graph form contains a closed subgraph of only holomorphic or anti-holomorphic edges, the sum over the momentum flowing in this subgraph can be performed explicitly by means of partial-fraction decomposition~\cite{DHoker:2016mwo}. For dihedral graphs with two holomorphic edges, this leads to
\begin{align}
  \cform{a_{+}&a_{-}&A\\0&0&B}&=\cform{a_{+}&a_{-}\\0&0}\cform{A\\B}-\binom{a_{+} + a_{-}}{a_{-}}\cform{a_{+}+a_{-}&A\\0&B}\notag\\
              &\quad+\sum_{k=4}^{a_{+}}\binom{a_{+}+a_{-}-1-k}{a_{+}-k}
              {\rm G}_k\cform{a_{+}+a_{-}-k&A\\0&B} \label{eq:37}\\
              &\quad+\sum_{k=4}^{a_{-}}\binom{a_{+}+a_{-}-1-k}{a_{-}-k}
              {\rm G}_k\cform{a_{+}+a_{-}-k&A\\0&B}\notag\\
              &\quad+\binom{a_{+}+a_{-}-2}{a_{+}-1}\left\{\widehat{\rm G}_{2}\cform{a_{+}+a_{-}-2&A\\0&B}+\frac{\pi}{\Im\tau}\cform{a_{+}+a_{-}-1&A\\-1&B}\right\} \, .\notag
\end{align}
The derivation of \eqref{eq:37} involves the following regularization of conditionally convergent sums, derived using the Eisenstein summation prescription~\cite{DHoker:2016mwo}:
\begin{align}
  \sum_{\substack{p\neq0\\p\neq q}} \frac{1}{p}=-\frac{1}{q} - \frac{\pi}{2\Im\tau}( q - \bar q).\hspace{5em}\sum_{\substack{p\neq0\\p\neq q}} \frac{1}{p^{2}}=-\frac{1}{q^{2}}+\widehat{\mathrm{G}}_{2}+\frac{\pi}{\Im\tau}\,.
\end{align}
Higher-point generalizations of these sums were worked out in~\cite{Gerken:2018zcy} and hence higher-point generalizations of \eqref{eq:37} can be obtained. A lengthy, but closed, formula for trihedral graphs is available in~\cite{Gerken:2018zcy}. 

\subsection{Verifying two-point Cauchy--Riemann equations}\label{sec:2ptMGFVrf}

As an example of the power of the identities outlined in the previous sections, we will prove the two-point Cauchy--Riemann equations \eqref{cr2ptcp} for component integrals in this section. The proof relies on applying momentum conservation~\eqref{eq:40} and holomorphic subgraph reduction~\eqref{eq:37} order by order in $\ap$.

The $\ap$-expansion of two-point component integrals \eqref{lapsec15} can be expressed in closed form (cf.\ \eqref{eq:7})
\begin{align}
  W^{\tau}_{(a|b)}=(-1)^{a}\sum_{k=0}^{\infty}s_{12}^{k}\frac{1}{k!}\Big(\frac{\Im\tau}{\pi}\Big)^{k}\cformS{a&0&1_{k}\\0&b&1_{k}} \, ,\hspace{4em} a,b>0,\quad(a,b)\neq(1,1)\,,\label{eq:41}
\end{align}
where $1_{k}$ denotes the row vector with $k$ entries of 1. In the case $a=0$, $b\geq0$, it is easy to check that~\eqref{cr2ptcp} is satisfied without using any identities for modular graph forms. The case $a>0$, $b=0$ follows along the lines of the derivation below by dropping the $[\begin{smallmatrix}0\\b\end{smallmatrix}]$ columns everywhere and adjusting the overall sign. The final case $a=b=1$ can be reduced to the $a=b=0$ case by using the integration-by-parts identity \eqref{ibp2b}.

Focusing on the coefficient of $s_{12}^{k}$ in~\eqref{eq:41}, we compute the action of the Cauchy--Riemann operator $\nabla^{(a)}$ by means of \eqref{eq:MFGRL}, leading to
\begin{align}
&(-1)^{a}k! \Big(\frac{\pi}{\Im\tau}\Big)^{k}\nabla^{(a)}W^{\tau}_{(a|b)}\Big|_{s_{12}^{k}}=a\cformS{a+1&0&1_{k}\\-1&b&1_{k}}+k \cformS{a&0&2&1_{k-1}\\0&b&0&1_{k-1}} \\
& \ \ = 
\,a\Big\{{-}\cformS{a+1&0&1_{k}\\0&b-1&1_{k}}-k \cformS{a+1&0&1&1_{k-1}\\0&b&0&1_{k-1}}\Big\}
 +k\Big\{\mathrm{G}_{a+2}\cformS{0&1_{k-1}\\b&1_{k-1}}-\frac{1}{2}(a{+}2)(a{+}1)\cformS{a+2&0&1_{k-1}\\0&b&1_{k-1}} 
\notag\\
&\hspace{3em}+  \sum_{\ell=4}^{a}(a{+}1{-}\ell)\mathrm{G}_{\ell}\cformS{a+2-\ell&0&1_{k-1}\\0&b&1_{k-1}}+a\Big(\widehat{\mathrm{G}}_{2}\cformS{a&0&1_{k-1}\\0&b&1_{k-1}}+\frac{\pi}{\Im\tau}\cformS{a+1&0&1_{k-1}\\-1&b&1_{k-1}}\Big)\Big\}\,,\notag
\end{align}
where we have used~\eqref{eq:39} to remove the -1 entry in the first term and~\eqref{eq:37} to simplify the second term. Using holomorphic subgraph reduction again in $\cformS{a+1&0&1&1_{k-1}\\0&b&0&1_{k-1}}$ and reorganizing the terms leads to (recall that $a,b>0$ and $(a,b)\neq(1,1)$)
\begin{align}
&(-1)^{a}k! \Big(\frac{\pi}{\Im\tau}\Big)^{k}\nabla^{(a)}W^{\tau}_{(a|b)}\Big|_{s_{12}^{k}}=
-a \cformS{a+1&0&1_{k}\\0&b-1&1_{k}}
\\
  &\hspace{1em}-k\Big({-}\frac{1}{2}(a{+}2)(a{-}1)\cformS{a+2&0&1_{k-1}\\0&b&1_{k-1}}+(a{+}1)\mathrm{G}_{a+2}\cformS{0&1_{k-1}\\b&1_{k-1}}+\sum_{\ell=4}^{a+1}(\ell{-}1)\mathrm{G}_{\ell}\cformS{a+2-\ell&0&1_{k-1}\\0&b&1_{k-1}}\Big) \ .\notag
\end{align}
Comparing this to the coefficient of $s_{12}^{k}$~in \eqref{cr2ptcp} shows that they agree.

\section{Component integrals versus $n$-point string amplitudes}
\label{app:string}

In this appendix, we specify the
component integrals $W_{(A|B)}^\tau(\sigma|\rho)$ (\ref{cptnpt1}) that 
enter $n$-point one-loop closed-string amplitudes of the
bosonic, heterotic and type-II theories in more detail.

Even though the zero modes of the world-sheet bosons couple the chiral
halves of the closed string at genus $g>0$, it is instructive to first review the analogous open-string
correlators:
\begin{itemize}
\item The $n$-point correlators of massless vertex operators of the open superstring are 
strongly constrained by its sixteen supercharges. This can be seen from the sum over spin structures
in the RNS formalism \cite{Ramond:1971gb, Neveu:1971rx} or from the fermionic zero modes in the pure-spinor formalism \cite{Berkovits:2000fe, Berkovits:2004px}.
As a result, these correlators comprise products $\prod_{k} f^{(a_k)}_{i_k j_k}$ of overall weight
$ \sum_k a_k = n-4$ as well as admixtures of holomorphic Eisenstein series 
${\rm G}_w\prod_{k} f^{(a_k)}_{i_k j_k}$ with $w+ \sum_k a_k = n-4$ and $w\geq 4$ 
\cite{Tsuchiya:1988va, Broedel:2014vla}.
When entering heterotic-string or type-II amplitudes as a chiral half,
open-superstring correlator introduce component integrals $W_{(A|B)}^\tau(\sigma|\rho)$
with holomorphic modular weights $|A| \leq n-4$.
\item In orbifold compactifications of the open superstring that preserve four or eight supercharges,
the RNS spin sums are modified by the partition function, see e.g.\ \cite{Blumenhagen:2006ci} for a review. 
The spin-summed $n$-point correlators of massless vertex operators may therefore depend on the punctures
via $\prod_{k} f^{(a_k)}_{i_k j_k}$ or ${\rm G}_{w\geq 4} \prod_{k} f^{(a_k)}_{i_k j_k}$
of weight $\sum_k a_k =n{-}2$ or $w{+} \sum_k a_k = n{-}2$ \cite{Bianchi:2015vsa, Berg:2016wux}.
The resulting component integrals $W_{(A|B)}^\tau(\sigma|\rho)$ in closed-string amplitudes
with such chiral halves have holomorphic modular weights $|A| \leq n{-}2$.
\item Open bosonic strings in turn allow for combinations $\prod_{k} f^{(a_k)}_{i_k j_k}$
and $\widehat {\rm G}_{w=2} \prod_{k} f^{(a_k)}_{i_k j_k}$ or
${\rm G}_{w\geq 4} \prod_{k} f^{(a_k)}_{i_k j_k}$ of weight $\sum_k a_k =n$ 
and $w+ \sum_k a_k = n$. The same is true for the $n$-point torus correlators of Kac--Moody currents
entering the gauge sector of the heterotic string \cite{Dolan:2007eh, Gerken:2018jrq}. Accordingly, closed-string 
amplitudes of the heterotic and bosonic theory comprise component integrals 
$W_{(A|B)}^\tau(\sigma|\rho)$ with modular weights $|A| \leq n$ or $|B| \leq n$ in one or two chiral halves.
\end{itemize}
The pattern of $f^{(a_k)}_{i_k j_k}$ obtained from a direct evaluation of the correlators may
not immediately line up with the integrands of $W_{(A|B)}^\tau(\sigma|\rho)$. First,
contractions among the world-sheet bosons introduce spurious derivatives $\partial_z f^{(1)}(z,\tau)$. Second,
one may encounter arrangements of the labels $i_j$ in the first argument 
in ``cycles'' $f^{(a_1)}_{i_1i_2}f^{(a_2)}_{i_2i_3}\ldots f^{(a_k)}_{i_ki_1}$
rather than the ``open chains'' $f^{(a_2)}_{i_1i_2}f^{(a_3)}_{i_2i_3} \ldots f^{(a_n)}_{i_{n-1}i_n}$
characteristic to (\ref{cptnpt1}). 

In both cases, combinations of integration by parts and Fay identities
of Appendix~\ref{app:Om} are expected to reduce any term in the above correlators to the integrands
of $W_{(A|B)}^\tau(\sigma|\rho)$. For instance, the methods in Appendix~D.1 of \cite{Gerken:2018jrq}
reduce the cycle $f^{(1)}_{12} f^{(1)}_{23} f^{(1)}_{34} f^{(1)}_{41}$
in a four-point current correlator to ${\rm G}_4$ as well as
$f^{(a_2)}_{12}f^{(a_3)}_{23}f^{(a_4)}_{34}$ with $a_2+a_3+a_4=4$ and permutations.

The above properties of chiral halves set upper bounds on the modular weights
$|A|,|B|$ seen in the component integrals $W_{(A|B)}^\tau(\sigma|\rho)$ of the respective
closed-string amplitudes. However, additional contributions with (non-negative)
weights $(|A|-k,|B|-k), \ k \in \NN$ arise from interactions between left and right-movers:
Both the direct contractions between left- and right-moving world-sheet bosons and the
contribution of (\ref{antipar}) to integrations by parts convert one unit of holomorphic
and non-holomorphic modular weight into a factor of $(\Im \tau)^{-1}$, see 
e.g.\ \cite{Richards:2008jg, Green:2013bza, Berg:2016wux}.

For instance, the six-point amplitude of the type-II superstring in the representation
of~\cite{Mafra:2016nwr} is composed of component integrals
$W_{(A|B)}^\tau(\sigma|\rho)$ with $(|A|,|B| ) \in \{ (2,2),(1,1),(0,0) \}$. The integrands 
associated with $|A| = |B| =2$ can take one of the forms 
\begin{align}
(|A|,|B| ) = (2,2) \ \ \leftrightarrow \ \ 
f^{(2)}_{ij}\, \overline{f^{(2)}_{pq}} \, , \ \
f^{(2)}_{ij}\, \overline{f^{(1)}_{pq}} \, \overline{f^{(1)}_{rs}} \, , \ \ 
f^{(1)}_{ij} \, f^{(1)}_{kl} \, \overline{f^{(2)}_{pq}} \, , \ \
f^{(1)}_{ij} \, f^{(1)}_{kl} \, \overline{f^{(1)}_{pq}}\, \overline{f^{(1)}_{rs}} \, ,
\end{align}
with at most one overlapping label among $f^{(1)}_{ij} \, f^{(1)}_{kl}$ (and separately among $\overline{f^{(1)}_{pq}}\, \overline{f^{(1)}_{rs}}$).


\section{Kinematic poles}


In this appendix, we spell out an example of the subtraction schemes that were
mentioned in Section~\ref{sec:alpha-expand-comp-ints} to capture kinematic poles of component
integrals in terms of modular graph forms.

\subsection{Subtraction scheme for a two-particle channel}
\label{app:poleA}

 For simplicity, we begin with the example
\begin{align}
W_{s_{12}}^{\tau} := \int \dd \mu_{n-1} \, f^{(1)}_{12} \, \overline{ f^{(1)}_{12}} 
\, \Phi(z_j,\bar z_j) \,
\KN^{\tau}_{n} \, ,
\label{subtr1}
\end{align}
where the integrand $ \Phi(z_j,\bar z_j) $ is tailored to admit no
kinematic pole different from $s_{12}^{-1}$. This can be reconciled with the general form
(\ref{cptnpt1}) of component integrals if
\begin{align} 
\Phi(z_j,\bar z_j)= f_{2i_3}^{(a_3)}\,f_{i_3i_4}^{(a_4)} \, \ldots  \, f_{i_{n-1},i_n}^{(a_n)} 
           \, \overline{ f_{2j_3}^{(b_3)}}  \, \overline{ f_{j_3,j_4}^{(b_4)}}\ldots  \overline{ f_{j_{n-1},j_n}^{(b_n)} } 
           \label{subtr2}
\end{align}
exhibits no singularities of the form $|z_{kl}|^{-2}$ which for instance imposes $i_3\neq j_3$.

The key idea of the subtraction scheme is to split the Koba--Nielsen factor of (\ref{subtr1}) into
\begin{align} 
W_{s_{12}}^{\tau} &= \int \dd \mu_{n-1} \, f^{(1)}_{12} \, \overline{ f^{(1)}_{12}} \, e^{s_{12} G(z_{12},\tau)}
\, \Phi(z_j,\bar z_j) \,  \label{subtr3} \\
& \ \ \ \times \bigg( \underbrace{ \prod_{j=3}^{n} e^{s_{1j} G(z_{1j},\tau)}
- \prod_{j=3}^{n} e^{s_{1j} G(z_{2j},\tau)} }_{(i)}
+ \underbrace{ \prod_{j=3}^{n} e^{s_{1j} G(z_{2j},\tau)} }_{(ii)}\bigg)
\widehat \KN^{\tau}_{n} \, ,
          \notag
\end{align}
where 
\begin{align} 
\widehat \KN^{\tau}_{n} :=  \prod_{2\leq i<j}^{n} \exp \big( s_{ij} G(z_{ij}, \tau) \big) \, .
           \label{subtr4}
\end{align}
The two terms in the integrand of (\ref{subtr3}) marked by $(i)$ vanish at the residue of the
pole $f^{(1)}_{12} \, \overline{ f^{(1)}_{12}} \sim |z_{12}|^{-2}$. As a result, the integrals over the punctures
$z_2,z_3,\ldots,z_n$ in $W_{s_{12}}^{\tau} \, \big|_{(i)}$ will be finite for each term of the Taylor-expanded 
Koba--Nielsen factor in (\ref{eq:20}) and yield modular graph forms by the discussion below that equation.

For the leftover term in (\ref{subtr3}) marked by $(ii)$, the integrand only depends on $z_1$
via $f^{(1)}_{12}  \overline{ f^{(1)}_{12}}  e^{s_{12} G(z_{12},\tau)}$ which can be conveniently integrated by parts
using
\begin{align}
f^{(1)}_{12}  \, \overline{ f^{(1)}_{12}}  \, e^{s_{12} G(z_{12},\tau)}
&=  -\frac{1}{s_{12}} \Big[ \partial_{z_1} \big( \overline{ f^{(1)}_{12}}  \, e^{s_{12} G(z_{12},\tau)} \big) + \frac{ \pi }{ \Im \tau}  e^{s_{12} G(z_{12},\tau)} \Big]\, .
\label{subtr5}
\end{align}
The modified Koba--Nielsen factor in (\ref{subtr3}, $ii$) with $e^{s_{1j} G(z_{2j},\tau)}$ in the place of 
$e^{s_{1j} G(z_{1j},\tau)}$ does not alter
the fact that the total $z_1$-derivative integrates to zero (there are by construction no poles
in $z_{1j},\bar z_{1j}$ with $j\geq 3$ that could contribute via Cauchy's theorem).
Even if the measure $\dd\mu_{n-1}$ is defined in (\ref{eq:dmu}) to not comprise an explicit integral over $z_1$,
one can still discard total derivatives w.r.t.\ $z_1$ after undoing the fixing $z_1=0$, setting $z_n=0$ and
integrating $z_1,\ldots,z_{n-1}$ over the torus instead of $z_2,\ldots,z_n$.

Hence, the splitting (\ref{subtr3}) along with the integration by parts due to (\ref{subtr5}) allow to
isolate the kinematic pole of the example (\ref{subtr1}),
\begin{align} 
W_{s_{12}}^{\tau} \, \big|_{(ii)}&= - \frac{1}{s_{12}} \frac{ \pi }{ \Im \tau}  \int \dd \mu_{n-1} \, 
\Phi(z_j,\bar z_j)\, e^{s_{12} G(z_{12},\tau)}\,\widehat \KN^{\tau}_{n}\,\prod_{j=3}^{n} e^{s_{1j} G(z_{2j},\tau)} 
  \, .
         \label{subtr6} 
\end{align}
The integral can again be performed as in (\ref{eq:20}), by Taylor-expanding the
(modified) Koba--Nielsen factor and identifying the terms at each order in $\ap$ as modular graph forms.


\subsection{Integration by parts at three points}
\label{app:poleB}

As an example of nested kinematic poles involving a three-particle channel
$\sim s_{123}^{-1}$, we shall study the three-point component integrals
\begin{align}
W^\tau_{(1,1|1,1)}(2{,}3|2{,}3) &= \int \dd \mu_2 \, f_{12}^{(1)} \, f_{23}^{(1)} \,\overline{f_{12}^{(1)}} \, \overline{f_{23}^{(1)}}\, \KN^{\tau}_{3} 
\notag \\
W^\tau_{(1,1|1,1)}(3{,}2|2{,}3) &= \int \dd \mu_2 \, f_{12}^{(1)} \, f_{23}^{(1)} \, \overline{f_{13}^{(1)}} \, \overline{f_{32}^{(1)}}\, \KN^{\tau}_{3}  \label{subtr20}
 \end{align}
in this section. In order to relegate their kinematic poles to the coefficients in an integration-by-parts relation,
the first step is to rewrite the chiral integrand by means of the Fay identity 
$f^{(1)}_{12}f^{(1)}_{23} + f^{(2)}_{12} + \te{cyc}(1,2,3)= 0$, cf.\ (\ref{eq:Fay}),
 \begin{align}
 f^{(1)}_{12} f^{(1)}_{23} = - \frac{ s_{13} }{s_{123}} ( f^{(2)}_{12} + f^{(2)}_{23}  + f^{(2)}_{31} )
 + \frac{ X_{12,3} }{s_{12} s_{123}} - \frac{ X_{23,1} }{s_{23} s_{123}}\, .
 \label{subtr21}
 \end{align}
The combinations on the right-hand side 
 \begin{align}
 X_{ij,k} := s_{ij}f^{(1)}_{ij} (s_{ik} f^{(1)}_{ik} + s_{jk} f^{(1)}_{jk} ) = s_{ij}f^{(1)}_{ij} \partial_{z_k} \log \KN^{\tau}_{3} 
 \label{subtr22}
 \end{align}
are Koba--Nielsen derivatives which we integrate by parts to act
on the anti-chiral factors
 \begin{align}
\int \dd \mu_2 \, f_{12}^{(1)} \, f_{23}^{(1)} \,\overline{f_{ij}^{(1)}} \, \overline{f_{jk}^{(1)}}\, \KN^{\tau}_{3} 
&= \int \dd \mu_2\, \KN^{\tau}_{3} \, \Big\{
 {-} \frac{ s_{13} }{s_{123}} ( f^{(2)}_{12} {+} f^{(2)}_{23}  {+} f^{(2)}_{31} )\,\overline{f_{ij}^{(1)}} \, \overline{f_{jk}^{(1)}} 
  \notag \\
  &\ \ \ \ \ \ \ \ \ \ \ - \frac{ f^{(1)}_{12} }{s_{123}} \, \partial_{z_3} (\overline{f_{ij}^{(1)}} \, \overline{f_{jk}^{(1)}})
  + \frac{ f^{(1)}_{23} }{s_{123}} \, \partial_{z_1} (\overline{f_{ij}^{(1)}} \, \overline{f_{jk}^{(1)}})
 \Big\} \, .
 \label{subtr23}
 \end{align}
Given that $\partial_z f^{(1)}(z,\tau) = - \frac{\pi }{\Im \tau}$ within Koba--Nielsen integrals, 
choosing $(i,j,k)=(1,2,3)$ and $(1,3,2)$ casts the examples (\ref{subtr20}) into the form
 \begin{align}
W^\tau_{(1,1|1,1)}(2{,}3|2{,}3)  &=  \int \dd \mu_2\, \KN^{\tau}_{3} \, \Big\{
 {-} \frac{ s_{13} }{s_{123}} ( f^{(2)}_{12} {+} f^{(2)}_{23}  {+} f^{(2)}_{31} )\,\overline{f_{12}^{(1)}} \, \overline{f_{23}^{(1)}} 
  \notag \\
  &\ \ \ \ \ \ \ \ \ \ \ -\frac{1}{s_{123}} \frac{ \pi }{\Im \tau} \big[  f_{12}^{(1)} \overline{f_{12}^{(1)}} 
  +f_{23}^{(1)} \overline{f_{23}^{(1)}}  \big]
 \Big\}
 \label{subtr24} \\
W^\tau_{(1,1|1,1)}(3{,}2|2{,}3) &= \int \dd \mu_2\, \KN^{\tau}_{3} \, \Big\{
 {-} \frac{ s_{13} }{s_{123}} ( f^{(2)}_{12} {+} f^{(2)}_{23}  {+} f^{(2)}_{31} )\,\overline{f_{13}^{(1)}} \, \overline{f_{32}^{(1)}} 
  \notag \\
  &\ \ \ \ \ \ \ \ \ \ \ + \frac{1}{s_{123}} \frac{ \pi }{\Im \tau} \big[  f_{12}^{(1)} ( \overline{f_{13}^{(1)}} + \overline{f_{23}^{(1)}} )
  +f_{23}^{(1)} \overline{f_{23}^{(1)}}  \big] \Big\}\, .
 \label{subtr25}
 \end{align}
The right-hand sides still feature integrals over $\frac{ \pi }{\Im \tau} f_{ij}^{(1)}  \overline{f_{ij}^{(1)}}$
with kinematic poles $s_{ij}^{-1}$ of lower complexity. The latter may either be addressed by the 
subtraction scheme (\ref{app:poleA}) or absorbed into integration-by-parts coefficients in the following
three-point generalization of (\ref{ibp2b})
\begin{align}
W^\tau_{(1,0|1,0)}(2{,}3|2{,}3) &=  \int \dd \mu_2 \, f_{12}^{(1)} \, \overline{f_{12}^{(1)}} \, \KN^{\tau}_{3} 
=  \int \dd \mu_2 \, \Big[ \frac{ s_{23} }{s_{12}} \, f_{23}^{(1)} \, \overline{f_{12}^{(1)}} - \frac{1}{s_{12}} \frac{ \pi }{\Im \tau} \Big] \, \KN^{\tau}_{3} \notag \\
&=  \frac{ s_{23} }{s_{12}}  W^\tau_{(0,1|1,0)}(2{,}3|2{,}3) - \frac{1}{s_{12}} \frac{ \pi }{\Im \tau}
W^\tau_{(0,0|0,0)}(2{,}3|2{,}3) \, .
 \label{subtr26}
 \end{align}
The three-point examples with integrands $f_{12}^{(1)} \, f_{23}^{(1)} \,\overline{f_{ij}^{(1)}} \, \overline{f_{jk}^{(1)}}$
illustrate the general strategy to successively manifest nested poles in $n$-point component integrals
$W_{(1,1,\ldots,1|1,1,\ldots,1)}(\sigma | \rho)$ via integration by parts:
Similar to (\ref{subtr21}), Fay identities can be used to rearrange the product of $f^{(1)}_{ij}$
into the Koba--Nielsen derivatives $\prod_{j=2}^n \sum_{i=1}^{j-1} s_{ij} f^{(1)}_{ij}$
that generalize (\ref{subtr22}). The less singular terms $\sim f^{(k>1)}_{ij}$ generated by Fay identities and
the contributions from $\partial_{z_i}$ acting on anti-chiral integrands give rise to fewer 
kinematic poles. In particular, these less singular terms can be further simplified by the 
lower-multiplicity analogues of the earlier manipulations.

\section{Proof of $s_{ij}$-form of product of Kronecker--Eisenstein series}
\label{app:sij}

In this appendix, we prove that the ubiquitous product of doubly-periodic Kronecker--Eisenstein series admits what we call an $s_{ij}$-form (for any choice of $1\leq i<j\leq n$) that is amenable to evaluating the differential operator given in~\eqref{eq:CRsimple2}. This form is given by
\begin{align}
  \label{eq:sijform}
  \prod_{p=2}^n \Omega(z_{p-1,p},\xi_p)
  &=(-1)^{j-i+1}  \sum_{k=i+1}^j \Omega(\{1,\ldots,i-1\}\shuffle \{k-1,\ldots,i+1\}) \Omega(z_{ij},\xi_k) \nn\\
  &\hspace{10mm}\times \Omega(\{j-1,\ldots,k\}\shuffle \{j+1,\ldots,n\})\,.
\end{align}
Note that $i<k\leq j$, so that $\{k-1,\ldots,i+1\}$ and $\{j-1,\ldots,k\}$ always have to be in descending order and thus are empty when $k=i+1$ or when $k=j$, respectively. Similarly, the sequences $\{1,\ldots,i-1\}$ and $\{j+1,\ldots,n\}$ are always in ascending and order and thus empty for $i=1$ or $j=n$, respectively. The first factor denotes a shuffle sum of products of a total of $k-2$ Kronecker--Eisenstein series
\begin{align}
  \label{eq:prodL}
  &\qquad \Omega(\{1,\ldots,i-1\}\shuffle \{k-1,\ldots,i+1\}) \nn\\ 
  &:= \sum_{(a_1,\ldots,a_{k-2})\in \{1,\ldots,i-1\}\shuffle \{k-1,\ldots,i+1\}} \left[\prod_{p=2}^{k-2} \Omega\bigg(z_{a_{p-1},a_p},\sum_{\ell=p}^{k-2} \eta_{a_\ell} + \xi_k\bigg) \right]\Omega(z_{a_{k-2},i},\eta_i+\xi_k)\, ,
\end{align}
while the last factor is a shuffle sum of products of $n-k$ Kronecker--Eisenstein series:
\begin{align}
  \label{eq:prodR}
  &\qquad \Omega(\{j-1,\ldots,k\}\shuffle \{j+1,\ldots,n\}) \nn\\ 
  &:= \sum_{(a_{k+1},\ldots,a_n)\in \{j-1,\ldots,k\}\shuffle \{j+1,\ldots,n\}} \Omega(z_{j,a_{k+1}},\xi_k-\eta_j)\left[\prod_{p=k+2}^{n} \Omega\bigg(z_{a_{p-1},a_p},\sum_{\ell=p}^{n} \eta_{a_\ell} \bigg)\right] \,.
\end{align}
Note that we use here
\begin{align}
  \eta_1 = -\sum_{p=2}^n \eta_p \ .
\end{align}

Since the notation is a bit involved, we also give a more intuitive description of~\eqref{eq:sijform}. For a fixed $k$ in the range $i<k\leq j$ we have a permutation $(a_1,\ldots,a_n)$ of the range $(1,\ldots,n)$ with the following boundary conditions:
\begin{enumerate}[(i)]
\item The indices $i$ and $j$ occur at positions $k-1$ and $k$: $a_{k-1}=i$ and $a_k=j$.
\item The subsequence $(a_1,\ldots,a_{k-2})$ to the left of $i$ (at position $k-1$) is a rearrangement of $\{1,\ldots,k-1\}\setminus\{i\}$ such that it is obtained as a shuffle in $\{1,\ldots,i-1\}\shuffle \{k-1,\ldots,i+1\}$. 
\item The subsequence $(a_{k+1},\ldots,a_n)$ to the right of $j$ (at position $k$) is a rearrangement of $\{k,\ldots,n\}\setminus\{j\}$ such that it is obtained as a shuffle in $\{j-1,\ldots,k\}\shuffle \{j+1,\ldots,n\}$. 
\end{enumerate}
The situation is also illustrated in Figure~\ref{fig:sijform}. To each such sequence $(a_1,\ldots,a_n)$ there is a product in~\eqref{eq:sijform}
\begin{align}
  \label{eq:Oseq}
  \prod_{p=2}^n \Omega\bigg( z_{a_{p-1},a_p} , \sum_{\ell=p}^n \eta_{a_\ell}\bigg)
\end{align}
and one is summing over all possible intermediate points $i<k\leq j$.

Note that we are using that 
\begin{align}
  \xi_k= \sum_{\ell=k}^n \eta_\ell = \sum_{\ell=k}^n \eta_{a_\ell}
\end{align}
since the subsequence $(a_k,\ldots, a_n)$ is a permutation of $\{k,\ldots,n\}$. Therefore also 
\begin{align}
\sum_{\ell=p}^{k-1} \eta_{a_\ell} + \xi_k = \sum_{\ell=p}^n \eta_{a_\ell}
\quad \text{and}\quad
\eta_i + \xi_k = \sum_{\ell=k-1}^n \eta_{a_\ell}\,,
\end{align}
such that the arrangement of $\eta$-arguments in~\eqref{eq:sijform} is correct to represent a different sequence of points with corresponding generating arguments.

We note that the change of variables at $n$ points
\begin{align}
  \xi_p  = \sum_{\ell=p}^n \eta_p 
  \qquad\Longleftrightarrow\qquad
  \eta_p = \left\{\begin{array}{cl}\xi_p-\xi_{p+1}&\text{for $1<p<n$}\\
                    \xi_n & \text{for $p=n$}\end{array}\right.
\end{align}
implies that for the differential operator one has
\begin{align}
  \label{eq:dopxik}
  \sum_{k=i+1}^j \partial_{\xi_k} = \partial_{\eta_j} - \partial_{\eta_i}
\end{align}
for all $1\leq i<j\leq n$ when setting $\partial_{\eta_1}=0$. Thus the differential operator  $\sum_{k=i+1}^j \partial_{\xi_k}$ acts only as $\partial_{\xi_k}$ on $\Omega(z_{ij},\xi_k)$ and vanishes on all other factors in~\eqref{eq:sijform}. This is true since $\xi_k$ only contains $\eta_j$ but not $\eta_i$. The term $\xi_k-\eta_{a_k}=\xi_k-\eta_j$ and all other terms to the right of $\Omega(z_{ij},\xi_k)$ are free of $\eta_j$ while $\xi_k+\eta_{a_{k-1}}= \xi_k+\eta_i$ and all other terms to the left of $\Omega(z_{ij},\xi_k)$ always contains $\eta_i+\eta_j$ and are therefore also annihilated by~\eqref{eq:dopxik}. Thus the $s_{ij}$-form~\eqref{eq:sijform} is the correct form for evaluating for $1\leq i<j\leq n$ that
\begin{align}
  \label{eq:sijdiff}
  & \left[f_{ij}^{(1)} \sum_{k=i+1}^j \partial_{\xi_k} - f_{ij}^{(2)} \right] \prod_{p=2}^n \Omega(z_{p-1,p},\xi_p,\tau)= \frac12  (\partial_{\eta_j}-\partial_{\eta_i})^2 \prod_{p=2}^n \Omega(z_{p-1,p},\xi_p,\tau)\nn\\
  & \ \ \ \ \ \  -(-1)^{j-i+1}\sum_{k=i+1}^j \wp(\xi_k,\tau) \sum_{(a_1,\ldots,a_n)\in S_n(i,j,k)} \prod_{p=2}^n \Omega\bigg( z_{a_{p-1},a_p} , \sum_{\ell=p}^n \eta_{a_\ell}\bigg)
\end{align}
using~\eqref{eq:25}. Here, we set again $\partial_{\eta_1}=0$ and have introduced the shorthand $S_n(i,j,k)$ defined in~\eqref{eq:Snijk} for the sequences obtained by all possibles shuffles occurring in~\eqref{eq:sijform} and illustrated in Figure~\ref{fig:sijform}.

\medskip 

The proof of the $s_{ij}$-form~\eqref{eq:sijform} of the product of Kronecker--Eisenstein series proceeds by several lemmata.

\subsection{$s_{1n}$-form at $n$ points}

We begin with establishing the extreme case when $i=1$ and $j=n$ for $n$ points. The formula~\eqref{eq:sijform} then specializes to
\begin{align}
  \label{eq:s1nform}
  \prod_{p=2}^n \Omega(z_{p-1,p},\xi_p)&= (-1)^n \sum_{k=2}^n \Omega(z_{k-1,k-2},\xi_k-\xi_{k-1})\cdots \Omega(z_{21},\xi_k-\xi_2)\Omega(z_{1n},\xi_k)\nn\\
                                       &\hspace{10mm}\times \Omega(z_{n,n-1},\xi_k-\xi_n)\cdots \Omega(z_{k+1,k},\xi_k-\xi_{k+1})\,,
\end{align}
with only descending parts to the left and right of $\Omega(z_{1n},\xi_k)$.

This form can be proved by induction on $n$. For $n=2$ there is nothing to do. Assume then that~\eqref{eq:s1nform} holds for $n-1$ points. Then
\begin{align}
  \prod_{p=2}^n \Omega(z_{p-1,p},\xi_p) &= \left[\prod_{p=2}^{n-1} \Omega(z_{p-1,p},\xi_p)\right] \Omega(z_{n-1,n},\xi_n)\nn\\
                                        &= \Bigg[(-1)^{n-1} \sum_{k=2}^{n-1} \Omega(z_{k-1,k-2},\xi_k-\xi_{k-1})\cdots \Omega(z_{21},\xi_k-\xi_2)\Omega(z_{1,n-1},\xi_k)\nn\\
                                        &\hspace{10mm}\times \Omega(z_{n-1,n-2},\xi_k-\xi_{n-1})\cdots \Omega(z_{k+1,k},\xi_k-\xi_{k+1})\Bigg] \Omega(z_{n-1,n},\xi_n)\nn\\
                                        &= (-1)^{n} \sum_{k=2}^{n-1} \Omega(z_{k-1,k-2},\xi_k-\xi_{k-1})\cdots \Omega(z_{21},\xi_k-\xi_2)\Omega(z_{1,n},\xi_k)\nn\\
                                        &\hspace{10mm}\times \Omega(z_{n,n-1},\xi_k-\xi_{n})\Omega(z_{n-1,n-2},\xi_k-\xi_{n-1})\cdots \Omega(z_{k+1,k},\xi_k-\xi_{k+1})\nn\\
                                        &\hspace{5mm}+\Bigg[(-1)^{n-1} \sum_{k=2}^{n-1} \Omega(z_{k-1,k-2},\xi_k-\xi_{k-1})\cdots \Omega(z_{21},\xi_k-\xi_2)\Omega(z_{1,n-1},\xi_k-\xi_n)\nn\\
                                        &\hspace{10mm}\times \Omega(z_{n-1,n-2},\xi_k-\xi_{n-1})\cdots \Omega(z_{k+1,k},\xi_k-\xi_{k+1})\Bigg] \Omega(z_{1n},\xi_n)\nn\\
                                        &= (-1)^{n} \sum_{k=2}^{n} \Omega(z_{k-1,k-2},\xi_k-\xi_{k-1})\cdots \Omega(z_{21},\xi_k-\xi_2)\Omega(z_{1,n},\xi_k)\nn\\
                                        &\hspace{10mm}\times \Omega(z_{n,n-1},\xi_k-\xi_{n})\cdots \Omega(z_{k+1,k},\xi_k-\xi_{k+1})\,,
\end{align}
where we have used the Fay identity~\eqref{eq:FayO} in the form
\begin{align}
  \Omega(z_{1,n-1},\xi_k) \Omega(z_{n-1,n},\xi_n) = - \Omega(z_{1n},\xi_k) \Omega(z_{n,n-1},\xi_k-\xi_n) + \Omega(z_{1n},\xi_n) \Omega(z_{1,n-1},\xi_k-\xi_n)
\end{align}
and have used the induction hypothesis again at shifted $\xi$-values in the last step to convert the expression into the missing summand for $k=n$.

A corollary of~\eqref{eq:s1nform} is obtained simply by shifting the indices 
\begin{align}
  \label{eq:sijext}
  \prod_{p=i+1}^j \Omega(z_{p-1,p},\xi_p)&= (-1)^{j-i+1} \sum_{k=i+1}^j \Omega(z_{k-1,k-2},\xi_k-\xi_{k-1})\cdots \Omega(z_{i+1,i},\xi_k-\xi_{i+1})\Omega(z_{ij},\xi_k)\nn\\
                                         &\hspace{10mm}\times \Omega(z_{j,j-1},\xi_k-\xi_j)\cdots \Omega(z_{k+1,k},\xi_k-\xi_{k+1})\,.
\end{align}
With~\eqref{eq:sijext} we have an expression for the product of Kronecker--Eisenstein series between $i$ and $j$. In order to obtain the full product from $1$ to $n$ we need to extend to the left and right. This can be done with the help of two little lemmata.

\subsection{Extending left and right}

We now want to prove~\eqref{eq:sijform} by induction on $n$ which means extending the product on the left and on the right, beginning with the right.

We first consider keeping $i$ and $j$ fixed and extend~\eqref{eq:sijform} by multiplying by $\Omega(z_{n,n+1},\xi_{n+1})$ on the right. If $k=n$ there is nothing to do since the factors just multiplies correctly at the end of the sequence: If the original sequence is $(a_1,\ldots,a_n)$ with $a_n=n$ the new sequence is $(a_1,\ldots,a_n,a_{n+1})$ with $a_{n+1}=n+1$ and the $\xi$-factors are correct for the action of the differential operator.

If $k<n$ then there is an $m<n$ such that $a_m=n$. Moreover, $m\geq k$. This means that in the product~\eqref{eq:Oseq} somewhere in the tail to the right of $\Omega(z_{ij},\xi_k)$ there is the subproduct
\begin{align}
  \Omega[n,\{a_{m+1},\ldots,a_n\}] &:= \prod_{p=m+1}^n \Omega\bigg(z_{a_{p-1},a_p},\sum_{\ell=p}^n \eta_{a_{\ell}}\bigg)\nn\\
                                   &\phantom{:}= \Omega\bigg(z_{n,a_{m+1}}, \sum_{\ell=m+1}^n \eta_{a_{\ell}}\bigg)\Omega[a_{m+1},\{a_{m+2},\ldots,a_n\}]
\end{align}
and the only place where the index $n$ appears is in the very first factor that has been made explicit in the second line. Multiplying such an product by $\Omega(z_{n,n+1},\xi_{n+1})$ leads to splicing in the index $n+1$ in all possible places:
\begin{align}
  \label{eq:spliceend}
  \Omega[n,\{a_{m+1},\ldots,a_n\}] \Omega(z_{n,n+1},\xi_{n+1}) &= \Omega[n,\{a_{m+1},\ldots,a_n\} \shuffle \{n+1\}]\nn\\
                                                               &= \Omega[n,\{n+1,a_{m+1},\ldots, a_n\}] \nn\\
                                                               &\hspace{10mm}+ \Omega[n,\{a_{m+1},n+1,a_{m+2},\ldots , a_n\}] \nn\\
                                                               &\hspace{10mm}+\ldots+ \Omega[n,\{a_{m+1},\ldots,a_n,n+1\}]\,.
\end{align}
The assertion~\eqref{eq:spliceend} is proved by induction on the length $n-m$ of the original product. For $n-m=0$ this is trivially true as already stated above.

Now assume the formula~\eqref{eq:spliceend} is correct for products of length $n-m$. Then 
\begin{align}
  &\qquad\Omega[n,\{a_{m},\ldots,a_n\}]\Omega(z_{n,n+1},\xi_{n+1}) = \Bigg[ \Omega\bigg(z_{n,n+1},\sum_{\ell=m}^{n+1}\eta_{a_\ell}\bigg)\Omega\bigg(z_{n+1,a_{m}},\sum_{\ell=m}^n\eta_{a_\ell}\bigg)\nn\\
  &\hspace{30mm}+\Omega\bigg(z_{n,a_{m}},\sum_{\ell=m}^{n+1}\eta_{a_\ell}\bigg)\Omega(z_{a_{m},n+1},\xi_{n+1})\Bigg] \Omega[a_{m},\{a_{m+1},\ldots,a_n\}]\nn\\
  &= \Omega[n,\{n+1,a_{m+1},\ldots,a_n\}] +\Omega\bigg(z_{n,a_{m}},\sum_{\ell=m}^{n+1}\eta_{a_\ell}\bigg) \Omega[a_{m},\{a_{m+1},\ldots,a_n\}\shuffle\{n+1\}]\nn\\
  &=  \Omega[n,\{a_{m},\ldots,a_n\}\shuffle\{n+1\}]\,,
\end{align}
where we have used a Fay identity in the first step and the induction hypotheses in the second step and collected terms in the last line. This proves~\eqref{eq:spliceend}. 

Note that the index $n+1$ always ends up to the right of $n$ in this product and so the order is preserved for them. This means that~\eqref{eq:prodR} is the correct shuffle prescription for extending from $j=n$ to any $j<n$ by shuffling in the additional indices into the reversed indices to the right of $\Omega(z_{ij},\xi_k)$ in~\eqref{eq:sijext}.

By similar methods one can also show that multiplying by $\Omega(z_{01},\xi_1)$ to extend on the left shuffles in the index $0$ in all possible places to the left of the index $1$. After renaming the indices, we conclude that~\eqref{eq:prodL} is the correct shuffle prescription for extending from $i=1$ to any $i>1$ by shuffling in the indices $\{1,\ldots,i-1\}$ into the reversed indices to the left of $\Omega(z_{ij},\xi_k)$ in~\eqref{eq:sijext}.

The sequences constructed by this inductive method constitute the set $S_n(i,j,k)$ defined in~\eqref{eq:Snijk} and illustrated in Figure~\ref{fig:sijform}. This concludes the proof of~\eqref{eq:sijform}.

\section{Derivation of component equations at three points}
\label{app:3pt}

In this appendix, we present additional details on the Cauchy--Riemann and Laplace equation for the 
component integrals of the generating series $W_{\vec\eta}^\tau(\sigma|\rho)$ at three points, cf.\ Sections~\ref{sec:CR3} and \ref{sec:Lap3}.

In order to infer Cauchy--Riemann equations for the component integrals from those for the generating series \eqref{eq:13}, we need to expand both sides of the equation in the same $\eta$ variables. If the starting permutation in the chiral sector is $\rho(2,3)=(2,3)$, we expand the corresponding $\Omega(z_{12},\eta_{23},\tau)\Omega(z_{23},\eta_{3},\tau)$ in $\eta_{23}=\eta_{2}+\eta_{3}$ and $\eta_{3}$.
However, the right-hand side of \eqref{eq:13} also introduces the permutation $\alpha(2,3)=(3,2)$ with parameters 
$\eta_{32}=\eta_2+\eta_3$ and $\eta_2$ in the integrand $\Omega(z_{13},\eta_{32},\tau)\Omega(z_{32},\eta_{2},\tau)$ 
that differ from the ones on the left-hand side of~\eqref{eq:13}. 

This means that we need to expand
\begin{align}
  W_{\eta_{2},\eta_{3}}^\tau(\sigma|3,2)&=\sum_{a_{2},a_{3}=0}^{\infty}\sum_{b_{2},b_{3}=0}^{\infty}\eta_{23}^{a_{2}-1}\eta_{2}^{a_{3}-1}\sigma[\bar{\eta}_{23}^{b_{2}-1}\bar{\eta}_{3}^{b_{3}-1}]W^{\tau}_{(a_{2},a_{3}|b_{2},b_{3})}(\sigma|3,2)\label{eq:14}
\end{align}
in $\eta_{23}$ and $\eta_{3}$ for the $\rho(2,3)=(2,3)$ component of \eqref{eq:13}. We do this separately for the negative and all non-negative powers of $\eta_{2}$,
\begin{align}
\label{eq:GSE}
  \frac{1}{\eta_{2}}&=\frac{1}{\eta_{23}-\eta_{3}}=\sum_{n=0}^{\infty}\eta_{23}^{-n-1}\eta_{3}^{n}\\
  \eta_{2}^{a_{3}-1}&=(\eta_{23}-\eta_{3})^{a_{3}-1}=\sum_{k=0}^{a_{3}-1}\binom{a_{3}-1}{k}\eta_{23}^{k}(-\eta_{3})^{a_{3}-1-k}\qquad a_{3}\geq1\ .
\end{align}
Plugging this back into \eqref{eq:14} yields
\begin{align}
  W_{\eta_{2},\eta_{3}}^\tau(\sigma|3,2)&=\sum_{a_{2}=-\infty}^{\infty}\sum_{a_{3}=0}^{\infty}\sum_{b_{2},b_{3}=0}^{\infty}\eta_{23}^{a_{2}}\eta_{3}^{a_{3}}\sigma[\bar{\eta}_{23}^{b_{2}-1}\bar{\eta}_{3}^{b_{3}-1}]\Bigg(W_{(a_{2}+a_{3}+2,0|b_{2},b_{3})}^{\tau}(\sigma|3,2)\notag\\
                                          &\hspace{3em}+(-1)^{a_{3}}\sum_{k=0}^{a_{2}+1}\binom{a_{3}+k}{k}W^{\tau}_{(a_{2}+1-k,a_{3}+1+k|b_{2},b_{3})}(\sigma|3,2)\Bigg)\ .\label{eq:15}
\end{align}
Here and in the following, we set $W^{\tau}_{(a_{2},a_{3}|b_{2},b_{3})}(\sigma|\rho)=0$ for any of $a_{2},a_{3},b_{2},b_{3}$ negative. Note that \eqref{eq:15} naively introduces infinitely high poles at $\eta_{23}=0$ due to the geometric-series expansion~\eqref{eq:GSE}. However, most of these higher poles in $\eta_{23}$ do not enter the right-hand side 
of the differential equation (\ref{eq:13}) at $\rho(2,3)=(2,3)$ since the pole of 
$\Omega(z_{32},\eta_2,\tau) = \frac{1}{\eta_2}+\ldots$ is always canceled by the accompanying factor of
\begin{align}
\wp(\eta_{23} ,\tau) - \wp(\eta_3,\tau) &= \eta_2 \bigg\{ {-}\frac{1}{\eta_3^2 \eta_{23} }
-\frac{1}{\eta_3 \eta_{23}^2 }  + \sum_{k=4}^{\infty} (k{-}1) {\rm G}_k(\tau) \sum_{\ell=0}^{k-3} {k-2\choose \ell+1} \eta_2^{\ell} \eta_3^{k-3-\ell} 
\bigg\} \, .\label{comp2} 
\end{align}
The residues of any pole in $\eta_2,\eta_3,\eta_{23}$ on the right-hand side of (\ref{eq:13}) which is
not present on the left-hand side can be shown to vanish by integration-by-parts relations. These relations generalize 
the ones given for two points in~\eqref{ibp2a}, and our method can therefore also be seen as automatically 
generating them.


\subsection{General Cauchy--Riemann component equations at three points}
\label{subappF.1}

Using \eqref{eq:15}, we can now expand the $\rho(2,3)=(2,3)$ component of \eqref{eq:13} in $\eta_{23}$ and $\eta_{3}$ and find the following general Cauchy--Riemann equations of the component integrals
\begin{align}
  &2\pi i \nabla^{(a_{2}+a_{3})} W^{\tau}_{(a_{2},a_{3}|b_{2},b_{3})}(\sigma|2{,}3)=2\pi i\Big(a_{2}W^{\tau}_{(a_{2}+1,a_{3}|b_{2}-1,b_{3})}(\sigma|2{,}3){+}a_{3}W^{\tau}_{(a_{2},a_{3}+1|b_{2},b_{3}-1)}(\sigma|2{,}3)\Big)\notag\\
  &\hspace{1.5em}{+}(\tau{-}\bar{\tau})s_{12}\left[\frac{1}{2}(a_{2}{-}1)(a_{2}{+}2)W^{\tau}_{(a_{2}+2,a_{3}|b_{2},b_{3})}(\sigma|2{,}3){-}\!\!\sum_{k=4}^{a_{2}+2}(k{-}1)\mathrm{G}_{k}W^{\tau}_{(a_{2}-k+2,a_{3}|b_{2},b_{3})}(\sigma|2{,}3)\right]\nonumber\\
  &\hspace{1.5em}{+}(\tau{-}\bar{\tau})s_{23}\left[\frac{1}{2}(a_{3}{-}1)(a_{3}{+}2)W^{\tau}_{(a_{2},a_{3}+2|b_{2},b_{3})}(\sigma|2{,}3){-}\!\!\sum_{k=4}^{a_{3}+2}(k{-}1)\mathrm{G}_{k}W^{\tau}_{(a_{2},a_{3}-k+2|b_{2},b_{3})}(\sigma|2{,}3)\right]\nonumber\\
  &\hspace{1.5em}{+}(\tau{-}\bar{\tau})s_{13}\,\Bigg[\frac{1}{2}(a_{3}{-}1)(a_{3}{+}2)W^{\tau}_{(a_{2},a_{3}+2|b_{2},b_{3})}(\sigma|2{,}3){+}\frac{1}{2}a_{2}(a_{2}{+}1)W^{\tau}_{(a_{2}+2,a_{3}|b_{2},b_{3})}(\sigma|2{,}3)\nonumber\\
  &\hspace{2.3em} {+} a_{2}a_{3}W^{\tau}_{(a_{2}+1,a_{3}+1|b_{2},b_{3})}(\sigma|2{,}3){+}(-1)^{a_{3}-1}\sum_{k=0}^{a_{2}+2}\binom{a_{3}{-}1{+}k}{k}W^{\tau}_{(a_{2}+2-k,a_{3}+k|b_{2},b_{3})}(\sigma|3{,}2)\nonumber\\
  &\hspace{2.3em}{+}(-1)^{a_{3}}\sum_{k=0}^{a_{2}}\binom{a_{3}{+}1{+}k}{k}W^{\tau}_{(a_{2}-k,a_{3}+2+k|b_{2},b_{3})}(\sigma|3{,}2){-}\!\!\sum_{k=4}^{a_{3}+2}(k{-}1)\mathrm{G}_{k}W^{\tau}_{(a_{2},a_{3}-k+2|b_{2},b_{3})}(\sigma|2{,}3)\nonumber\\
  &\hspace{2.3em}{+}\Theta(a_{3}{-}1)\sum_{k=\text{max}(a_{3}+2,4)}^{a_{2}+a_{3}+2}(k{-}1)\mathrm{G}_{k}W^{\tau}_{(a_{2}+a_{3}+2-k,0|b_{2},b_{3})}(\sigma|3{,}2)\nonumber\\
  &\hspace{2.3em}{+}\Theta(a_{3}{-}1)(-1)^{a_{3}-1}\sum_{j=4}^{a_{2}+2}\sum_{k=0}^{a_{2}+2-j}\binom{a_{3}{-}1{+}k}{k}(j{-}1)\mathrm{G}_{j}W^{\tau}_{(a_{2}+2-j-k,a_{3}+k|b_{2},b_{3})}(\sigma|3{,}2)\nonumber\\
  &\hspace{2.3em}{+}\left.\sum_{j=4}^{a_{3}+1}\sum_{k=0}^{a_{2}}(-1)^{a_{3}-j}\binom{a_{3}{+}1{-}j{+}k}{k}(j{-}1)\mathrm{G}_{j}W^{\tau}_{(a_{2}-k,a_{3}+2+k-j|b_{2},b_{3})}(\sigma|3{,}2)\right]\label{eq:16}\ ,
\end{align}
where the Heaviside function $\Theta(x)= 1$ ($x\geq 0$) and $\Theta(x)=0$ ($x<0$) imposes intermediate selection criteria from re-expanding in the correct $\eta$-variables. 


\subsection{Examples of Cauchy--Riemann component equations at three points}
\label{subappF.2}

At low weights, \eqref{eq:16} implies for example the Cauchy--Riemann equations (\ref{eq:17main}) and
\begin{align}
  2\pi i \nabla^{(1)} W^{\tau}_{(0,1|b_{2},b_{3})}(\sigma|2{,}3)&=2\pi iW^{\tau}_{(0,2|b_{2},b_{3}-1)}(\sigma|2{,}3)-(\tau-\bar{\tau})s_{12}W^{\tau}_{(2,1|b_{2},b_{3})}(\sigma|2{,}3)\notag\\
                                                                    &\hspace{-9em}+(\tau-\bar{\tau})s_{13}\Big(W^{\tau}_{(1,2|b_{2},b_{3})}(\sigma|3{,}2)+W^{\tau}_{(2,1|b_{2},b_{3})}(\sigma|3{,}2)\Big)\notag\\[1em]
  2\pi i \nabla^{(2)} W^{\tau}_{(2,0|b_{2},b_{3})}(\sigma|2{,}3)&=4\pi iW^{\tau}_{(3,0|b_{2}-1,b_{3})}(\sigma|2{,}3)-(\tau-\bar{\tau})s_{23}W^{\tau}_{(2,2|b_{2},b_{3})}(\sigma|2{,}3)\notag\\
                                                                    &\hspace{-9em}+(\tau-\bar{\tau})s_{12}\Big(2W^{\tau}_{(4,0|b_{2},b_{3})}(\sigma|2{,}3)-3 \mathrm{G}_{4}W^{\tau}_{(0,0|b_{2},b_{3})}(\sigma|2{,}3)\Big)\notag\\                                                                   &\hspace{-9em}+(\tau-\bar{\tau})s_{13}\Big(W^{\tau}_{(2,2|b_{2},b_{3})}(\sigma|3{,}2)-W^{\tau}_{(2,2|b_{2},b_{3})}(\sigma|2{,}3)+3W^{\tau}_{(4,0|b_{2},b_{3})}(\sigma|2{,}3)\notag\\
                                                                    &\hspace{-3em}+3W^{\tau}_{(0,4|b_{2},b_{3})}(\sigma|3{,}2)+2W^{\tau}_{(1,3|b_{2},b_{3})}(\sigma|3{,}2)-W^{\tau}_{(4,0|b_{2},b_{3})}(\sigma|3{,}2)\Big)\notag\\[1em]
  2\pi i \nabla^{(2)} W^{\tau}_{(0,2|b_{2},b_{3})}(\sigma|2{,}3)&=4\pi iW^{\tau}_{(0,3|b_{2},b_{3})}(\sigma|2{,}3)-(\tau-\bar{\tau})s_{12}W^{\tau}_{(2,2|b_{2},b_{3})}(\sigma|2{,}3)\notag\\
                                                                    &\hspace{-9em}+(\tau-\bar{\tau})s_{23}\Big(2W^{\tau}_{(0,4|b_{2},b_{3})}(\sigma|2{,}3)-3 \mathrm{G}_{4}W^{\tau}_{(0,0|b_{2},b_{3})}(\sigma|2{,}3)\Big)\notag\\                                                                   &\hspace{-9em}+(\tau-\bar{\tau})s_{13}\Big(2W^{\tau}_{(0,4|b_{2},b_{3})}(\sigma|2{,}3)-2W^{\tau}_{(0,4|b_{2},b_{3})}(\sigma|3{,}2)-2W^{\tau}_{(1,3|b_{2},b_{3})}(\sigma|3{,}2)\notag\\
                                                                    &\hspace{-3em}-W^{\tau}_{(2,2|b_{2},b_{3})}(\sigma|3{,}2)+3 \mathrm{G}_{4}W^{\tau}_{(0,0|b_{2},b_{3})}(\sigma|3{,}2)\Big)\label{eq:17}\\[1em]
  2\pi i \nabla^{(2)} W^{\tau}_{(1,1|b_{2},b_{3})}(\sigma|2{,}3)&=2\pi i\Big(W^{\tau}_{(2,1|b_{2}-1,b_{3})}(\sigma|2{,}3)+W^{\tau}_{(1,2|b_{2},b_{3}-1)}(\sigma|2{,}3)\Big)\notag\\
                                                                    &\hspace{-9em}+(\tau-\bar{\tau})s_{13}\Big(W^{\tau}_{(2,2|b_{2},b_{3})}(\sigma|2{,}3)+W^{\tau}_{(3,1|b_{2},b_{3})}(\sigma|2{,}3)-2W^{\tau}_{(0,4|b_{2},b_{3})}(\sigma|3{,}2)\notag\\
                                                                    &\hspace{-3em}+W^{\tau}_{(2,2|b_{2},b_{3})}(\sigma|3{,}2)+W^{\tau}_{(3,1|b_{2},b_{3})}(\sigma|3{,}2)+W^{\tau}_{(3,1|b_{2},b_{3})}(\sigma|3{,}2)\notag\\
                                                                    &\hspace{-3em}+3 \mathrm{G}_{4}W^{\tau}_{(0,0|b_{2},b_{3})}(\sigma|3{,}2)\Big)\notag\\[1em]
  2\pi i \nabla^{(3)} W^{\tau}_{(3,0|b_{2},b_{3})}(\sigma|2{,}3)&=6\pi i W^{\tau}_{(4,0|b_{2}-1,b_{3})}(\sigma|2{,}3)-(\tau-\bar{\tau})s_{23}W^{\tau}_{(3,2|b_{2},b_{3})}(\sigma|2{,}3)\notag\\
                                                                    &\hspace{-9em}+(\tau-\bar{\tau})s_{12}\Big(5W^{\tau}_{(5,0|b_{2},b_{3})}-3 \mathrm{G}_{4}W^{\tau}_{(1,0|b_{2},b_{3})}(\sigma|2{,}3)\Big)\notag\\
                                                                    &\hspace{-9em}+(\tau-\bar{\tau})s_{13}\Big(2W^{\tau}_{(2,3|b_{2},b_{3})}(\sigma|3{,}2)+W^{\tau}_{(3,2|b_{2},b_{3})}(\sigma|3{,}2)-W^{\tau}_{(3,2|b_{2},b_{3})}(\sigma|2{,}3)\notag\\
                                                                    &\hspace{-3em}+6W^{\tau}_{(5,0|b_{2},b_{3})}(\sigma|2{,}3)+4W^{\tau}_{(0,5|b_{2},b_{3})}(\sigma|3{,}2)-W^{\tau}_{(5,0|b_{2},b_{3})}(\sigma|3{,}2)\notag\\
                                                                    &\hspace{-3em}+3W^{\tau}_{(1,4|b_{2},b_{3})}(\sigma|3{,}2)\Big)\notag\\[1em]
  2\pi i \nabla^{(4)} W^{\tau}_{(4,0|b_{2},b_{3})}(\sigma|2{,}3)&=8\pi i W^{\tau}_{(5,0|b_{2}-1,b_{3})}(\sigma|2{,}3)-(\tau-\bar{\tau})s_{23}W^{\tau}_{(4,2|b_{2},b_{3})}(\sigma|2{,}3)\notag\\
                                                                    &\hspace{-9em}+(\tau-\bar{\tau})s_{12}\Big(9W^{\tau}_{(6,0|b_{2},b_{3})}(\sigma|2{,}3)-3 \mathrm{G}_{4}W^{\tau}_{(2,0|b_{2},b_{3})}(\sigma|2{,}3)-5 \mathrm{G}_{6}W^{\tau}_{(0,0|b_{2},b_{3})}(\sigma|2{,}3)\Big)\notag\\
                                                                    &\hspace{-9em}+(\tau-\bar{\tau})s_{13}\Big(W^{\tau}_{(4,2|b_{2},b_{3})}(\sigma|3{,}2)-W^{\tau}_{(4,2|b_{2},b_{3})}(\sigma|2{,}3)+3W^{\tau}_{(2,4|b_{2},b_{3})}(\sigma|3{,}2)\notag\\
                                                                    &\hspace{-3em}+10W^{\tau}_{(6,0|b_{2},b_{3})}(\sigma|2{,}3)+5W^{\tau}_{(0,6|b_{2},b_{3})}(\sigma|3{,}2)-W^{\tau}_{(6,0|b_{2},b_{3})}(\sigma|3{,}2)\notag\\
                                                                    &\hspace{-3em}+4W^{\tau}_{(1,5|b_{2},b_{3})}(\sigma|3{,}2)+2W^{\tau}_{(3,3|b_{2},b_{3})}(\sigma|3{,}2)\Big)\notag\ .
\end{align}
In Section \ref{sec:CR3.2} we discuss one particular instance of this involving 
trihedral modular graph forms at the lowest order in $\ap$.


\subsection{Further examples for Laplace equations at three points}\label{sec:3ptLaExpls}

In this section, we provide two more examples of Laplace equations for three-point component integrals similar to (\ref{eq:LapW3}) for $W_{(0,0|0,0)}^{\tau }(2{,}3|2{,}3)$. By expanding both sides of~\eqref{eq:Lapn3} in the $\eta$ and $\bar{\eta}$ variables as discussed in Section~\ref{sec:Lap3}, we obtain
\begin{align}
&(2\pi i)^2 \Delta^{(1,1)} W_{(1,0|0,1)}^{\tau }(2{,}3|2{,}3) =
 4\pi^2 W_{(1, 0|0, 1)}(2{,}3|2{,}3) \notag \\
 & \ \ 
 + 2\pi i (\tau {-} \bar \tau) \Big\{
 {-}   s_{12}   W_{(2, 0|1, 1)}(2{,}3|2{,}3)
-    s_{23}   W_{(1, 1|0, 2)}(2{,}3|2{,}3)
+ s_{13}   W_{(2, 0|1, 1)}(3{,}2|2{,}3)
 \notag \\
& \ \ \ \ 
-  s_{13}  \big[ W_{(0, 2|0, 2)}(2{,}3|2{,}3)
{+} W_{(1, 1|0, 2)}(2{,}3|2{,}3)
{+}  W_{(1, 1|0, 2)}(2{,}3|3{,}2)
{+} W_{(2, 0|0, 2)}(2{,}3|3{,}2)
\big]
\Big\} \notag \\
& \ \ + (\tau {-}\bar \tau)^2 \Big\{
  s_{13}^2   W_{(3, 0|1, 2)}(3{,}2|2{,}3)
+  s_{13}^2   W_{(3, 0|2, 1)}(3{,}2|2{,}3)
+  s_{13}^2   W_{(1, 2|1, 2)}(3{,}2|3{,}2)\notag \\
& \ \  \ \ 
+  s_{13}^2   W_{(1, 2|2, 1)}(3{,}2|3{,}2)
-  s_{13}^2   W_{(3, 0|1, 2)}(3{,}2|3{,}2)
-  s_{13}^2   W_{(3, 0|2, 1)}(3{,}2|3{,}2)
 \\
& \ \  \ \ 
 -  s_{13}^2   W_{(1, 2|1, 2)}(3{,}2|2{,}3)
-  s_{13}^2   W_{(1, 2|2, 1)}(3{,}2|2{,}3)
 - 2 s_{13}^2   W_{(0, 3|1, 2)}(3{,}2|2{,}3)
\notag \\
& \ \  \ \ 
-  2 s_{13}^2   W_{(0, 3|2, 1)}(3{,}2|2{,}3)
-  s_{12} s_{13}   W_{(3, 0|2, 1)}(2{,}3|2{,}3)
-  s_{12} s_{13}   W_{(1, 2|2, 1)}(2{,}3|3{,}2)\notag \\
& \ \  \ \ 
+  s_{12} s_{13}   W_{(3, 0|2, 1)}(2{,}3|3{,}2)
+  s_{12} s_{13}   W_{(1, 2|2, 1)}(2{,}3|2{,}3)
+ 2 s_{12} s_{13}   W_{(0, 3|2, 1)}(2{,}3|2{,}3)\notag \\
& \ \  \ \ 
+  s_{12} s_{23}   W_{(1, 2|2, 1)}(2{,}3|2{,}3)
-  s_{13} s_{23}   W_{(1, 2|1, 2)}(3{,}2|2{,}3)
-  s_{13} s_{23}   W_{(1, 2|2, 1)}(3{,}2|2{,}3)
\Big\} \notag
\end{align}
and
\begin{align}
&(2\pi i)^2\Delta^{(2,0)} W_{(2,0|0,0)}^{\tau }(2{,}3|2{,}3) =
-  4 \pi i (\tau {-} \bar \tau)\big[ s_{12}   W_{(3, 0|1, 0)}(2{,}3|2{,}3)
     + s_{13}   W_{(3, 0|1, 0)}(3{,}2|2{,}3)  \big] \notag \\
& \ \ +
3 s_{12} {\rm G}_4 (\tau {-} \bar \tau)^2 \big[ s_{23}   W_{(0, 0|0, 2)}(2{,}3|2{,}3)
+   s_{12}    W_{(0, 0|2, 0)}(2{,}3|2{,}3)
+   s_{13}    W_{(0, 0|2, 0)}(3{,}2|2{,}3) \big] \notag \\
& \ \ + (\tau {-} \bar \tau)^2     \Big\{
 s_{23}^2   W_{(2, 2|0, 2)}(2{,}3|2{,}3)
 -  2 s_{12}^2   W_{(4, 0|2, 0)}(2{,}3|2{,}3)
-  s_{13}^2   W_{(2, 2|2, 0)}(3{,}2|3{,}2)\notag \\
& \ \  \ \ 
+  s_{13}^2   W_{(4, 0|2, 0)}(3{,}2|3{,}2)
+  s_{13}^2   W_{(2, 2|2, 0)}(3{,}2|2{,}3)
-  2 s_{13}^2   W_{(1, 3|2, 0)}(3{,}2|3{,}2)\notag \\
& \ \  \ \ 
-  3 s_{13}^2   W_{(4, 0|2, 0)}(3{,}2|2{,}3)
-  3 s_{13}^2   W_{(0, 4|2, 0)}(3{,}2|2{,}3)
+  s_{12} s_{13}   W_{(2, 2|2, 0)}(2{,}3|2{,}3)  \label{eq:42}\\
& \ \  \ \ 
-  s_{12} s_{13}   W_{(2, 2|2, 0)}(2{,}3|3{,}2)
+  s_{12} s_{13}   W_{(4, 0|2, 0)}(2{,}3|3{,}2)
-  2 s_{12} s_{13}   W_{(4, 0|2, 0)}(3{,}2|2{,}3)\notag \\
& \ \  \ \ 
 -  2 s_{12} s_{13}   W_{(1, 3|2, 0)}(2{,}3|3{,}2)
-  3 s_{12} s_{13}   W_{(0, 4|2, 0)}(2{,}3|2{,}3)
 -  3 s_{12} s_{13}   W_{(4, 0|2, 0)}(2{,}3|2{,}3)\notag \\
& \ \  \ \ 
+  s_{12} s_{23}   W_{(2, 2|2, 0)}(2{,}3|2{,}3)
-  2 s_{12} s_{23}   W_{(4, 0|0, 2)}(2{,}3|2{,}3)
+  s_{13} s_{23}   W_{(2, 2|0, 2)}(2{,}3|2{,}3)\notag \\
& \ \  \ \ 
-  s_{13} s_{23}   W_{(2, 2|0, 2)}(2{,}3|3{,}2)
+  s_{13} s_{23}   W_{(4, 0|0, 2)}(2{,}3|3{,}2)
+  s_{13} s_{23}   W_{(2, 2|2, 0)}(3{,}2|2{,}3)\notag \\
& \ \  \ \ 
-  2 s_{13} s_{23}   W_{(1, 3|0, 2)}(2{,}3|3{,}2)
 -  3 s_{13} s_{23}   W_{(0, 4|0, 2)}(2{,}3|2{,}3)
-  3 s_{13} s_{23}   W_{(4, 0|0, 2)}(2{,}3|2{,}3) 
\Big\}\,. \notag
\end{align}

\bibliographystyle{JHEP}

\providecommand{\href}[2]{#2}\begingroup\raggedright\endgroup

\end{document}